\documentclass{article}

\usepackage{arxiv}

\usepackage[utf8]{inputenc} 
\usepackage[T1]{fontenc}    
\usepackage{hyperref}       
\usepackage{url}            
\usepackage{booktabs}       
\usepackage{amsfonts}       
\usepackage{nicefrac}       
\usepackage{microtype}      
\usepackage{enumerate}      
\usepackage{graphicx}
\usepackage{natbib}
\usepackage{doi}

\usepackage{xcolor}
\hypersetup{
  colorlinks=true,
  linkcolor=red,
  citecolor=orange,
  urlcolor=blue,
}

\graphicspath{{../figure}}

\title{Pattern wavelengths and transport characteristics in three--dimensional bioconvection 
generated by chemotactic bacteria
}

\author{\href{https://orcid.org/0000-0002-4875-8174}{\includegraphics[scale=0.06]{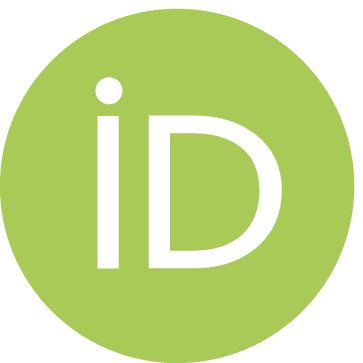}\hspace{1mm}Hideki Yanaoka}
\thanks{Email address for correspondence: yanaoka@iwate-u.ac.jp} \\
	Department of Systems Innovation Engineering, \\
    Faculty of Science and Engineering, Iwate University, \\
    4-3-5 Ueda, Morioka1, Iwate 202-8551, Japan \\
	\And
	Tomomu Nishimura \\
    Kawasaki Heavy Industries, Ltd., \\
    1-14-5 Kaigan, Minato-ku, Tokyo 105‐8315, Japan
	\texttt{} \\
}

\renewcommand{\shorttitle}{Pattern wavelengths and transport characteristics in three--dimensional bioconvection}

\makeatletter
\hypersetup{
pdfusetitle,
bookmarksnumbered,
pdftitle={\@title},
pdfsubject={},
pdfauthor={Hideki Yanaoka, Tomomu Nishimura},
pdfkeywords={Bioconvection, Pattern formation, Nonlinear instability},
}

\date{4 July 2022}
\makeatother

\begin{document}

\maketitle

\begin{abstract}
We conducted a three-dimensional numerical simulation of bioconvection 
generated by oxygen-reactive chemotactic bacteria. 
This study investigated the bioconvection patterns, interference between plumes, 
and the wavelength of bioconvection patterns. 
In addition, we clarified the transport characteristics of cells and oxygen 
in the bioconvection. 
Multiple plumes occur in the suspension and 
three-dimensional bioconvection is formed around the plumes by the cells with 
vortex rings arising around the plumes. 
Even if bioconvection at a high Rayleigh number is disturbed, 
the bioconvection is strongly stable with respect to disturbances, 
and the pattern does not change due to disturbances. 
Bioconvection changes depending on the physical properties of bacteria and oxygen, 
and, in particular, the rate of oxygen consumption by bacteria 
significantly affects the strength of bioconvection. 
Bioconvection patterns with different plume arrangements and shapes 
are formed for different Rayleigh numbers or initial disturbances 
of the cell concentration. 
As a result, the wavelengths of the patterns also vary. 
As the Rayleigh number increases, interference between plumes is strengthened 
by the shortening of the pattern wavelength, 
so the velocities of both upward and downward flows increase. 
Many cells are located under the plumes and a strong shear flow occurs in these regions. 
As the pattern wavelength decreases, the cells are affected 
by high shear stress. 
Then, the convective transport of the entire suspension strengthens 
and the transport characteristics of cells and oxygen improve. 
When the chamber boundary is changed to side walls, 
bacteria adhere to the wall surface, 
and plumes are regularly arranged along the side walls.
\end{abstract}



\section{Introduction}

Microorganisms inhabit the Earth in a vast variety of environments \citep{Omori_et_al_2003}. 
Some of these microorganisms respond to outside stimuli 
and move in selected directions. 
These responses are called taxis. 
Taxis responding to gravity, light, and chemicals are respectively called 
gravitaxis, phototaxis, and chemotaxis. 
In suspension, when a certain quantity of microorganisms accumulates 
near a free surface due to taxis, 
cells of the microorganisms fall, and thus bioconvection is generated \citep{Platt_1961} 
because the cells are denser than water \citep{Hart&Edwards_1987}.

Many studies have been made on bioconvection 
\citep{Pedley&Kessler_1992, Hillesdon_et_al_1995, Hillesdon&Pedley_1996, 
Bees&Hill_1997, Metcalfe&Pedley_1998, Czirok_et_al_2000, Ghorai&Hill_2002, 
Metcalfe&Pedley_2001, Yanaoka_et_al_2007, Yanaoka_et_al_2008, 
Williams&Bees_2011, Chertock_et_al_2012, Kage_et_al_2013, Karimi&Paul_2013}. 
Microorganisms have been used for environmental cleanup in various fields 
\citep{Omori_et_al_2000, Hirooka&Nagase_2003, Omori_et_al_2003}. 
Bioconvection can be applied to driving micromechanical systems 
\citep{Itoh_et_al_2001, Itoh_et_al_2006}, 
mixing chemicals \citep{Geng&Kuznetsov_2005}, 
detecting toxicity \citep{Noever&Matsos_1991a, Noever&Matsos_1991b, Noever_et_al_1992}, 
controlling microorganisms in biochips, as well as other applications. 
Furthermore, the production of biofuels by microorganisms has attracted 
attention from the viewpoint of being clean and environmentally friendly. 
\citet{Bees&Croze_2014} discussed the possibility that microorganisms with taxis 
may be able to mix biofuel-producing microorganisms efficiently. 
Therefore, for the efficient utilization of microorganisms 
in various fields, it is important to determine the behavior of microorganisms 
and the mass transfer characteristics in bioconvection generated 
by microorganisms with taxis.

Regarding bioconvection generated by chemotactic bacteria responding to oxygen, 
theoretical \citep{Hillesdon&Pedley_1996, Metcalfe&Pedley_1998, Metcalfe&Pedley_2001}, 
experimental \citep{Pedley&Kessler_1992, Hillesdon_et_al_1995, Czirok_et_al_2000}, 
and numerical studies \citep{Yanaoka_et_al_2007, Yanaoka_et_al_2008, Chertock_et_al_2012, Lee&Kim_2015} 
have been conducted. 
However, earlier numerical studies were two-dimensional analyses or 
three-dimensional analyses on a single plume with a fixed wavelength \citep{Yanaoka_et_al_2007, Yanaoka_et_al_2008}. 
Hence, it was not possible to capture complicated three-dimensional phenomena 
that multiple plumes exhibit in the chamber. 
Although basic studies on the wavelength of bioconvection patterns 
have been carried out 
\citep{Bees&Hill_1997, Czirok_et_al_2000, Ghorai&Hill_2002, Karimi&Paul_2013}, 
no studies have been reported on the influences of wavelength variations 
on the transport characteristics and interference between plumes.

Studies on the control of bioconvection have been performed to utilize microorganisms 
for engineering purposes 
\citep{Itoh_et_al_2001, Itoh_et_al_2006, Kuznetsov_2005a}. 
Furthermore, various studies have been conducted on nano-bioconvection 
in suspensions containing nanoparticles and bacteria 
\citep{Kuznetsov_2011, Geng&Kuznetsov_2005, Uddin_et_al_2016, Zadeha_et_al_2020}. 
Recently, bioconvection in suspensions containing microorganisms 
and nanoparticles under an applied magnetic field has been investigated 
\citep{Naseem_et_al_2017, Khan_et_al_2020, Shi_et_al_2021}. 
New applicative research on bioconvection is underway. 
However, in the previous studies, the phenomenon of three-dimensional bioconvection 
has not been captured because a stability analysis was performed 
or the fundamental equations were solved by similarity transformation. 
Bioconvection with multiple microbial plumes is complex, 
and the details of the transport characteristics in bioconvection have not been clarified.

Bioconvection is a three-dimensional phenomenon and multiple plumes will exist 
in a chamber. 
Under such circumstances, the plumes interfere with each other 
and the accompanying change in the wavelength of the bioconvection pattern 
affects the transport characteristics. 
From the above viewpoints, we simulate three-dimensional bioconvection 
generated by oxygen-reactive chemotactic bacteria, 
and then clarify the bioconvection patterns, 
interference between plumes, the wavelengths of each pattern, 
and the transport characteristics of cells and oxygen when multiple plumes arise.

\section{Numerical Procedures}

\begin{figure}
\centering
\includegraphics[trim=0mm 0mm 0mm 0mm, clip, width=80mm]{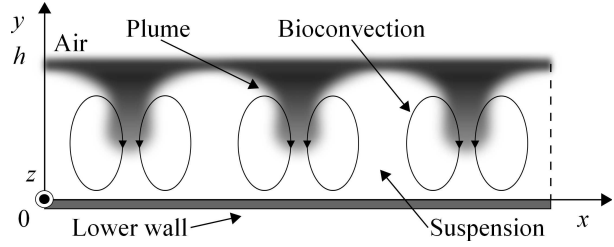} \\
\caption{Flow configuration and coordinate system}
\label{flow_model}
\end{figure}

Figure \ref{flow_model} shows the flow configuration and coordinate system. 
A suspension in a chamber contains bacterial cells, 
and the depth of the suspension is $h$. 
The origin is at the bottom wall of the chamber. 
The $x$- and $y$-axes are in the horizontal and vertical direction, respectively, 
and the $z$-axis is in the direction perpendicular to the page.

We assume that the suspension is sufficiently dilute 
for hydrodynamic cell--cell interactions to be negligible, 
and consider an incompressible viscous fluid. 
The fundamental equations are the continuity equation, 
the momentum equation under the Boussinesq approximation, 
and the conservation equations for cells and oxygen 
\citep{Hillesdon_et_al_1995, Hillesdon&Pedley_1996}, given as
\begin{equation}
   \nabla \cdot {\bf u}=0
\end{equation}
\begin{equation}
   \frac{\partial {\bf u}}{\partial t} + \nabla \cdot ({\bf u} \otimes {\bf u}) 
   = - \frac{1}{\rho} \nabla p + \nu \nabla^{2} {\bf u} 
     + \frac{g n V (\rho_n - \rho) {\bf e}}{\rho}
\end{equation}
\begin{equation}
   \frac{\partial n}{\partial t} 
   + \nabla \cdot ({\bf u} n + {\bf V} n - {\bf D}_n \nabla n) = 0
\end{equation}
\begin{equation}
   \frac{\partial c}{\partial t} 
   + \nabla \cdot ({\bf u} c - D_c \nabla c) = - K n
\end{equation}
where $t$ is time, ${\bf u}$ is flow velocity, $p$ is pressure, 
$\rho$ and $\nu$ are the density and kinematic viscosity of the fluid, respectively, 
$\rho_n$ is the density of the cell, $V$ is the volume of a cell, 
$g$ is the gravity acceleration, 
${\bf e}=-\hat{\bf y}$ is the unit vector in the direction of gravity, 
$n$ is cell concentration, $c$ is oxygen concentration, 
${\bf V}$ is the average cell swimming velocity, 
${\bf D}_n$ is the cell diffusivity tensor, $D_c$ is the oxygen diffusivity, 
and $K$ is the rate of oxygen consumption by cells.

In this study, the swimming of microorganisms is modelled similarly to 
that of \citet{Hillesdon_et_al_1995}. 
\citet{Berg&Brown_1972} found that the swimming velocity of microorganisms 
has both directional and random components. 
The random swimming of the cell is modelled as cell diffusion. 
The cell diffusion is assumed to be isotropic and the cell diffusivity tensor 
${\bf D}_n$ is modelled as ${\bf D}_n=D_{n0} H(c^*) {\bf I}$. 
Here, $c^*$ is the non-dimensional oxygen concentration, 
$H(c^*)$ is a step function, and ${\bf I}$ is the identity tensor. 
$c^*$ is defined as 
$c^*=(c-c_\mathrm{min})/(c_0-c_\mathrm{min})$, 
where $c_0$ is the initial oxygen concentration, 
and $c_\mathrm{min}$ is the minimum oxygen concentration required 
for the cell to be active. 
Next, the directional swimming of the cell is modelled as an average swimming velocity. 
The average cell swimming velocity vector ${\bf V}$ is modelled as being proportional 
to the oxygen concentration gradient and is defined as 
${\bf V}=b V_s H(c^*) \nabla c^*$. 
The oxygen consumption rate $K$ by cells is modelled as $K=K_0 H(c^*)$ 
like that of \citet{Hillesdon_et_al_1995}. 
Here, $D_{n0}$, $b$, $V_s$, and $K_0$ are constants. 
In this study, $H(c^*)$ is modelled with an approximate equation, 
$H(c^*)=1-\exp(-c^*/c^*_1)$, to suppress discontinuous changes 
due to the step function \citep{Hillesdon_et_al_1995}. 
Here, $c^*_1$ is 0.01 \citep{Hillesdon_et_al_1995, Yanaoka_et_al_2007, Yanaoka_et_al_2008}.
\citet{Hillesdon_et_al_1995} investigated the effect of $c_1^*$ 
on the concentration distribution in a stationary field, 
and the qualitative trend did not change. 
In the present study, $c_1^*$ is set to a small value. 
Therefore, in a shallow chamber treated in the present study, 
since oxygen is transported to the bottom, 
the step function $H(c^*)$ is approximately 1.0, 
and the modelling of $H(c^*)$ does not affect the calculation results.

A non-slip boundary condition is set 
at the bottom wall 
and the gradients perpendicular to the wall are to be assumed zero 
for the cell and oxygen concentrations. 
At the free surface, a slip boundary condition is imposed for the velocity field, 
the flux of cell concentration is zero, 
and the oxygen concentration is constant. 
Further, periodic boundary conditions are imposed in the $x$- and $z$-directions 
for the velocity and concentration fields.

The variables of the fundamental equations are non-dimensionalized 
in the same way as in previous studies 
\citep{Hillesdon_et_al_1995, Yanaoka_et_al_2007, Yanaoka_et_al_2008} as follows.
\begin{equation}
   {\bf x}^* = \frac{{\bf x}}{h}, \quad 
   {\bf u}^* = \frac{{\bf u}}{D_{n0}/h}, \quad 
   p^* = \frac{p}{\rho (D_{n0}/h)^2}, \quad 
   t^* = \frac{t}{h^2/D_{n0}}, \quad 
   n^* = \frac{n}{n_0},
\end{equation}
where the superscript $*$ represents a non-dimensional variable, 
and $n_0$ is the initial cell concentration. 
Thus, the non-dimensional fundamental equations are written as
\begin{equation}
   \nabla \cdot {\bf u}^* = 0
   \label{continuity}
\end{equation}
\begin{equation}
   \frac{1}{Sc} \left[ \frac{\partial {\bf u}^*}{\partial t^*} 
   + \nabla \cdot ({\bf u}^* \otimes {\bf u}^*) \right] 
   = - \nabla p^* + \nabla^{2} {\bf u}^* + Ra n^* {\bf e}
   \label{navier-stokes}
\end{equation}
\begin{equation}
   \frac{\partial n^*}{\partial t^*} 
   + \nabla \cdot \Bigl[{\bf u}^* n^* + H(c^*) \gamma n^* \nabla c^* - H(c^*) \nabla n^*\Bigr] = 0
   \label{cell_concentration}
\end{equation}
\begin{equation}
   \frac{\partial c^*}{\partial t^*} 
   + \nabla \cdot ({\bf u}^* c^* - \delta \nabla c^*) = - H(c^*) \delta \beta n^*
   \label{oxygen_concentration}
\end{equation}
where the non-dimensional parameters are given by equation (\ref{parameter}). 
The parameter $\beta$ represents the strength of oxygen consumption 
relative to its diffusion, 
$\gamma$ is a measure of the relative strengths of directional 
and random swimming, 
and $\delta$ is the ratio of oxygen diffusivity to cell diffusivity. 
$Ra$ is the Rayleigh number and $Sc$ is the Schmidt number.
\begin{eqnarray}
   \beta = \frac{K_0 n_0 h^2}{D_c (c_0 - c_{\mathrm{min}})}, \, 
   \gamma = \frac{b V_s}{D_{n0}}, \, 
   \delta = \frac{D_c}{D_{n0}}, \, 
   Ra = \frac{V n_0 g h^3 (\rho_n - \rho)}{\nu D_{n0} \rho}, \, 
   Sc = \frac{\nu}{D_{n0}}
   \label{parameter}
\end{eqnarray}

The governing equations are solved using the SMAC method \citep{Amsden&Harlow_1970}. 
This study uses the Euler implicit method for the time differentials 
and the second-order central difference scheme for the space differentials.

\section{Calculation conditions}

This study considers chemotactic bacteria responding to oxygen 
as microorganisms with taxis, 
and investigates the characteristics of three-dimensional bioconvection 
in the chamber. 
As the initial condition, the suspension is stationary, 
and the cell and oxygen concentrations are constant. 
Similar to the previous study \citep{Ghorai&Hill_2002}, 
low initial disturbances are added to the cell concentration 
in the suspension to model the initial state in an experiment of bioconvection. 
The initial concentration is defined as
\begin{equation}
   n = n_0 \left[ 1 + \varepsilon A(x,y) \right]
   \label{random}
\end{equation}
where $\varepsilon=10^{-2}$ and 
$A(x,y)$ is a random number generated in the range of $-1$ to 1. 
The random number is generated by the Mersenne twister method 
\citep{Matsumoto&Nishimura_1998} 
and the initial value is given by the linear congruent method.
To investigate the stability of bioconvection with respect to disturbance, 
we use the obtained calculation result as an initial condition.

The $x$- and $z$-dimensions of the computational region are set to $10h$. 
To clarify the effect of the length of the calculation area 
on the formation of bioconvection, 
we also performed a calculation for the calculation area $L=20h$. 
This study used four uniform grids of $91 \times 31 \times91$ (grid1), 
$101 \times 51 \times 101$ (grid2), $111 \times 81 \times 111$ (grid3), 
and $141 \times 101 \times 141$ (grid4) for $L=10h$ 
to examine the grid dependency of the numerical results. 
We confirmed that the grid resolution of grid2 was suitable 
for obtaining valid results, so the results of grid2 are shown in the following. 
In the calculation for $L=20h$, we used a uniform grid of 
$201 \times 51 \times 201$, which has the same resolution as grid2.

Non-dimensional parameters are given using the properties of 
{\it Bacillus subtilis} that responds to oxygen gradients. 
The diffusion coefficient of bacteria $D_{n0}$ is different in each study, 
and the range is $10^{-6}$ to $10^{-5}$ cm$^2$/s 
\citep{Hillesdon_et_al_1995, Hillesdon&Pedley_1996, Tuval_et_al_2005}. 
This study takes $D_{n0}=10^{-5}$ cm$^2$/s 
and adopts values for the other properties from \citet{Hillesdon_et_al_1995} 
and \citet{Hillesdon&Pedley_1996}. 
Therefore, the non-dimensional base parameters are 
$\beta=1$, $\gamma=2$, $\delta=2$, and $Sc=1000$. 
We change some base parameters to investigate the effects of the physical properties 
of bacteria and oxygen on bioconvection. 
The calculation was conducted for eight different Rayleigh numbers, 
$Ra=0$, 250, 500, 1000, 2000, 3000, 4000, and 5000. 
The bioconvection occurs above the critical Rayleigh number. 
Variations of the bioconvection patterns and the wavelengths 
depending on the initial condition have been observed 
in earlier experimental research \citep{Bees&Hill_1997}. 
At $Ra=500$ and 5000, greater than the critical Rayleigh number, 
we have investigated whether the same tendency 
as the experimental result is observed 
for 10 kinds of initial disturbance to the cell concentration. 
The results for all Rayleigh numbers are convergent solutions 
and represent steady-state flow and concentration fields.

\section{Numerical results and discussion}

\subsection{Comparison with the theoretical solution}

\begin{figure}
\begin{minipage}{0.485\linewidth}
\begin{center}
\includegraphics[trim=2mm 0mm 0mm 0mm, clip, width=65mm]{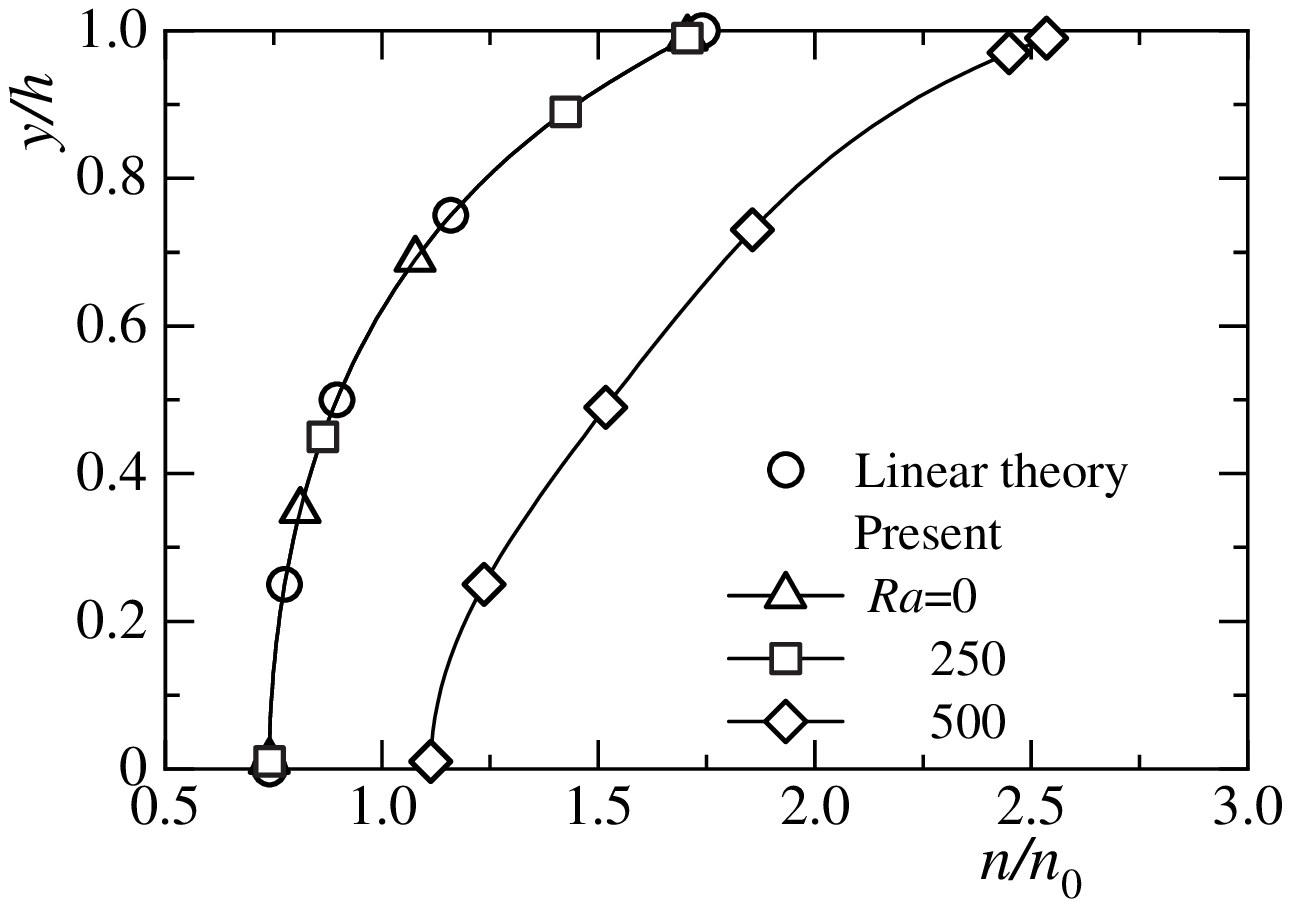} \\
(a) bacteria
\end{center}
\end{minipage}
\hspace{0.005\linewidth}
\begin{minipage}{0.485\linewidth}
\begin{center}
\includegraphics[trim=2mm 0mm 0mm 0mm, clip, width=65mm]{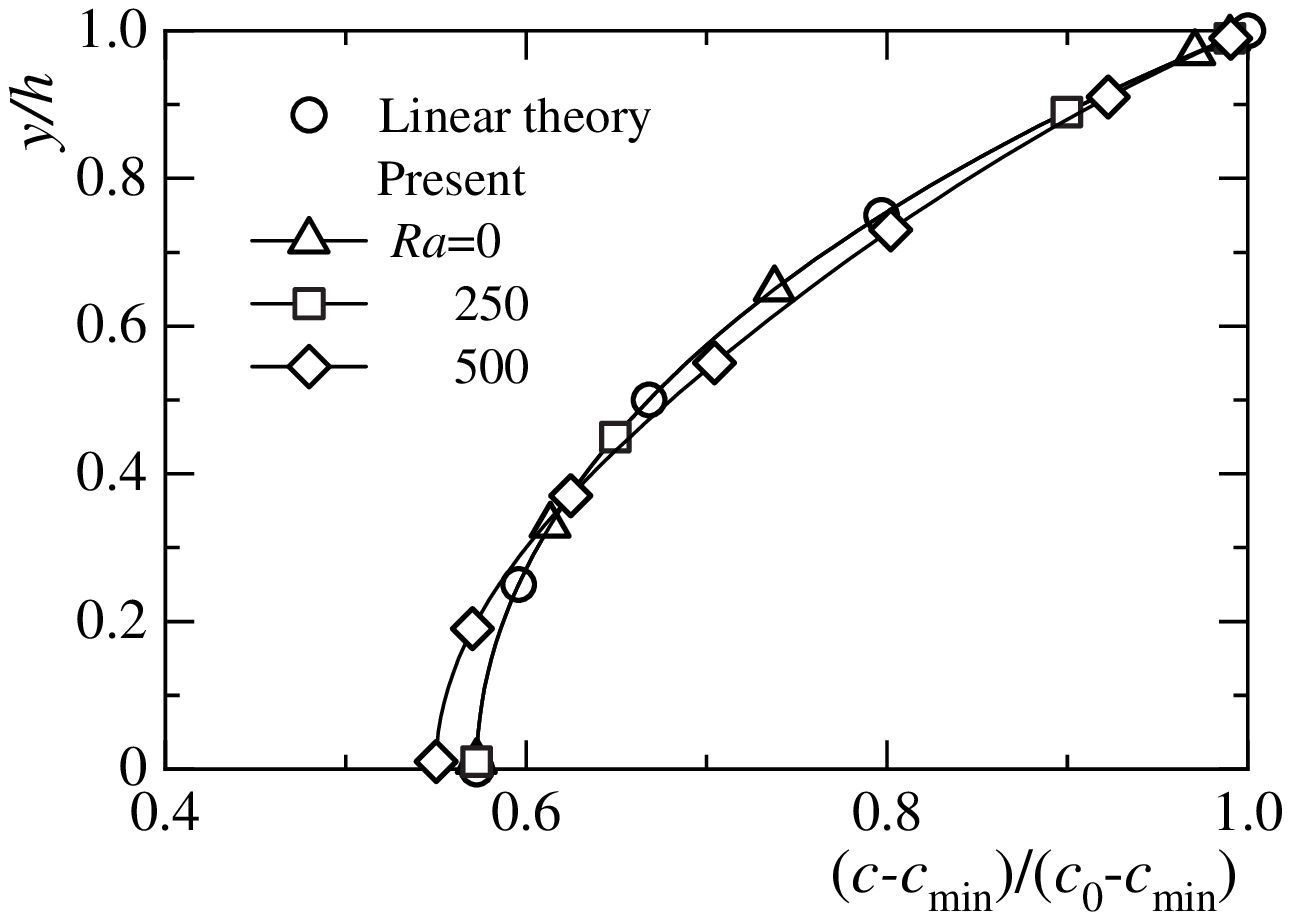} \\
(b) oxygen \\
\end{center}
\end{minipage}
\caption{Comparison of concentration distributions 
with linear theoretical solution \citep{Hillesdon_et_al_1995}.}
\label{linear_theory}
\end{figure}

The present numerical result is compared with the linear theoretical solution 
\citep{Hillesdon_et_al_1995} to confirm its validity. 
The theoretical value is a solution for a shallow chamber. 
Figure \ref{linear_theory} shows the cell and oxygen concentration distributions 
in the $y$-direction for $Ra=0$, 250, and 500. 
Here, the cross-section are at $x/h=5.0$, $z/h=5.0$ 
for $Ra=0$ and $Ra=250$, and $x/h=5.9$, $z/h=4.1$ 
at the centre of the plume for $Ra=500$. 
The plume is shown below in figures \ref{Ra500} and \ref{Ra5000}. 
We define the plume centre as the position 
where the cell concentration on the free surface is maximum. 
These calculation results for $Ra=0$, 250 agree well with 
the theoretical solution. 
However, it is confirmed that a large difference appears 
between the result at $Ra=500$ and the theoretical result, 
suggesting the occurrence of bioconvection, 
which is a non-linear phenomenon. 
The previous study \citep{Yanaoka_et_al_2008} showed 
that bioconvection occurs at a Rayleigh number greater than the critical number. 
In this study, since bioconvection occurs at $Ra=500$, 
it is considered that the critical Rayleigh number lies between 
$Ra=250$ and 500. 

\begin{figure}
\begin{center}
\includegraphics[trim=0.0mm 10.0mm 0.0mm 5.0mm, clip, width=95mm]{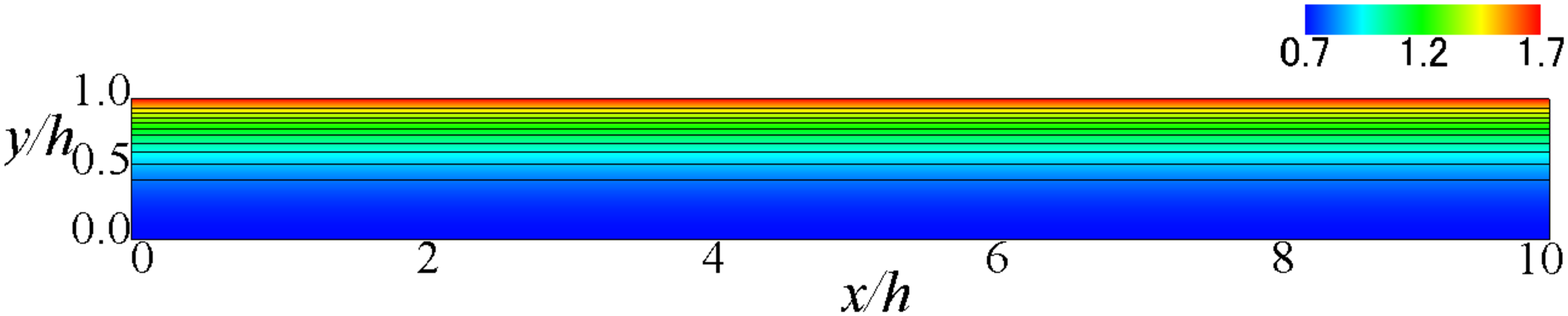} \\
(a) bacteria \\
\includegraphics[trim=0.0mm 10.0mm 0.0mm 5.0mm, clip, width=95mm]{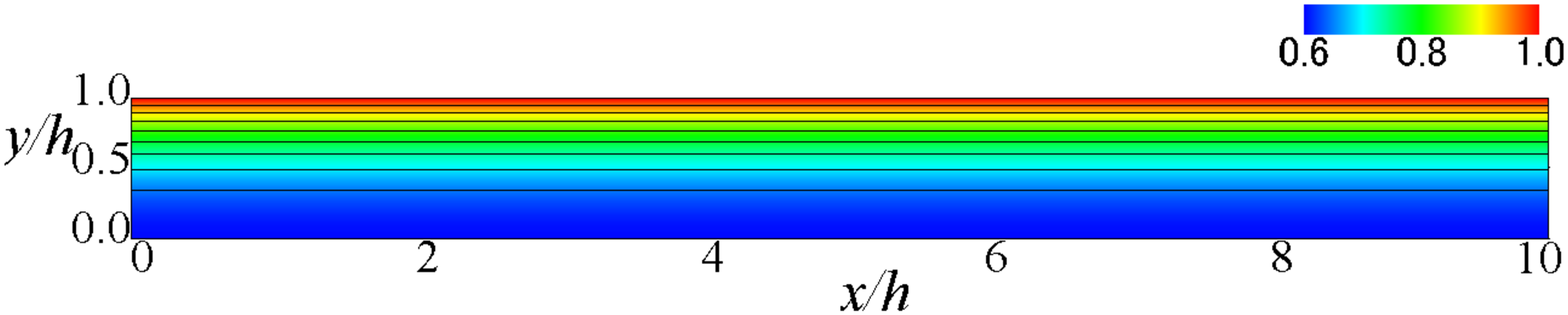} \\
(b) oxygen \\
%
%
\end{center}
\caption{Bacterial and oxygen concentration contours 
for $Ra=250$ at $z/h=5.0$.}
\label{Ra250}
\end{figure}

\begin{figure}
\begin{center}
\includegraphics[trim=0mm 10mm 0mm 0mm, clip, width=90mm]{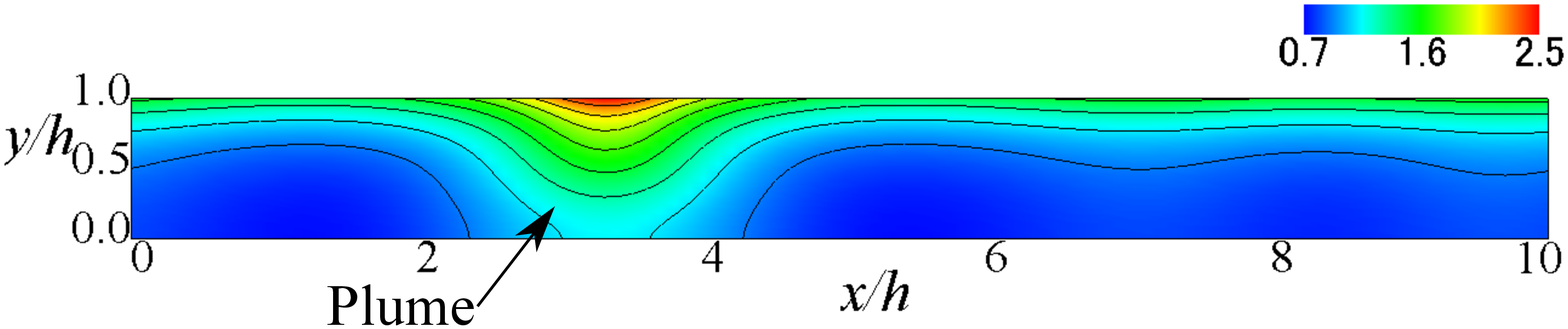} \\
(a) bacteria \\
\vspace*{1.0\baselineskip}
\includegraphics[trim=0mm 10mm 0mm 10mm, clip, width=90mm]{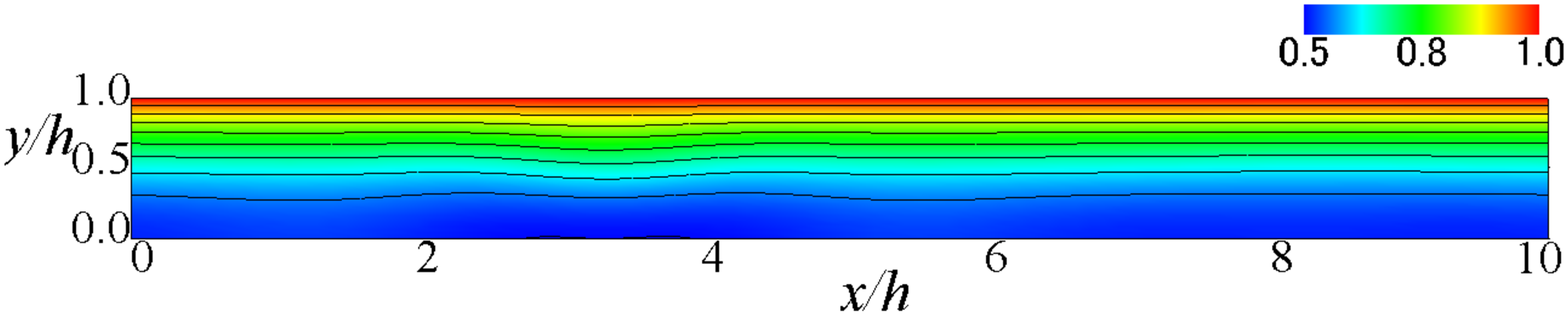} \\
(b) oxygen \\
\vspace*{1.0\baselineskip}
\includegraphics[trim=0mm 0mm 6mm 0mm, clip, width=90mm]{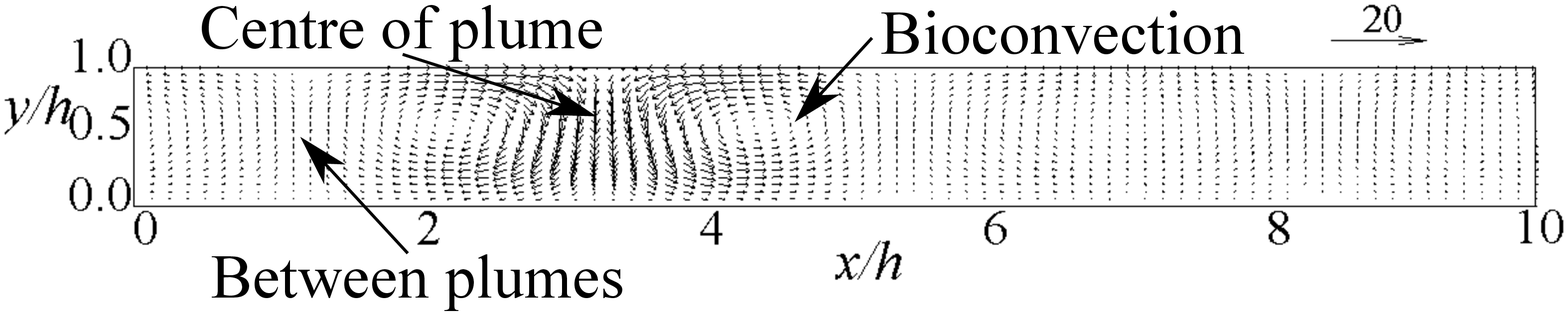} \\
(c) velocity vectors \\
\vspace*{1.0\baselineskip}
\includegraphics[trim=0mm 0mm 0mm 0mm, clip, width=90mm]{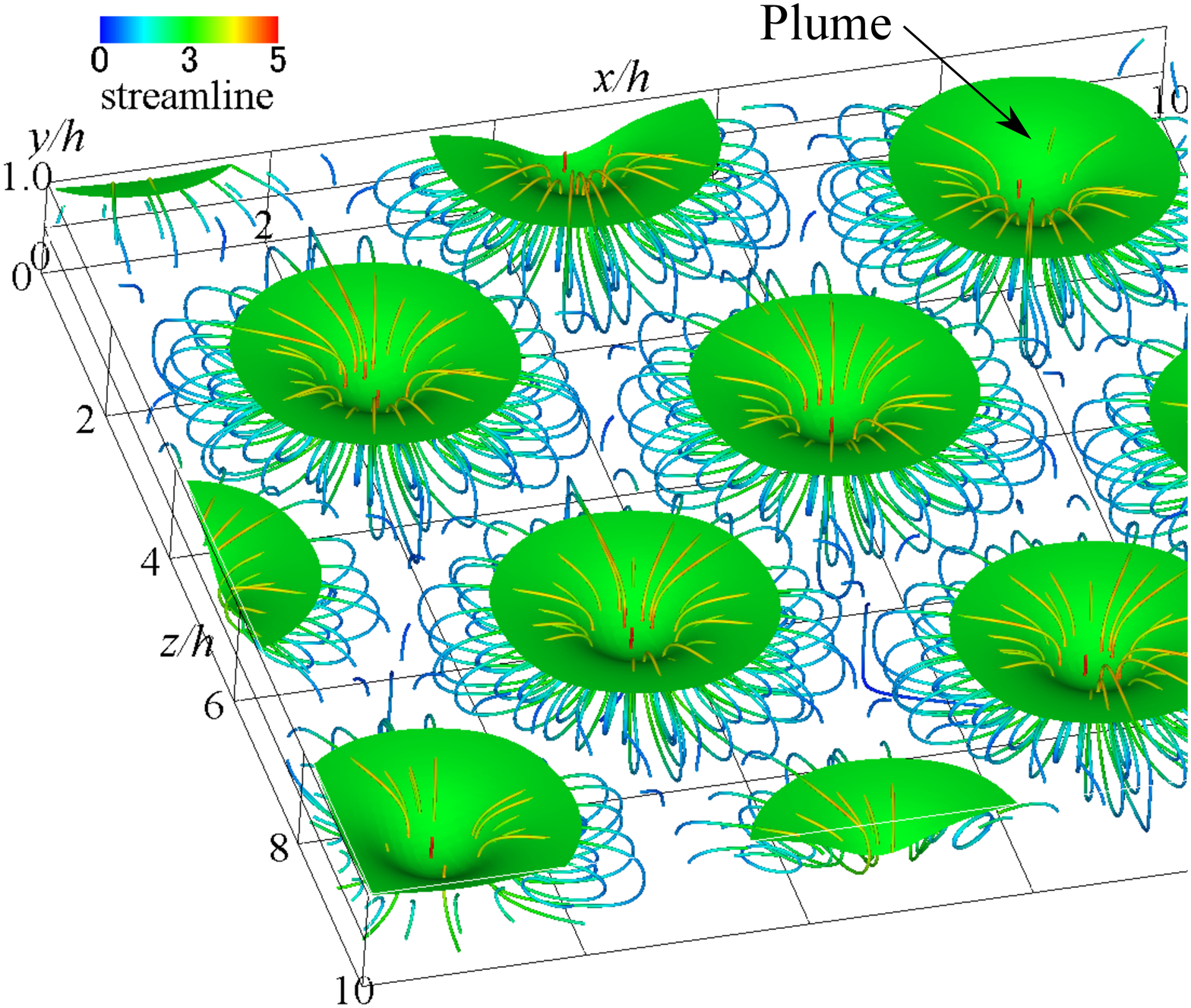} \\
(d) isosurface and streamlines
\end{center}
\caption{Bacterial and oxygen concentration contours 
and velocity vectors at $z/h=6.5$, 
isosurface of bacterial concentration, 
and streamlines for $Ra=500$: Isosurface value is 1.5.}
\label{Ra500}
\end{figure}

\subsection{Occurrence of bioconvection}

First, we investigate the occurrence of bioconvection 
when the Rayleigh number is changed. 
Figures \ref{Ra250} and \ref{Ra500} show the flow and concentration fields 
for $Ra=250$, 500. 
At $Ra=250$, the bacterial cells are concentrated near the water surface 
because they react with the oxygen supplied from the water surface 
and move upward due to chemotaxis. 
As the fluid is almost stationary and is not transported by convection, 
the oxygen concentration distribution is two-dimensional. 
The oxygen concentration decreases from the water surface to the lower wall. 
From above, because the suspension is stable up to $Ra=250$ 
and bioconvection does not occur, 
it is considered that the critical Rayleigh number is at $Ra>250$. 
In the cell concentration field of $Ra=500$ shown in figure \ref{Ra500} (a), 
high-concentration cells concentrated near the water surface settle 
toward the lower wall at $x/h=3.3$. 
This falling region of cells is called a plume, 
and this result suggests the occurrence of bioconvection. 
As shown in figure \ref{Ra500} (d), the plumes form in a staggered manner, 
so the cross-section of $z/h=6.5$ is located across one plume. 
Therefore, in figure \ref{Ra500} (a), 
the influence of one plume appears significantly. 
Still, even around $x/h=7.0$ and 9.5, the cell concentration is distorted 
by the influence of the surrounding plumes. 
Slight fluctuations are observed in the oxygen concentration distribution, 
and weak convection occurs in the velocity vector. 
In the oxygen concentration field for $Ra=500$, because the Rayleigh number is low, 
the effect of diffusion is greater than that of convection. 
In the cell concentration field, even if the Rayleigh number is low, 
transport by directional swimming works in addition to random swimming, 
so the effect of convection becomes large. 
Since bacteria consume oxygen, 
oxygen consumption increases in the plume, where the cell concentration is high. 
Therefore, the oxygen concentration does not have a distribution similar to 
the cell concentration. 
We can see from figure \ref{linear_theory} that the oxygen concentration for $Ra=500$ 
is similar to that for $Ra=250$. 
The effect of the interference between bacteria and oxygen appears 
in the oxygen concentration distribution.

When $\beta=0$, bacteria do not consume oxygen. 
We calculated the condition of $\beta=0$ at $Ra=500$, 
and bioconvection did not occur. 
The suspension became a uniform concentration field. 
This result suggests that oxygen consumption due to bacteria causes 
interference between bacteria and oxygen, forming bioconvection. 
Bioconvection is similar to the Rayleigh--B\'{e}nard convection. 
However, the properties of microorganisms and their interaction with oxygen 
determine the stability of the suspension and convection pattern. 
Chemotactic bacteria respond to the oxygen gradient and consume oxygen. 
Therefore, the bacteria themselves create an oxygen gradient. 
As the bacteria move, the oxygen concentration field changes, 
and the bacterial diffusion (random swimming) 
and swimming velocity (directional swimming) also change. 
In addition, double diffusion occurs due to the difference in 
oxygen diffusion and bacterial diffusion (random swimming). 
However, it is not a double-diffusion convective problem 
because only bacteria contribute to the density 
(Metcalfe \& Pedley 2001). 
This bacterial-oxygen interaction complicates the phenomenon 
(Metcalfe \& Pedley 2001). 
Such interference between bacteria and oxygen affects 
suspension stability and bioconvective transport characteristics. 
In addition, when $\beta>0$, the difference in density between water and cell causes 
the instability of suspension and the generation of bioconvection. 
However, we believe that the instability also changes 
depending on the difference in diffusion coefficient between cells and oxygen. 
In subsection \ref{physical_properties}, 
we investigate the effect of the oxygen diffusion coefficient on bioconvection.

\begin{figure}
\begin{center}
\includegraphics[trim=0mm 10mm 0mm 0mm, clip, width=90mm]{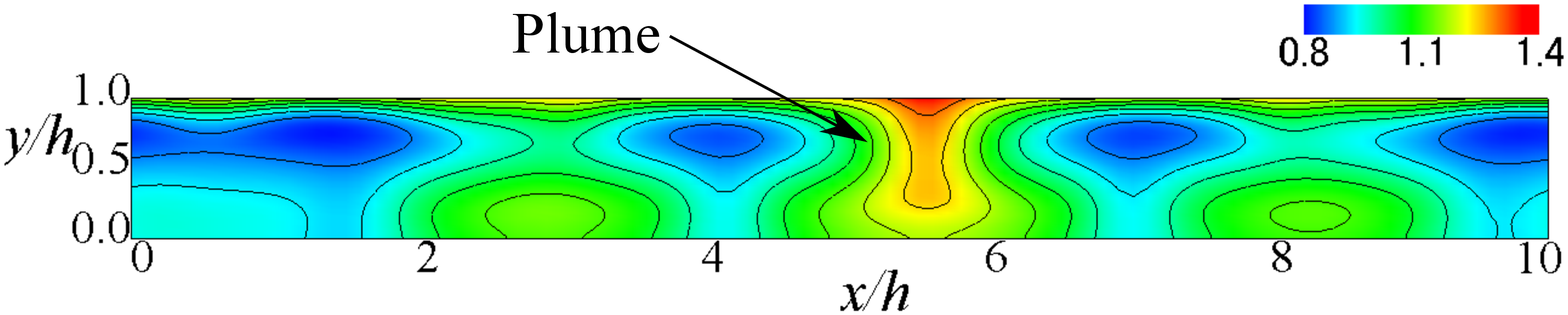} \\
(a) bacteria \\
\vspace*{1.0\baselineskip}
\includegraphics[trim=0mm 10mm 0mm 10mm, clip, width=90mm]{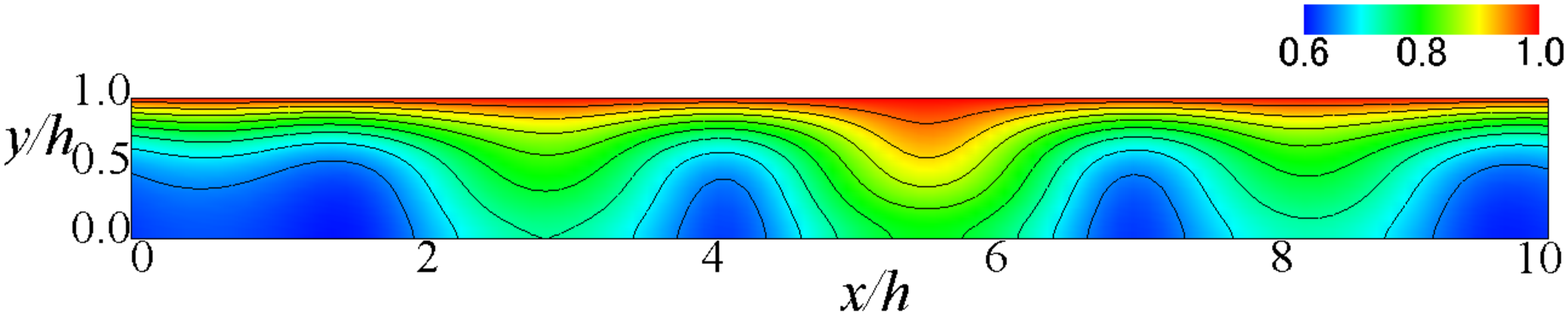} \\
(b) oxygen \\
\vspace*{1.0\baselineskip}
\includegraphics[trim=0mm 0mm 6mm 0mm, clip, width=90mm]{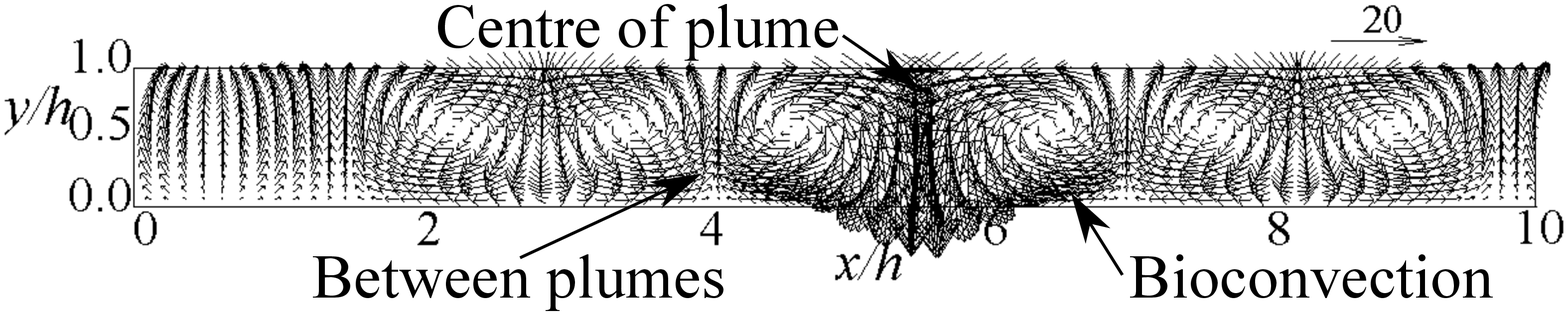} \\
(c) velocity vectors \\
\vspace*{1.0\baselineskip}
\includegraphics[trim=0mm 0mm 0mm 0mm, clip, width=90mm]{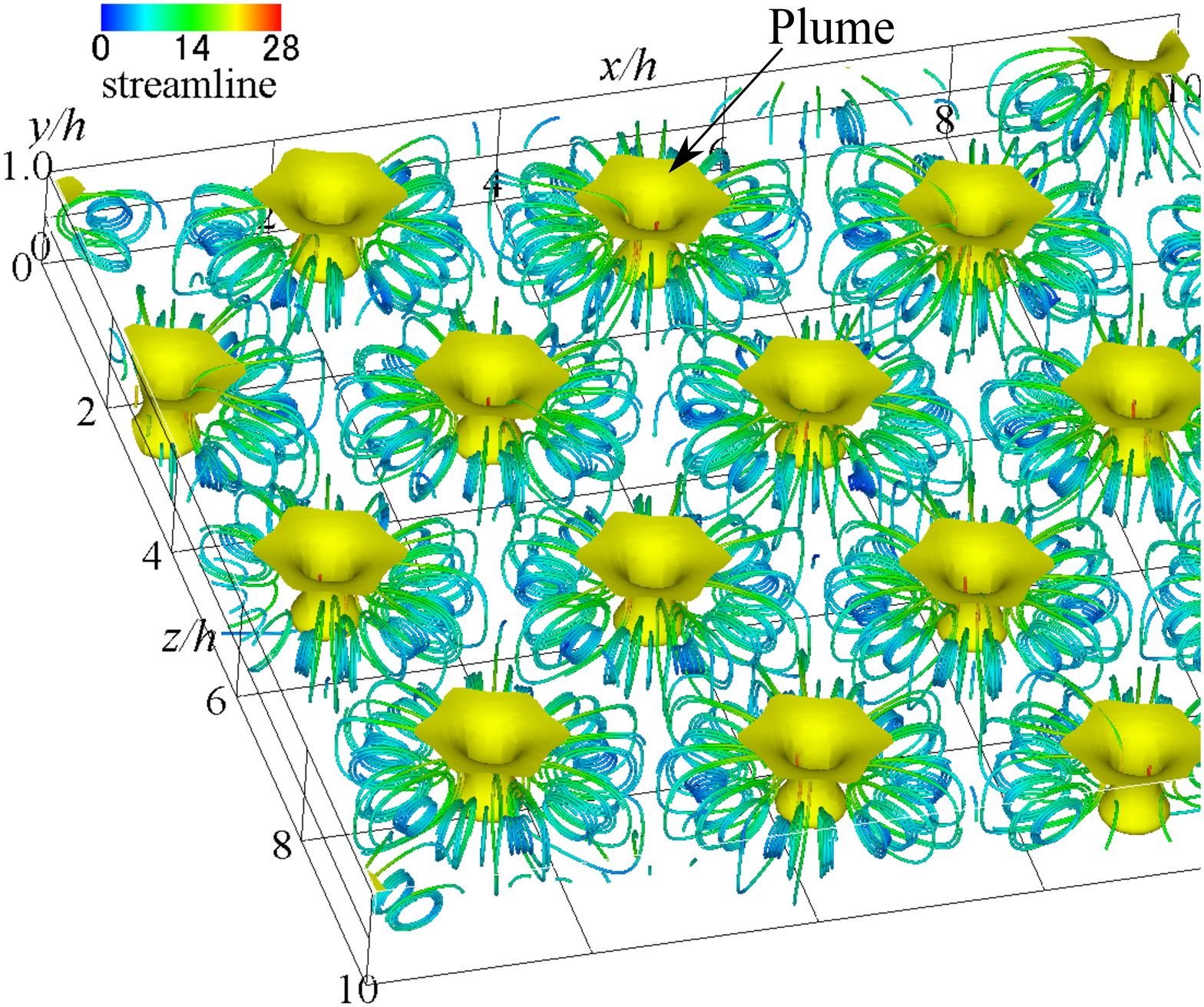} \\
(d) isosurface and streamlines
\end{center}
\caption{Bacterial and oxygen concentration contours 
and velocity vectors at $z/h=4.0$, isosurface of bacterial concentration, 
and streamlines for $Ra=5000$: Isosurface value is 1.2.}
\label{Ra5000}
\end{figure}

Next, we discuss the transport phenomena of cells and oxygen 
in three-dimensional bioconvection at high Rayleigh number. 
Figure \ref{Ra5000} shows the flow field and concentration field 
for $Ra=5000$. 
In the cell concentration contours shown in figure \ref{Ra5000}(a), 
it is observed that the cells accumulating near the free surface fall 
toward the bottom at $x/h=5.7$ and diffuse to the surroundings. 
This area of high cell concentration is a plume, 
and bioconvection occurs around this plume. 
In addition, as can be seen from figure \ref{Ra5000}(d), 
multiple plumes are formed in the suspension. 
The formation of plumes has also been confirmed 
in earlier experimental studies 
\citep{Bees&Hill_1997, Janosi_et_al_1998, Czirok_et_al_2000} 
and numerical analyses \citep{Ghorai&Hill_2000, Chertock_et_al_2012, Karimi&Paul_2013}. 
\citet{Chertock_et_al_2012} and \citet{Ghorai&Hill_2000} 
conducted two-dimensional numerical analysis, 
and \citet{Karimi&Paul_2013} performed three-dimensional numerical analysis. 
However, these works did not show the three-dimensional behavior of bioconvection. 
As shown in figure \ref{Ra5000}(d), this study was able to capture 
multiple three-dimensional plumes. 
In figure \ref{Ra5000}(b), more oxygen is transported toward the bottom wall 
in the region with the plume. 
This is because the bioconvection draws the fluid around the plume downward, 
increasing the oxygen transport due to the convection. 
It can be seen from figure \ref{Ra5000}(d) that the number of plumes 
generated in the suspension increased compared to the flow field 
at $Ra=500$ and that the distance between the plumes decreased. 
It is found from the velocity vectors that the downward flow 
toward the bottom wall occurs at the centre of a plume. 
Inversely, it is observed that the upward flow 
toward the free surface occurs between the plumes 
and that the velocity is slower than that of the downward flow. 
Near the water surface, surrounding fluid concentrates toward the plume centre, 
so cells gather in the plume. 
As a result, a descending flow of high-concentration suspension occurs 
toward the bottom wall. 
Near the lower wall, the flow diffuses from the plume to the surroundings, 
so the ascending flow velocity between the plumes is slower 
than the downward flow velocity. 
This phenomenon can also be explained by the mathematical model used. 
It can be seen that if more cells are transported downward by the plume, 
then the downward flow velocity will increase 
because the downward volume force in the fundamental equation will increase. 
The descending and ascending flow velocities are faster than those 
at $Ra=500$, and the bioconvection is stronger. 
It is observed from the streamlines that three-dimensional vortex rings 
arise around the plumes 
because the cells falling toward the bottom wall draw in the surrounding fluid. 

For $Ra=5000$, using the random numbers used in equation (\ref{random}), 
we added 1\% and 10\% disturbances to the obtained steady-state cell concentration 
and continued the calculation. 
The converged results were the same as those before the disturbance, 
and the pattern of bioconvection did not change. 
The pattern did not change even with different initial disturbances. 
The suspension is stable with respect to the disturbances. 
Next, we perturbed the parameters related to the properties of bacteria and oxygen 
and investigated the effects of parameter changes on the pattern. 
In the present study, parameters $\beta$, $\gamma$, and $\delta$ were increased by 1\%. 
We calculated three cases of ($\beta=1.01$, $\gamma=\delta=2$), 
($\beta=1$, $\gamma=2.02$, $\delta=2$), and ($\beta=1$, $\gamma=2$, $\delta=2.02$). 
As a result, the pattern did not change, 
and there was no significant change in the flow or concentration field. 
From these results, 
we found that steady bioconvection is strongly stable with respect to disturbances.

\citet{Metcalfe&Pedley_1998} investigated the patterns formed 
at the onset of bioconvection 
and clarified that the pattern is determined by the physical properties 
of bacteria and oxygen, such as the strength of oxygen consumption, 
the strength of directional swimming, and the diffusion coefficient of oxygen. 
Existing numerical simulations 
(Chertock {\it et al.} 2012; Lee \& Kim 2015) 
have shown that changing the physical properties of bacteria and oxygen causes 
various bioconvection patterns. 
In the present study, we investigated the pattern of bioconvection by fixing these parameters 
and changing the Rayleigh number and initial disturbance. 
As will be shown later, at a high Rayleigh number, various patterns occurred 
when only the initial disturbance was changed. 
Therefore, since the stable steady-state bioconvection pattern varies 
depending on the initial disturbance, 
we can speculate that boundary conditions determine the pattern. 
Since the present study uses periodic boundary conditions, 
the initial disturbance can freely change not only the cell concentration 
at the boundary, but also the velocity and oxygen at the boundary, 
so that the bioconvection pattern can also vary. 
In subsection \ref{comp_region}, we present calculation results 
for changing the chamber boundary to the side wall 
to investigate the effect of boundary conditions on bioconvection patterns.

\subsection{Effect of physical properties of bacteria and oxygen}
\label{physical_properties}

Theoretically, the cell and oxygen concentration distributions 
in a stationary fluid in a shallow chamber do not depend on 
the parameter $\delta$ (oxygen versus cell diffusion), 
but rather change only with the parameters $\beta$ (oxygen consumption versus oxygen diffusion) 
and $\gamma$ (directed versus random cell swimming) 
\citep{Hillesdon&Pedley_1996}. 
In shallow chambers, there is enough oxygen for bacteria to be active. 
\citet{Hillesdon&Pedley_1996} called $\beta$ (or $\beta \gamma$) depth parameters 
and investigated the effects of these parameters, $\delta$ and $\beta \gamma$, 
on the instability of suspension. 
As a result, when $\beta \gamma$ was changed with fixed $\delta$, 
the minimum value of the critical Rayleigh number that made the suspension unstable appeared. 
In addition, when $\delta$ increased with fixed $\beta \gamma$, 
the critical Rayleigh number increased. 
In the present study, to clarify the effect of parameters on bioconvection, 
we investigate the bioconvection when the parameters $\delta$ and $\beta \gamma$ are changed 
under the conditions $Ra=500$ and 5000 in which bioconvection occurs.

\begin{figure}
\begin{minipage}{0.485\linewidth}
\begin{center}
\includegraphics[trim=2mm 0mm 0mm 0mm, clip, width=65mm]{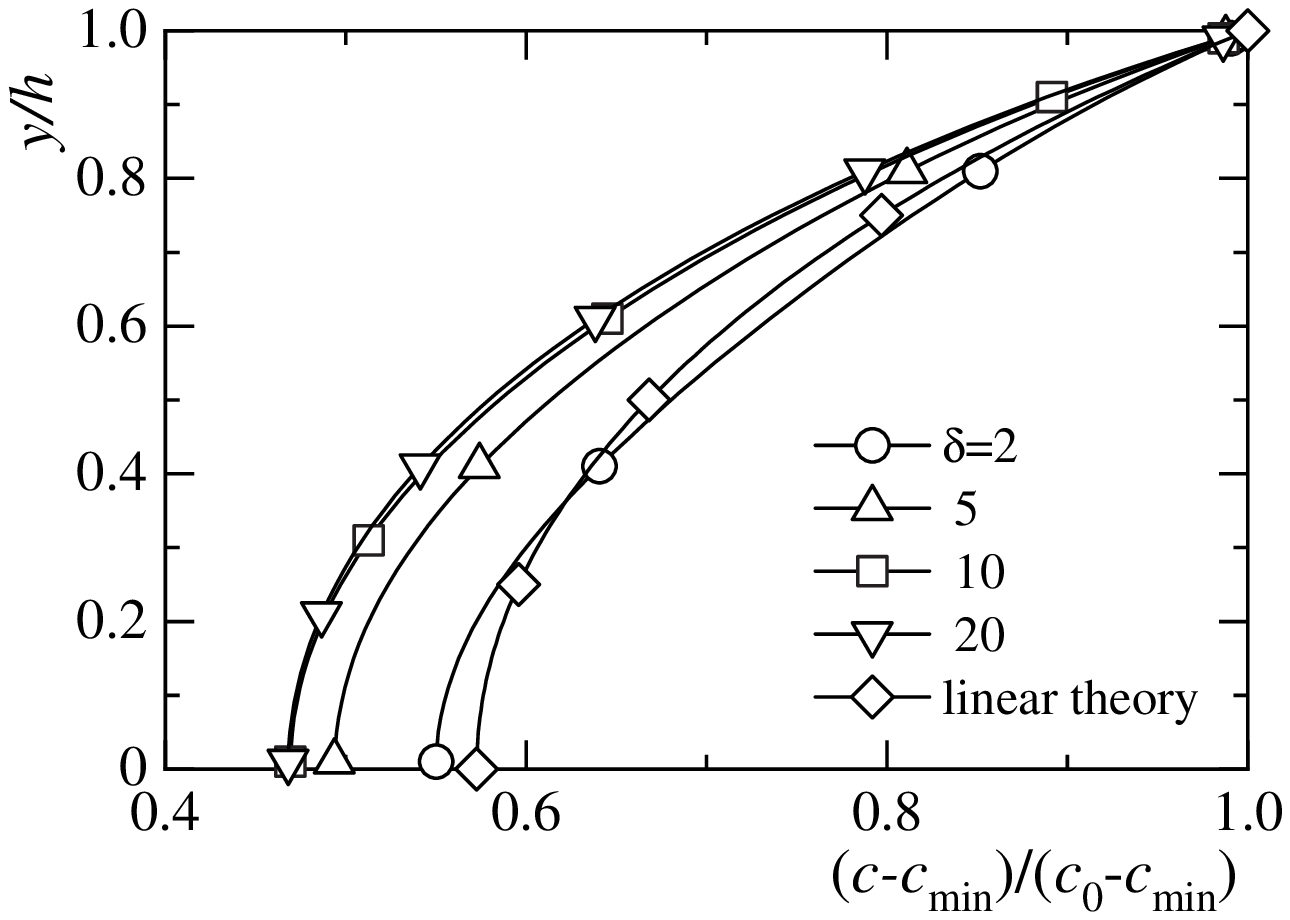} \\
(a) oxygen
\end{center}
\end{minipage}
\hspace{0.005\linewidth}
\begin{minipage}{0.485\linewidth}
\begin{center}
\includegraphics[trim=2mm 0mm 0mm 0mm, clip, width=65mm]{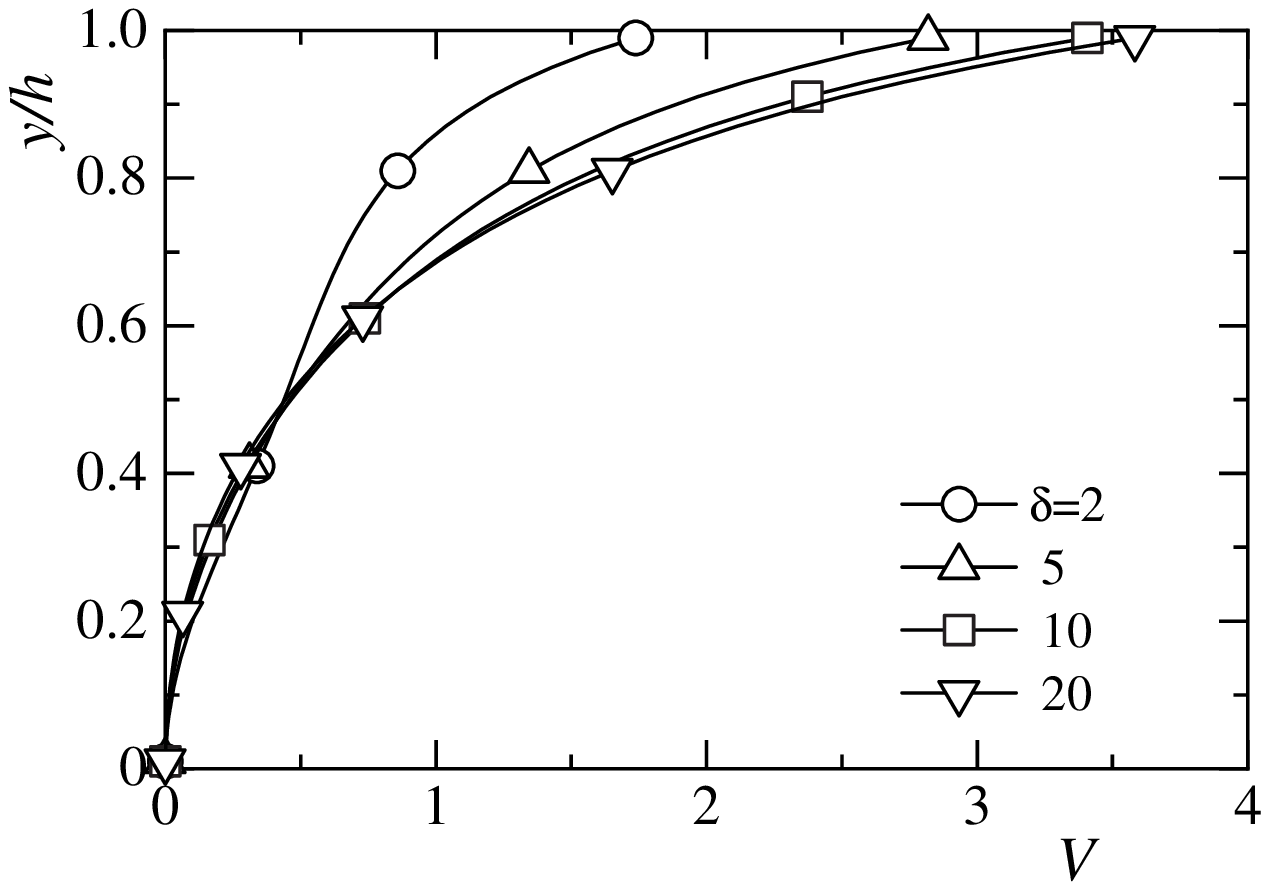} \\
(b) directional velocity
\end{center}
\end{minipage}
\caption{Distributions of oxygen concentration and directional velocity.}
\label{direc_vel}
\end{figure}

First, we consider the condition that the diffusion coefficient of oxygen 
to bacteria is large (the diffusion coefficient of bacteria is small). 
For $Ra=500$, which is close to the critical Rayleigh number, 
$\delta$ was increased with fixed $\beta \gamma$. 
Figure \ref{direc_vel} shows the oxygen concentration distribution 
and directional swimming velocity magnitude 
$V=|H(c)\gamma \nabla c|$ at $\beta=1$, $\gamma=2$, and $\delta=5$, 10, 20. 
The theoretical value and the result of $\delta=2$ are also shown for comparison. 
The cross-section is at the centre of the plume, and its position is 
$(x/h, z/h)=(6.6, 8.0)$ for $\delta=5.0$, 
$(x/h, z/h)=(9.2, 2.6)$ for $\delta=10$, 
and $(x/h, z/h)=(1.7, 7.6)$ for $\delta=20$. 
As $\delta$ increases, more oxygen is transported to the bottom, 
and the gradient of oxygen concentration in the $y$-direction increases, 
thus increasing the upward directional swimming velocity. 
This result leads to a decrease in cell transport due to convection 
(downward flow and upward directional swimming) inside plumes, 
which also decreases nonlinear effects. 
Therefore, the bioconvection decays. 
To compare the strength of convection, 
we calculated the average kinetic energy in the computational domain. 
The average kinetic energies are $K_{av}=1.56$, 1.03, 0.90, and 0.84 
for $\delta=2$, 5, 10, and 20, respectively, 
and the kinetic energy attenuates as $\delta$ increases. 
Although $\delta$ was increased to $\delta=20$, the bioconvection did not disappear. 
The directional swimming velocity in figure \ref{direc_vel} is 
the result of the plume centre. 
We calculated the average value $V_{av}$ of the magnitude 
of the directional swimming velocity in the computational domain. 
For $\delta=2$, 5, 10, and 20, $V_{av}=0.542$, 0.540, 0.538, and 0.537, respectively. 
As $\delta$ increases, $V_{av}$ decreases slightly but remains almost unchanged. 
This result suggests that the increase in directional swimming velocity 
in the plume significantly affects the attenuation of bioconvection. 
The numbers of plumes are 8, 5, 4, and 4 for $\delta=2$, 5, 10, and 20, respectively, 
and the number of plumes decreases as $\delta$ increases. 
From the above investigation, we find that increasing $\delta$ attenuates bioconvection.

\begin{figure}
\begin{center}
\includegraphics[trim=0mm 10mm 0mm 0mm, clip, width=90mm]{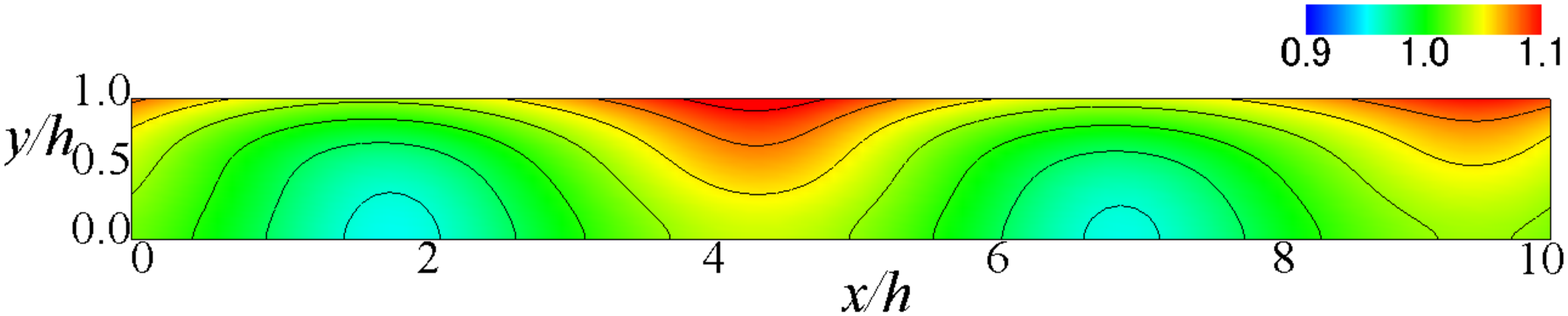} \\
(a) bacteria \\
\vspace*{1.0\baselineskip}
\includegraphics[trim=0mm 10mm 0mm 10mm, clip, width=90mm]{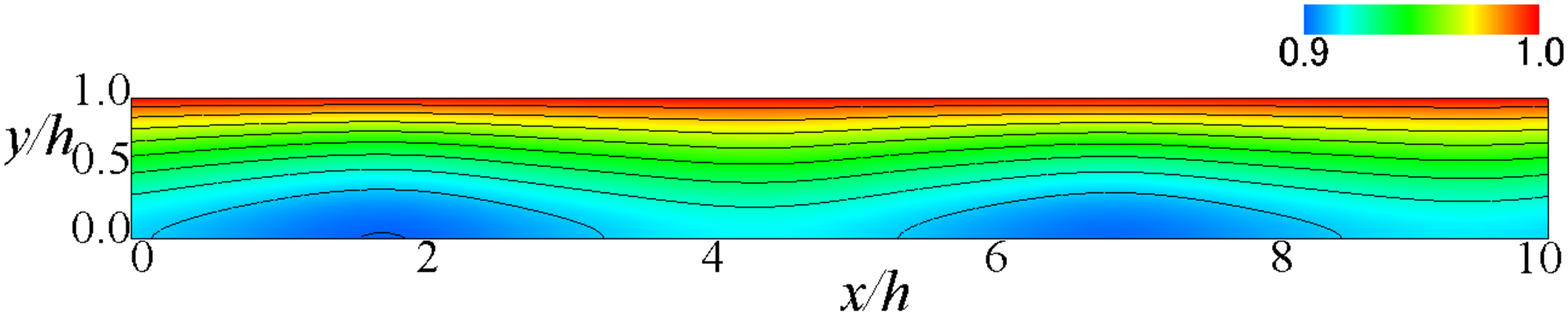} \\
(b) oxygen \\
\vspace*{1.0\baselineskip}
\includegraphics[trim=0mm 0mm 6mm 0mm, clip, width=90mm]{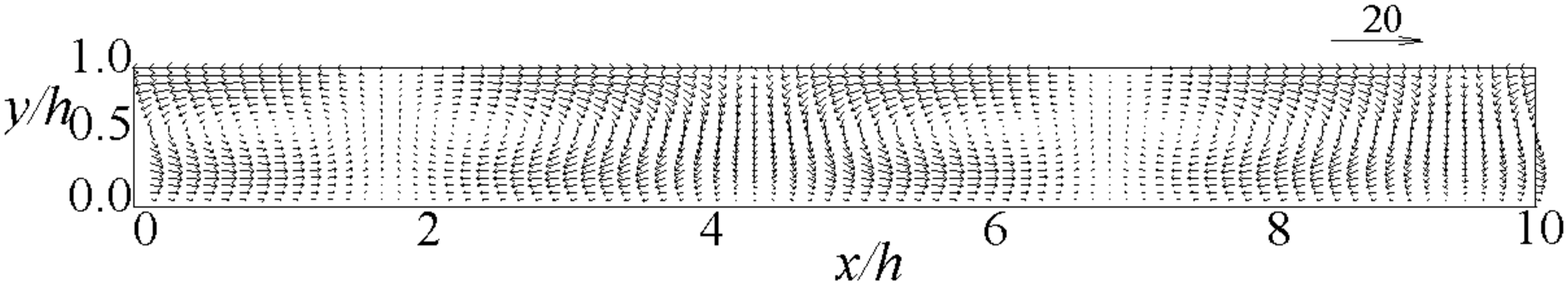} \\
(c) velocity vectors \\
\vspace*{1.0\baselineskip}
\includegraphics[trim=0mm 0mm 0mm 0mm, clip, width=90mm]{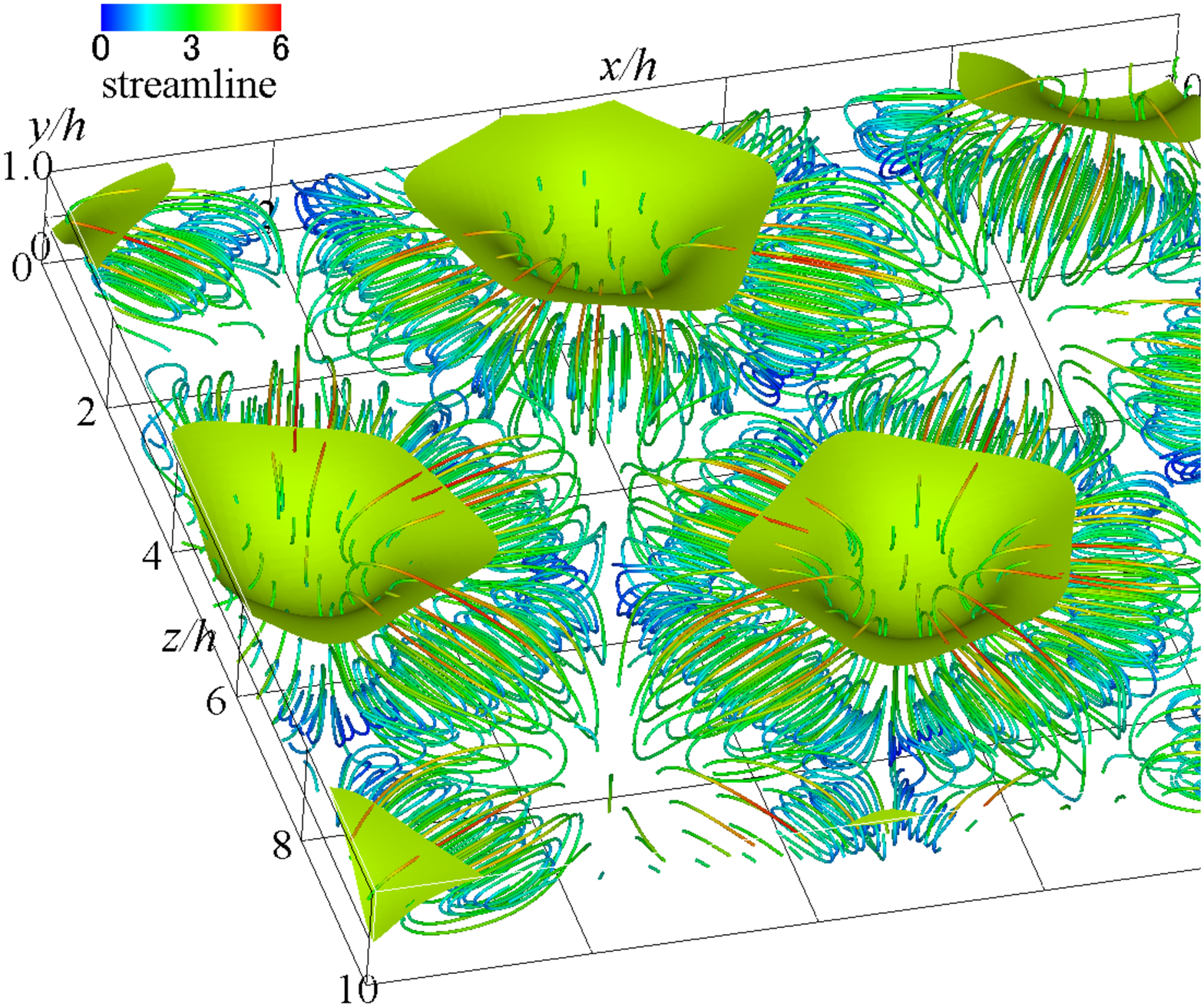} \\
(d) isosurface and streamlines
\end{center}
\caption{Bacterial and oxygen concentration contours 
and velocity vectors at $z/h=0.8$, isosurface of bacterial concentration, 
and streamlines for $\beta=0.1$ and $Ra=5000$: Isosurface value is 1.06.}
\label{Ra5000bt01}
\end{figure}

\begin{figure}
\begin{center}
\includegraphics[trim=0mm 10mm 0mm 0mm, clip, width=90mm]{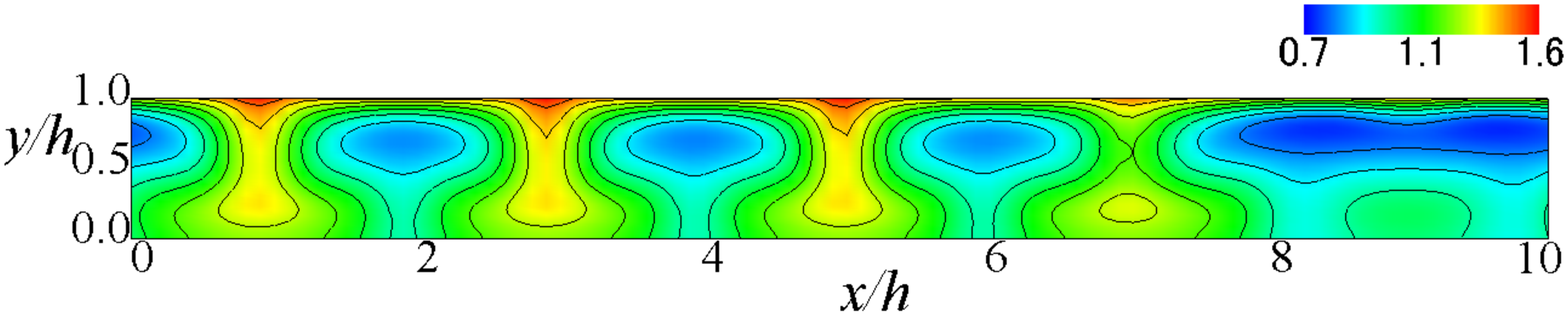} \\
(a) bacteria \\
\vspace*{1.0\baselineskip}
\includegraphics[trim=0mm 10mm 0mm 10mm, clip, width=90mm]{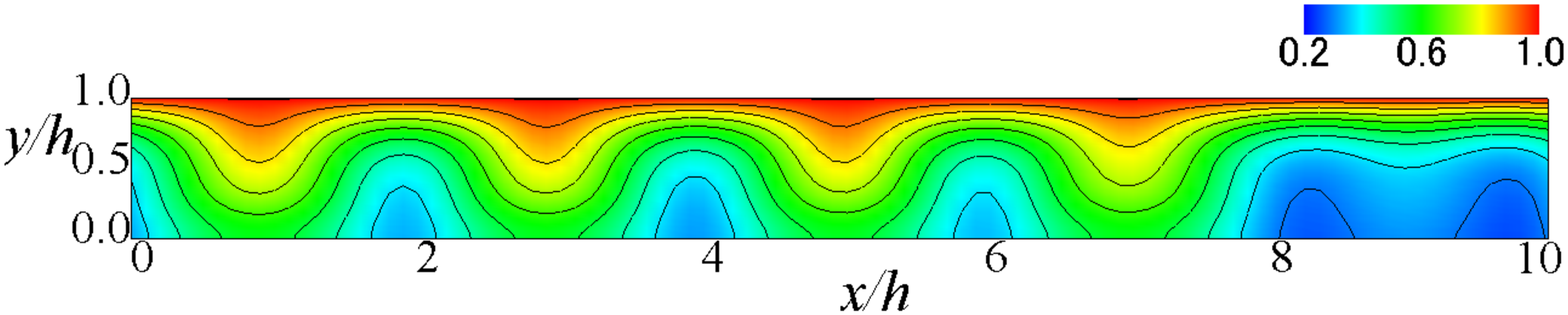} \\
(b) oxygen \\
\vspace*{1.0\baselineskip}
\includegraphics[trim=0mm 0mm 6mm 0mm, clip, width=90mm]{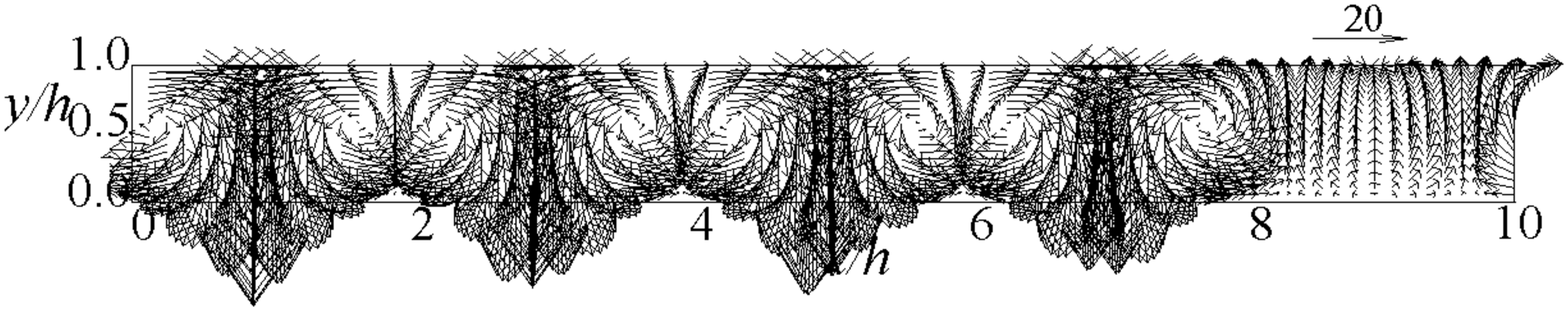} \\
(c) velocity vectors \\
\vspace*{1.0\baselineskip}
\includegraphics[trim=0mm 0mm 0mm 0mm, clip, width=90mm]{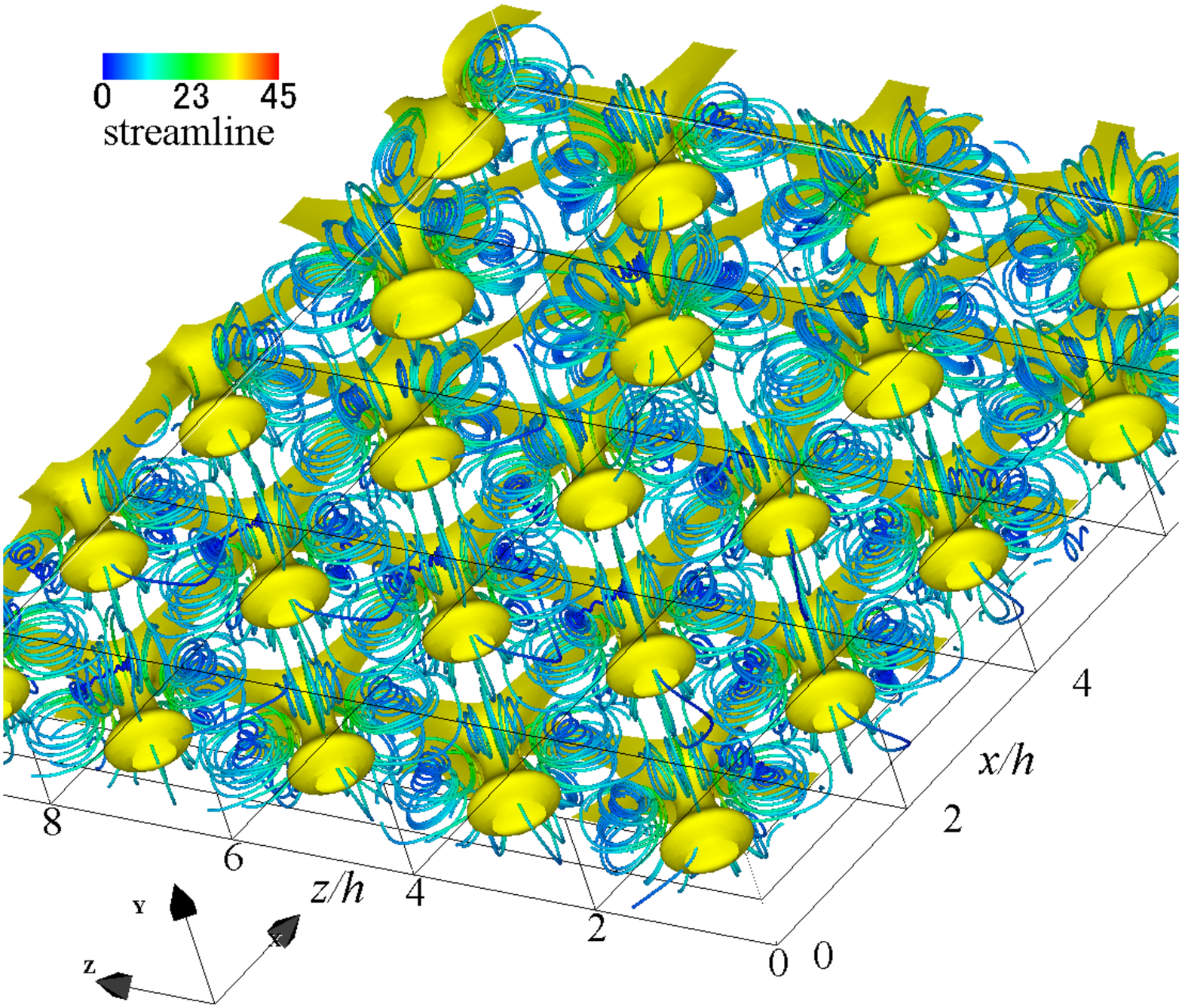} \\
(d) isosurface and streamlines
\end{center}
\caption{Bacterial and oxygen concentration contours 
and velocity vectors at $z/h=1.4$, isosurface of bacterial concentration, 
and streamlines for $\beta=2$ and $Ra=5000$: Isosurface value is 1.25.}
\label{Ra5000bt2}
\end{figure}

Next, we investigate the effect of bacterial-oxygen interaction 
on bioconvection at a high Rayleigh number. 
Figures \ref{Ra5000bt01} and \ref{Ra5000bt2} show the results 
for $\beta=0.1$ ($\beta \gamma=0.2$) and $\beta=2$ ($\beta \gamma=4$) 
at $\delta=2$, respectively. 
In figure \ref{Ra5000bt2} (d), the plume is observed from below. 
When $\beta=0.1$, oxygen consumption by bacteria is suppressed. 
Chemotactic bacteria respond to oxygen gradients and consume oxygen. 
When oxygen consumption decreases, bacterial activity declines 
because the bacteria themselves do not generate oxygen gradients. 
Therefore, the bioconvection at $\beta=0.1$ is weaker than 
the result at $\beta=1$ in figure \ref{Ra5000}. 
For $\beta=2$, the convective velocity is higher than the result for $\beta=1$, 
and more bacteria and oxygen are transported to the vicinity of the bottom surface. 
An increase in $\beta$ means that the oxygen consumption by bacteria strengthens. 
As can be seen from the oxygen concentration distribution in figure \ref{Ra5000bt2}, 
the oxygen consumption by bacteria increases the oxygen concentration gradient 
and strengthens directional swimming. 
At $Ra=5000$, the upward and downward flows are strong, 
and if the directional swimming velocity further increases, 
the convection is strengthened on average in the suspension. 
To compare the strength of directional swimming, 
we calculated the average magnitude $V_{av}$ of the directional swimming velocity 
in the computational domain. 
For $\beta=0.1$, 1, and 2, $V_{av}=6.4\times10^{-3}$, $4.8\times10^{-1}$, and 1.7, respectively, 
and the directional swimming velocity at $\beta=2$ is 273 times faster than 
that at $\beta=0.1$. 
This result shows that the bioconvection at $\beta=0.1$ is considerably attenuated 
even at the high Rayleigh number. 
The numbers of plumes are 8, 15, and 23 for $\beta=0.1$, 1, and 2, respectively, 
and the number of plumes increases as $\beta$ increases. 
From the above results, we found that the rate of oxygen consumption by bacteria 
significantly affects the strength of bioconvection.

\begin{figure}
\begin{center}
\includegraphics[trim=0mm 10mm 0mm 0mm, clip, width=90mm]{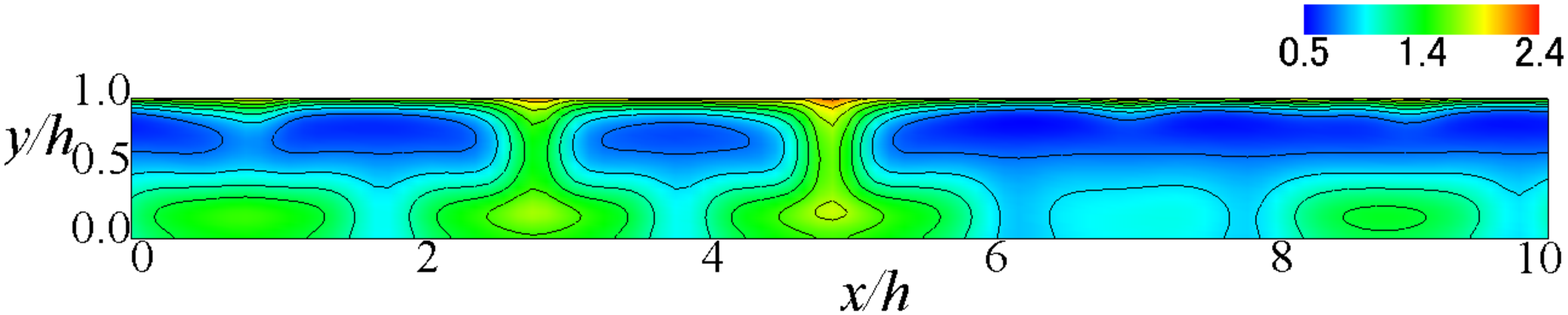} \\
(a) bacteria \\
\vspace*{1.0\baselineskip}
\includegraphics[trim=0mm 10mm 0mm 10mm, clip, width=90mm]{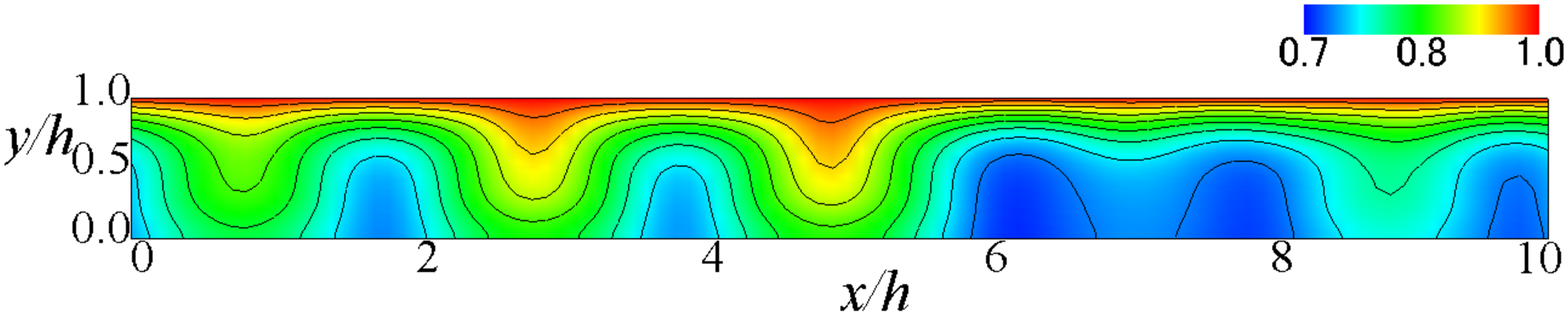} \\
(b) oxygen \\
\vspace*{1.0\baselineskip}
\includegraphics[trim=0mm 0mm 6mm 0mm, clip, width=90mm]{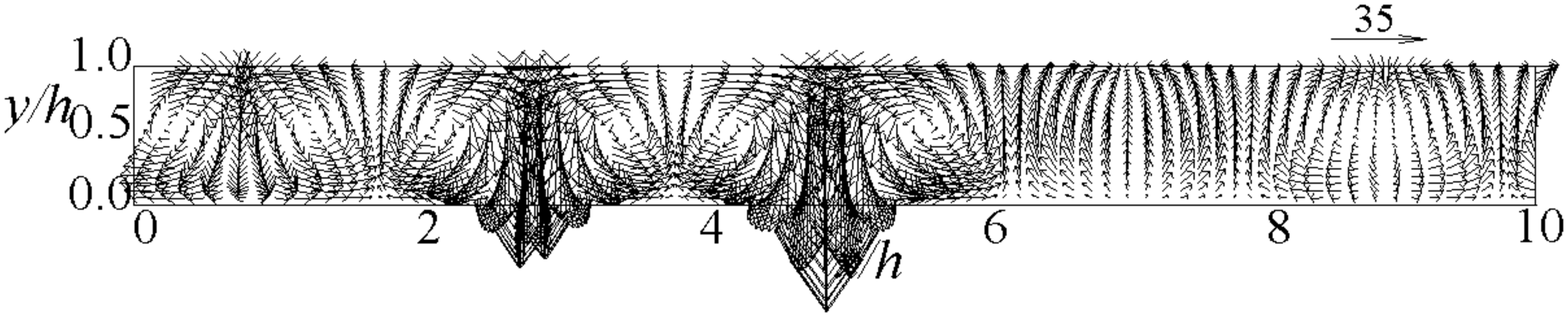} \\
(c) velocity vectors \\
\vspace*{1.0\baselineskip}
\includegraphics[trim=0mm 0mm 0mm 0mm, clip, width=90mm]{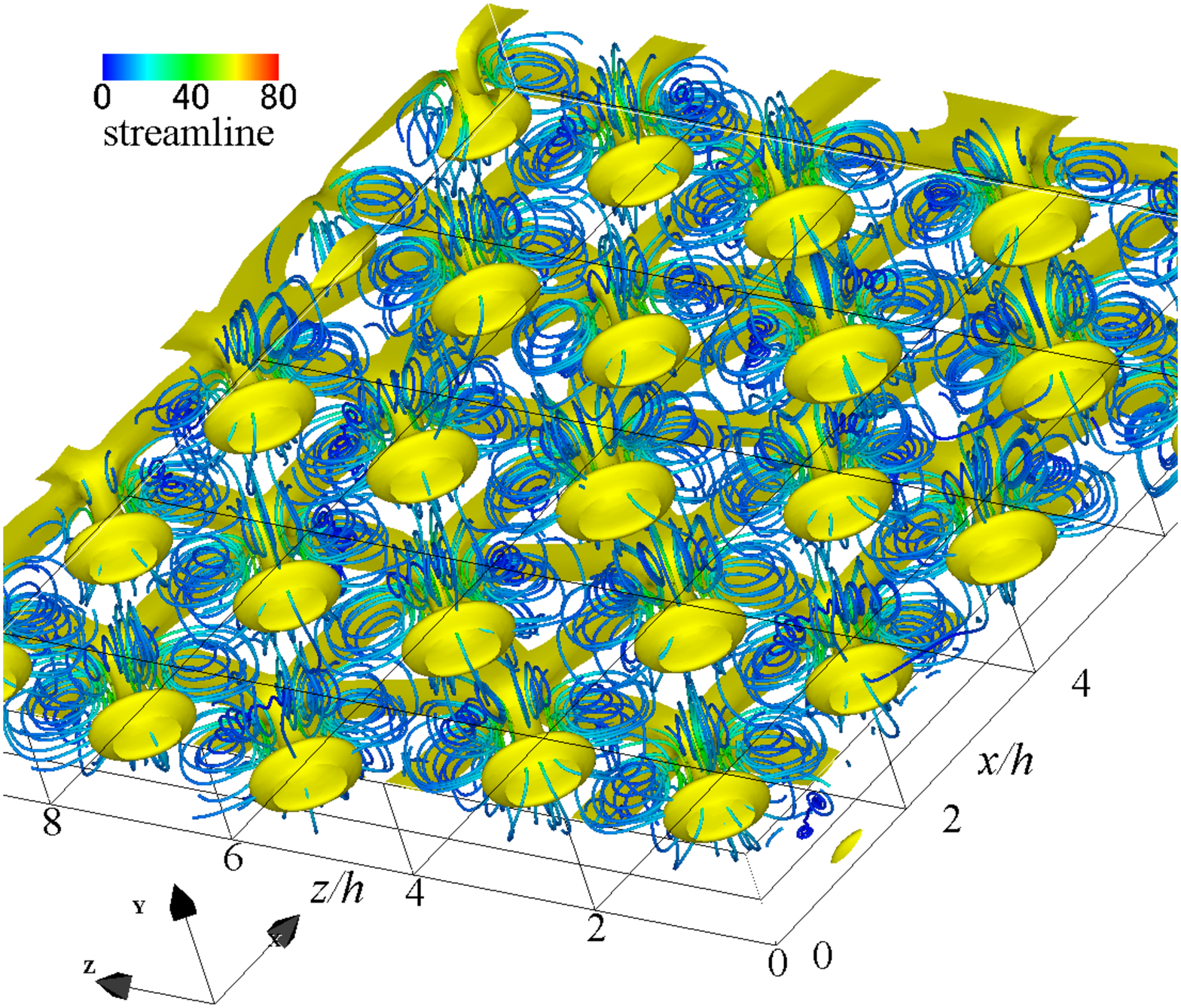} \\
(d) isosurface and streamlines
\end{center}
\caption{Bacterial and oxygen concentration contours 
and velocity vectors at $z/h=4.0$, isosurface of bacterial concentration, 
and streamlines for $\gamma=10$ and $Ra=5000$: Isosurface value is 1.46.}
\label{Ra5000gm10}
\end{figure}

\begin{figure}
\begin{minipage}{0.485\linewidth}
\begin{center}
\includegraphics[trim=2mm 0mm 0mm 0mm, clip, width=65mm]{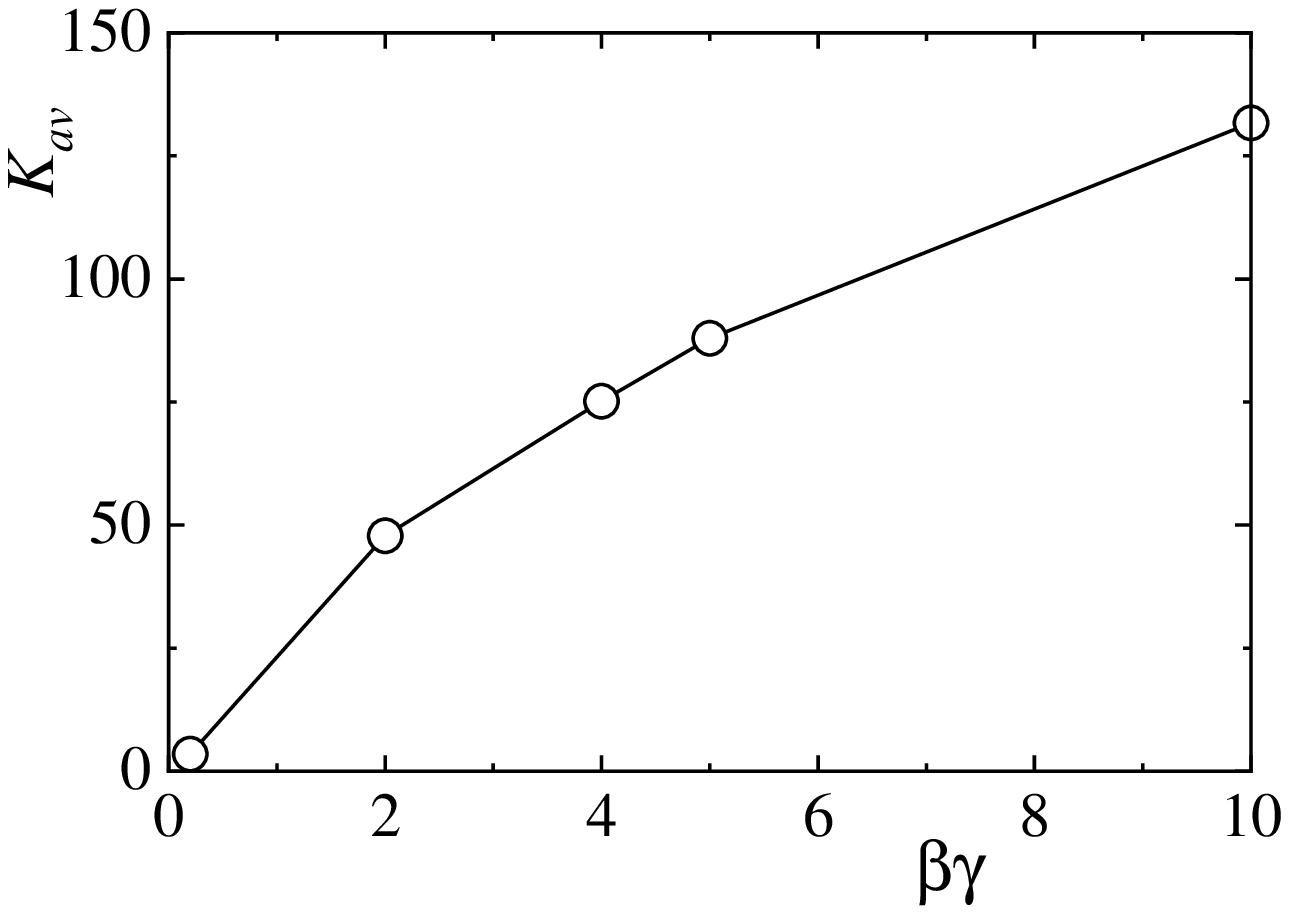} \\
(a) average kinetic energy
\end{center}
\end{minipage}
\hspace{0.005\linewidth}
\begin{minipage}{0.485\linewidth}
\begin{center}
\includegraphics[trim=2mm 0mm 0mm 0mm, clip, width=65mm]{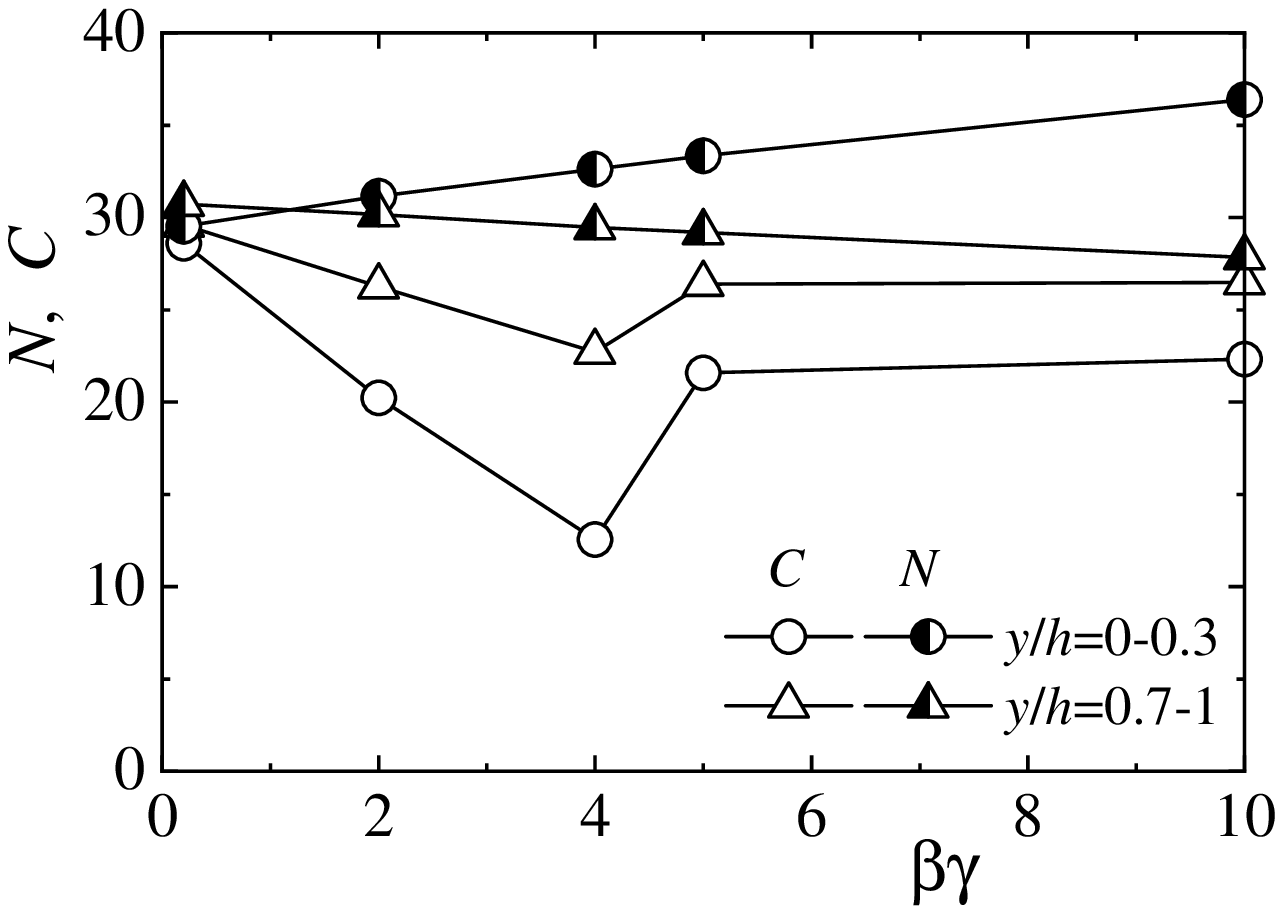} \\
(b) integral values of bacteria and oxygen
\end{center}
\end{minipage}
\caption{Average kinetic energy and integral values of bacteria and oxygen.}
\label{Kav_N_C}
\end{figure}

We investigate the enhancement of convection when directional swimming is increased 
without changing the rate of oxygen consumption by bacteria. 
Figure \ref{Ra5000gm10} shows the results for $\delta=2$ and $\gamma=10$ 
($\beta \gamma=10$). 
An increase in $\gamma$ means an increase in the directional swimming velocity of bacteria. 
The bioconvection is stronger than the results for $\gamma=2$ ($\beta \gamma=2$) 
in figure \ref{Ra5000} and $\beta=2$ ($\beta \gamma=4$) in figure \ref{Ra5000bt2}. 
The same effect as enhancement of convection with increasing $\beta$ appears. 
The plumes arrange randomly, and the number of plumes is 25. 
As the number of plumes increases, 
it is considered that the interference between plumes increases. 
The average magnitude of directional swimming velocity is $V_{av}=9.8$, 
and the transport by directional swimming increases more than 
the results for $\beta=0.1$ ($\beta \gamma=0.2$), $\beta=1$ ($\beta \gamma=2$), 
and $\beta=2$ ($\beta \gamma=4$).

We confirmed that the bioconvection and its pattern 
observed in figures \ref{Ra5000bt01} to \ref{Ra5000gm10} were different from 
the result in figure \ref{Ra5000}. 
Although the Rayleigh number and initial disturbance are the same, 
the different bioconvections and patterns occur 
even if the parameters related to the physical properties of bacteria and oxygen are changed. 
This trend is the same as the result of the existing numerical simulations 
\citep{Chertock_et_al_2012, Lee&Kim_2015}.

We clarify the effects of changes in the physical properties of bacteria and oxygen 
on the enhancement of bioconvection and the amount of transport. 
For $\beta=0.1$ ($\beta \gamma=0.2$), $\beta=1$ ($\beta \gamma=2$), 
$\beta=2$ ($\beta \gamma=4$), $\gamma=5$ ($\beta \gamma=5$), and $\gamma=10$ ($\beta \gamma=10$), 
figure \ref{Kav_N_C} shows the average kinetic energy $K_{av}$ 
and the integral values, $N$ and $C$, of the amounts of cell and oxygen, respectively. 
The depth parameter $\beta \gamma$ is a parameter of the chamber depth 
and, at the same time, represents the parameters of the oxygen consumption rate by bacteria 
and the magnitude of the directional swimming velocity. 
In a convection-free concentration field in a shallow chamber, 
theoretically, the cell concentration depends only on $\beta \gamma$, 
while the oxygen concentration varies with $\beta \gamma$ and $\gamma$ 
\citep{Hillesdon&Pedley_1996}. 
The average kinetic energy increases with the increase of $\beta \gamma$, 
indicating that bioconvection strengthens. 
The oxygen consumption rate by bacteria or directional swimming velocity 
increases non--linearity and enhances convection. 
At this time, cell transport is improved, 
so many bacteria move to the vicinity of the lower wall. 
The cell amount decreases slightly near the water surface. 
Since an increase in $\beta \gamma$ increases the consumption of oxygen, 
the oxygen amount near the water surface and lower wall decreases. 
For $\beta \gamma=5$ ($\beta=1$, $\gamma=5$) and 
$\beta \gamma=10$ ($\beta=1$, $\gamma=10$), 
the consumption rate of oxygen by bacteria is low, 
so the oxygen amount does not decrease monotonically. 
Unlike the cell concentration, 
the oxygen concentration does not depend only on $\beta \gamma$, 
and this trend is the same as the theory 
\citep{Hillesdon&Pedley_1996}. 

\subsection{Bioconvection pattern and wavelength of the pattern}

\begin{figure}
\begin{minipage}{.48\linewidth}
\begin{center}
\includegraphics[trim=0mm 0mm 0mm 0mm, clip, width=65mm]{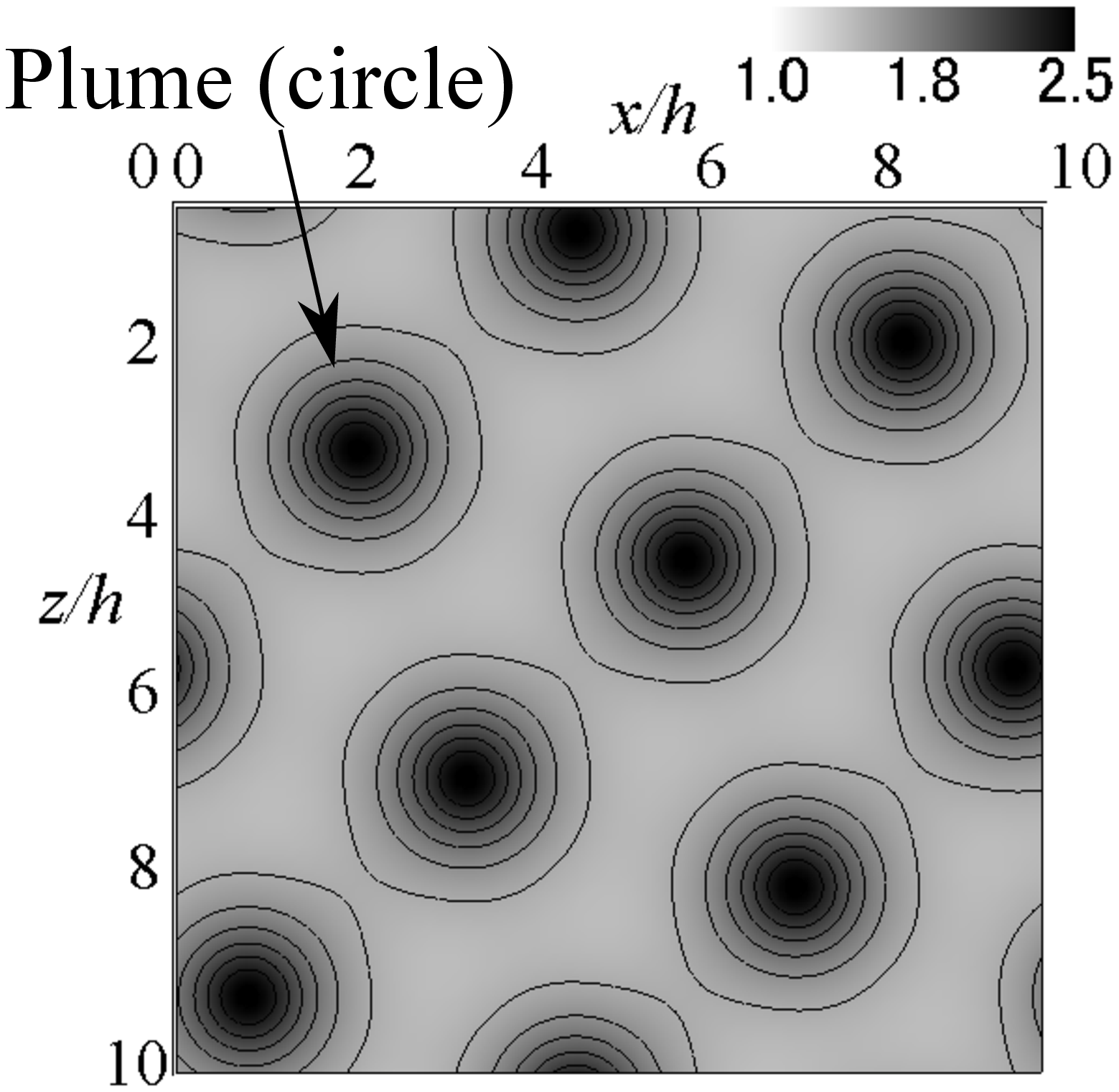} \\
(a) No.2 ($Ra=500$)
\end{center}
\end{minipage}
\begin{minipage}{.48\linewidth}
\begin{center}
\includegraphics[trim=0mm 0mm 0mm 0mm, clip, width=65mm]{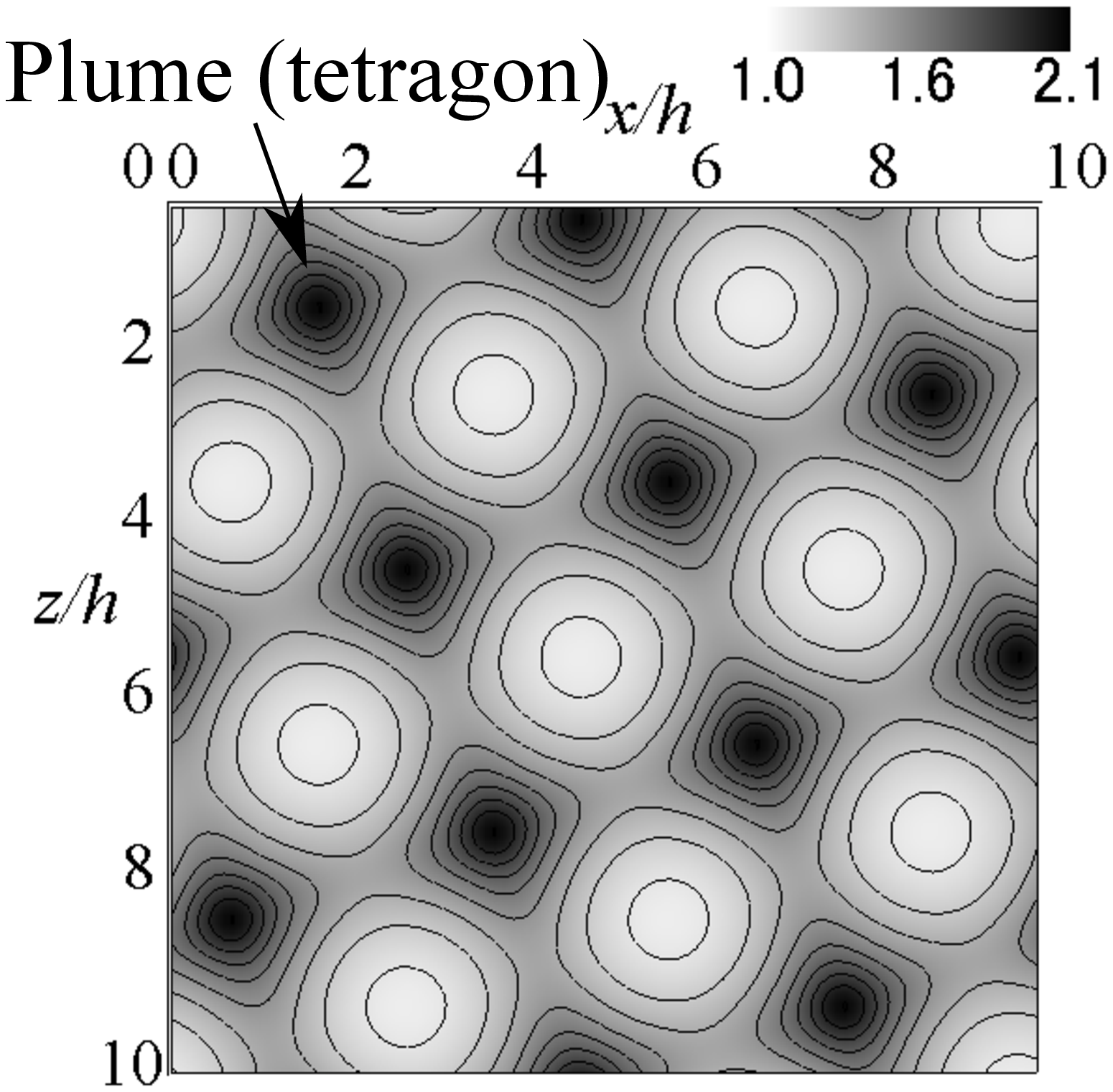} \\
(b) No.11 ($Ra=1000$) \\
\end{center}
\end{minipage}

\vspace*{1.0\baselineskip}
\begin{minipage}{.48\linewidth}
\begin{center}
\includegraphics[trim=0mm 0mm 0mm 0mm, clip, width=65mm]{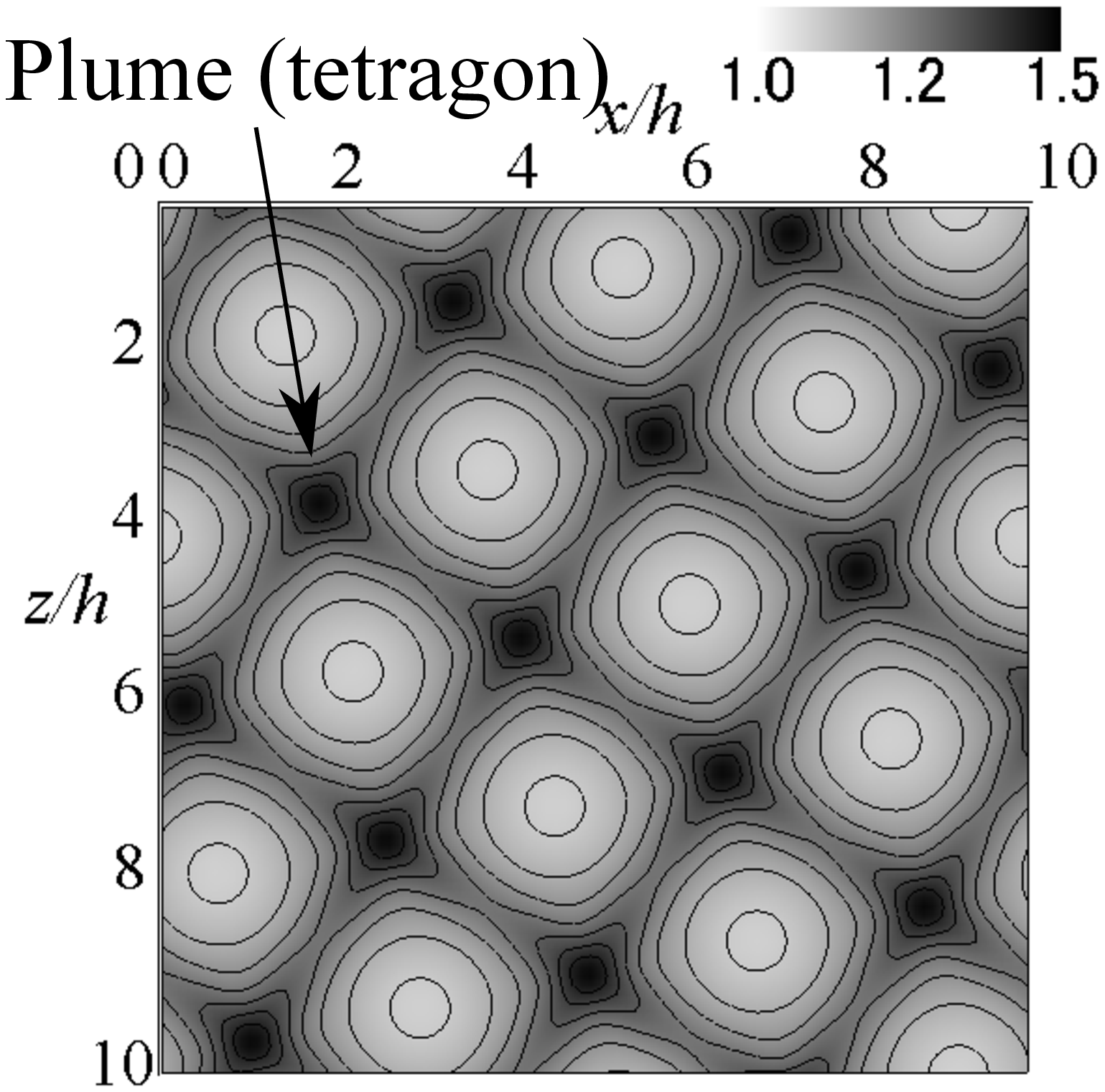} \\
(c) No.13 ($Ra=3000$) \\
\end{center}
\end{minipage}
\begin{minipage}{.48\linewidth}
\begin{center}
\includegraphics[trim=0mm 0mm 0mm 0mm, clip, width=65mm]{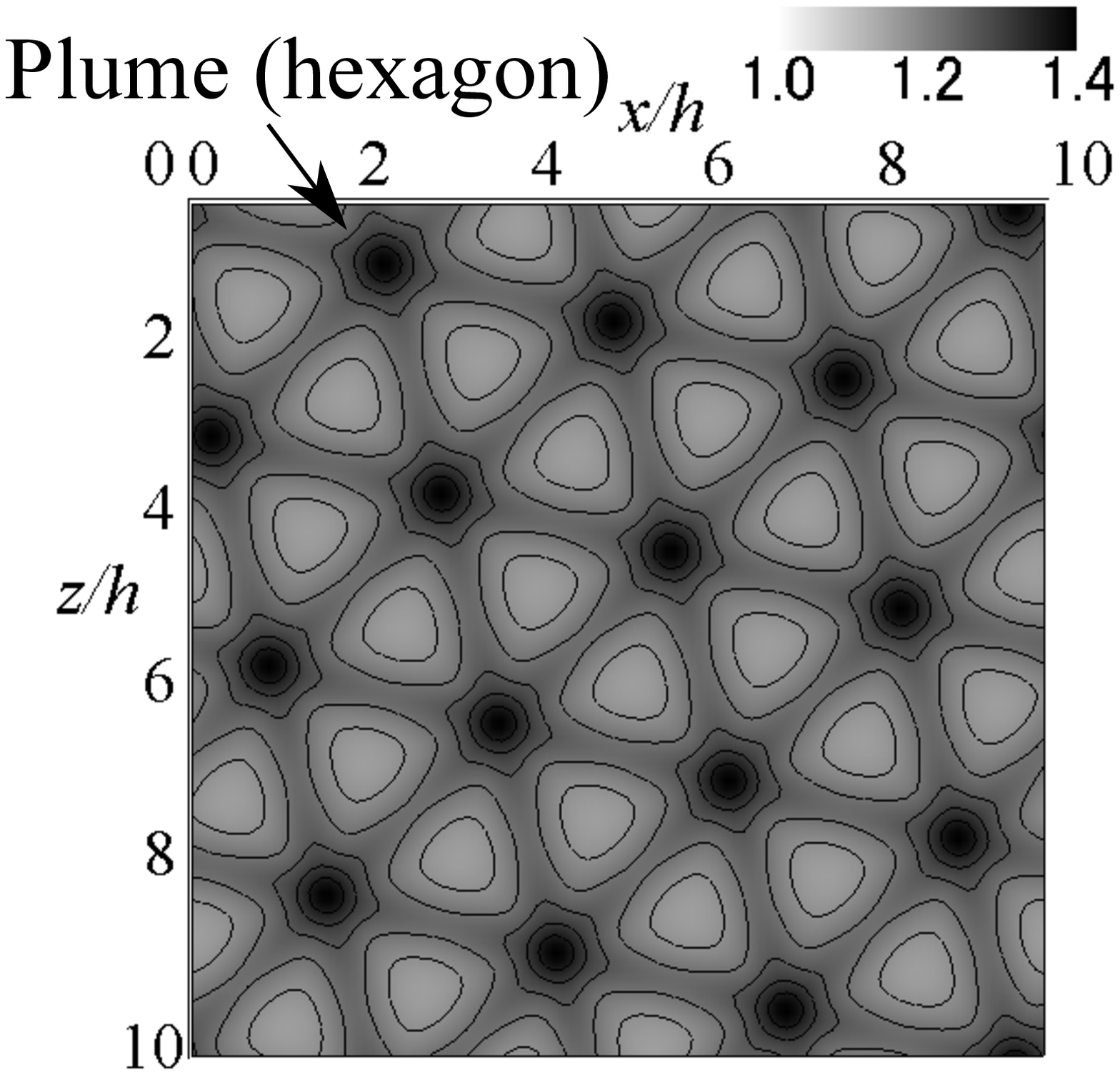} \\
(d) No.16 ($Ra=5000$) \\
\end{center}
\end{minipage}

\vspace*{1.0\baselineskip}
\begin{minipage}{.48\linewidth}
\begin{center}
\includegraphics[trim=0mm 0mm 0mm 0mm, clip, width=65mm]{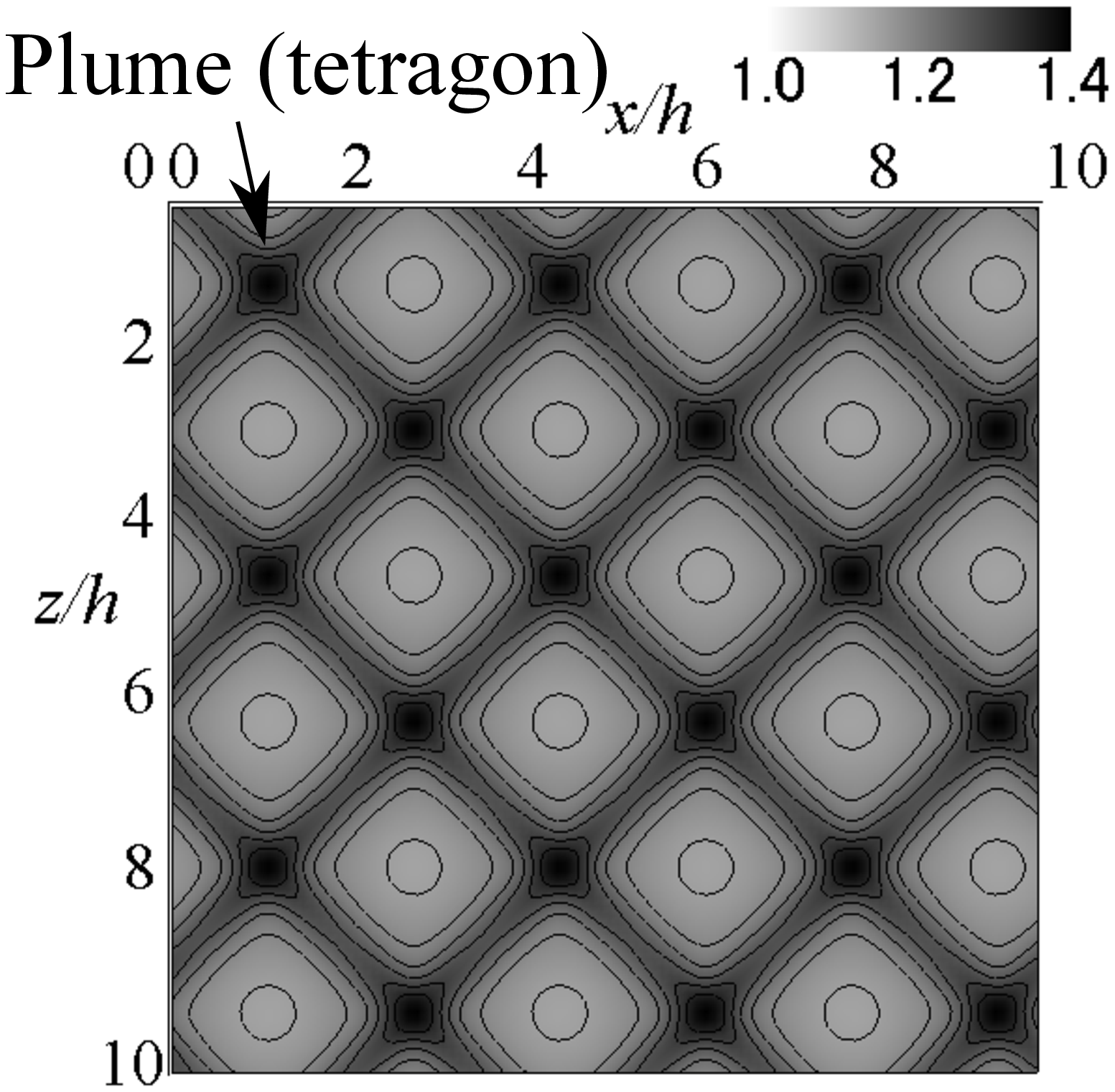} \\
(e) No.17 ($Ra=5000$) \\
\end{center}
\end{minipage}
\begin{minipage}{.48\linewidth}
\begin{center}
\includegraphics[trim=0mm 0mm 0mm 0mm, clip, width=65mm]{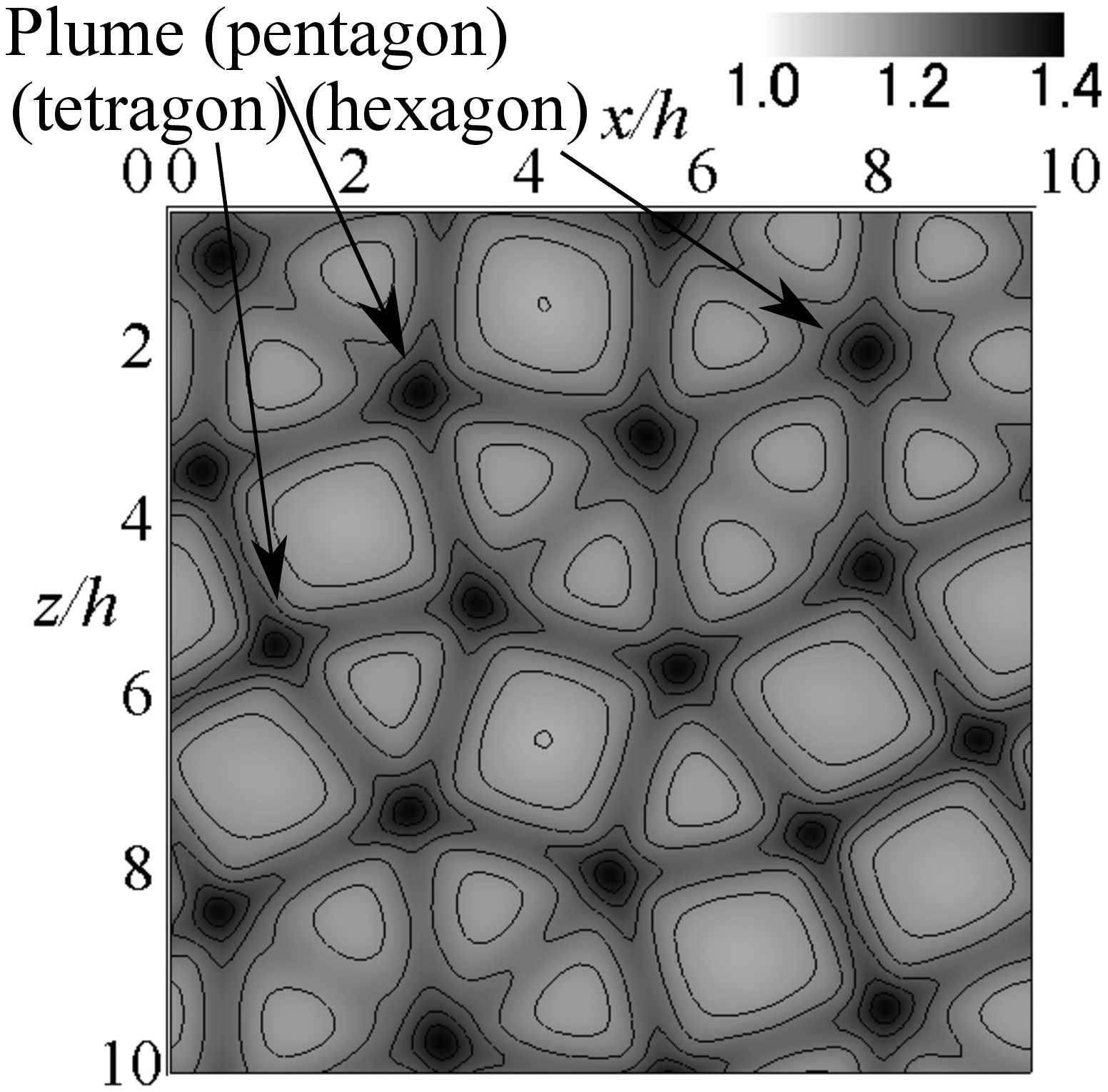} \\
(f) No.21 ($Ra=5000$) \\
\end{center}
\end{minipage}
\caption{Bacterial concentration contours in $x$--$z$ plane at $y/h=1.0$ 
for $Ra=500$, 1000, 3000, 5000.}
\label{pattern}
\end{figure}

In this subsection, we discuss the influence of the initial disturbance 
on the bioconvection pattern 
and the effect of the Rayleigh number $Ra$ on the wavelength of the pattern. 
Figure \ref{pattern} shows cell concentration contours on the free surface 
for $Ra=500$, 1000, 3000, and 5000.
The No. written with each figure corresponds to the pattern name described 
in table \ref{pattern&wavelength},
with a higher contour level indicating 
a higher cell concentration. 
The plume is formed in the dark-coloured region. 
In figures \ref{pattern}(b), (c), (d), (e), and (f), 
the high concentration regions appear to connect adjacent plumes. 
Since such a pattern looks like spokes, it was called a spoke pattern 
in an earlier study \citep{Mazzoni_et_al_2008}. 
\citet{Mazzoni_et_al_2008} reported that the spoke patterns 
have been observed in previous studies on bioconvection 
\citep{Platt_1961, Janosi_et_al_1998, Czirok_et_al_2000}. 
The present study was able to confirm the spoke pattern as shown 
in previous studies. 

Next, the arrangement of the plumes in the bioconvection patterns is compared. 
As shown in figures \ref{pattern}(a) and (d), the plumes appear 
in a ``staggered arrangement". 
The plumes in figures \ref{pattern}(b), (c), and (e) appear 
to have a ``lattice arrangement". 
In both the staggered and lattice arrangements, 
the plumes are regularly arranged. 
This regular plume arrangement has also been observed in a previous experiment 
on bioconvection \citep{Bees&Hill_1997} 
and a numerical analysis \citep{Karimi&Paul_2013}. 
In contrast, since the plumes appear irregularly in figure \ref{pattern}(f), 
this pattern is defined as a ``random arrangement". 

We also compare the plume shapes of each bioconvection pattern. 
This study defines the shape of the plumes by the number of spokes 
extending from one plume to the adjacent plumes in the spoke pattern. 
When the number of spokes extending from one plume was 4, 5, or 6, 
the plume shape is called a tetragon, pentagon, or hexagon, respectively. 
Since no spoke pattern was observed for $Ra=500$, 
the plume shape is called a circle. 
Focusing on the relation between the arrangement and shape of the plumes, 
the staggered arrangement shown forms 
a circular or hexagonal plume, 
while the lattice arrangement gives 
rise to a tetragonal plume. 
As can be observed in figure \ref{pattern}(f), tetragonal, pentagonal, 
and hexagonal plumes exist in an erratic mixture in a random arrangement. 
Such plume shapes have been confirmed in a previous experiment 
and numerical studies on bioconvection. 
The experimental result by \citet{Bees&Hill_1997} indicated 
that tetragonal or hexagonal plumes regularly form arrays in a long-term pattern. 
This result corresponds to the tetragonal and hexagonal plumes observed in this study. 
The experiment of \citet{Czirok_et_al_2000} showed 
that hexagonal plumes occur at the final pattern mode of bioconvection. 
A numerical analysis by \citet{Karimi&Paul_2013} confirmed 
the formation of a star-shaped plume. 
As can be seen by observing this plume in detail, 
a region of high cell concentration extends from one plume to an adjacent plume 
in five directions. 
This star-shaped plume is similar to the pentagonal plume defined in this study.

\begin{table}
\begin{minipage}{.48\linewidth}
\begin{center}
\begin{tabular}{ccccc}
 pattern's name & $Ra$ & arrangement & shape & $\lambda/h$ \\ \hline
  No.1 & 500 & staggered & circle & 3.54 \\
  No.2 & 500 & staggered & circle & 3.54 \\
  No.3 & 500 & staggered & circle & 3.54 \\
  No.4 & 500 & staggered & circle & 3.54 \\
  No.5 & 500 & staggered & circle & 3.54 \\
  No.6 & 500 & random & circle & 3.54 \\
  No.7 & 500 & staggered & circle & 3.54 \\
  No.8 & 500 & lattice & tetragon & 3.33 \\
  No.9 & 500 & staggered & circle & 3.16 \\
  No.10 & 500 & random & circle & 3.54 \\
  No.11 & 1000 & lattice & tetragon & 3.16 \\
  No.12 & 2000 & staggered & tetragon & 2.77 \\
\end{tabular}
\end{center}
\end{minipage}
\hspace*{0.2\baselineskip}
\begin{minipage}{.48\linewidth}
\begin{center}
\begin{tabular}{ccccc}
 pattern's name & $Ra$ & arrangement & shape & $\lambda/h$ \\ \hline
  No.13 & 3000 & lattice & tetragon & 2.77 \\
  No.14 & 4000 & lattice & tetragon & 2.50 \\
  No.15 & 5000 & random & mixture & 2.50 \\
  No.16 & 5000 & staggered & hexagon & 2.43 \\
  No.17 & 5000 & lattice & tetragon & 2.36 \\
  No.18 & 5000 & random & mixture & 2.24 \\
  No.19 & 5000 & random & mixture & 2.24 \\
  No.20 & 5000 & random & mixture & 2.43 \\
  No.21 & 5000 & random & mixture & 2.24 \\
  No.22 & 5000 & lattice & tetragon & 2.43 \\
  No.23 & 5000 & lattice & tetragon & 2.43 \\
  No.24 & 5000 & random & mixture & 2.50 \\
\end{tabular}
\end{center}
\end{minipage}
\caption{Rayleigh number $Ra$, arrangement of plumes, 
shape of plumes and pattern wavelength $\lambda/h$.}
\label{pattern&wavelength}
\end{table}

Table \ref{pattern&wavelength} shows the plume arrangement, plume shape, 
and wavelength $\lambda/h$ of the bioconvection pattern for each Rayleigh number. 
Here, the wavelength of the pattern was calculated 
using a two-dimensional fast Fourier transform of the cell concentration 
on the free surface as in previous studies 
\citep{Bees&Hill_1997, Czirok_et_al_2000, Williams&Bees_2011, Kage_et_al_2013}. 
First, we investigate the influence of the given initial disturbance 
in the cell concentration on the bioconvection patterns. 
This study used the initial cell concentration with ten different disturbances 
for $Ra=500$ and 5000. 
At $Ra=500$, the circular plumes showed a staggered arrangement 
for many of the calculation conditions, 
and no significant changes in the bioconvection pattern 
due to the initial disturbance were observed. 
In contrast, for $Ra=5000$, various bioconvection patterns 
including the staggered, lattice, and random arrangements occur 
due to the initial disturbances. 
This result indicates that even for the same Rayleigh number, 
bioconvection patterns with different plume arrangements, shapes, 
and numbers can arise due to the effects of different kinds of initial disturbances 
and that the wavelengths of the bioconvection patterns are different. 
In a previous experiment \citep{Bees&Hill_1997}, 
it was reported that it is difficult to achieve a homogeneous cell concentration 
as the ideal initial state of the suspension 
because fluid motion after mixing the suspension remains. 
\citet{Bees&Hill_1997} found that the wavelengths 
of the bioconvection patterns also varied, accompanying the pattern change, 
even though the same cell concentration and suspension depth were set. 
In this study, initial disturbances are added to the initial conditions 
of the cell concentration to simulate the initial state 
in the previous experiment \citep{Bees&Hill_1997}. 
From this result, it can be concluded that the present study confirms the phenomenon observed 
in the previous study. 
As mentioned above, it is found that more significant changes in the bioconvection pattern 
due to the initial disturbance appears 
at high Rayleigh number 
because the increase in the Rayleigh number enhances the non--linearity. 
Finally, comparing the wavelengths of the patterns in table \ref{pattern&wavelength}, 
it is confirmed that the wavelengths of the bioconvection patterns are changed 
depending on the plume arrangement. 
In addition, the wavelength of the pattern becomes shorter 
as the Rayleigh number increases.

\subsection{Wavelength comparison with previous results}

Next, we compare this numerical result with previous results 
for the wavelengths of bioconvection patterns. 
\citet{Bees&Hill_1997} conducted experiments on bioconvection formed by 
single-celled alga {\it Chlamydomonas nivalis} 
and investigated the wavelengths of the bioconvection patterns 
at the onset of bioconvection and in a long-term pattern. 
They found that the wavelength decreases with increasing cell concentration 
and that the cell concentration has a significant effect on the wavelength 
of the pattern. 
\citet{Czirok_et_al_2000} experimentally investigated the bioconvection 
formed by {\it Bacillus subtilis} 
and measured the wavelength of the bioconvection pattern 
at the onset of bioconvection. 
They clarified that the wavelength decreases with increasing 
in cell concentration measured by optical density measurements. 
To compare our numerical results with earlier experimental results 
using the Rayleigh number, we calculated the Rayleigh number 
of the experimental data. 
The volume of a cell $V$, density ratio of a cell to water 
$(\rho_n-\rho)/\rho$, water kinematic viscosity $\nu$, 
and cell diffusivity $D_{n0}$ were not described in Bees and Hill's paper, 
so we used values from \citet{Ghorai&Hill_2002}. 
Likewise, $V$, $(\rho_n-\rho)/\rho$, $\nu$, and $D_{n0}$ 
were not mentioned in \citet{Czirok_et_al_2000}, 
so we took $\nu$ from \citet{Hillesdon&Pedley_1996}, 
and other values from \citet{Janosi_et_al_1998}. 
The gravity acceleration $g$ was not reported in these papers 
\citep{Hillesdon&Pedley_1996, Bees&Hill_1997, Janosi_et_al_1998, Czirok_et_al_2000, 
Ghorai&Hill_2002}, and thus we took $g$ from \citet{Pedley_et_al_1988}. 
The numerator of the Rayleigh number equation (\ref{parameter}) 
defined in this paper includes the initial cell concentration $n_0$. 
Thus, it can be considered that increasing the Rayleigh number increases 
the initial cell concentration.

\begin{figure}
\begin{minipage}{0.485\linewidth}
\begin{center}
\includegraphics[trim=2mm 0mm 0mm 2mm, clip, width=65mm]{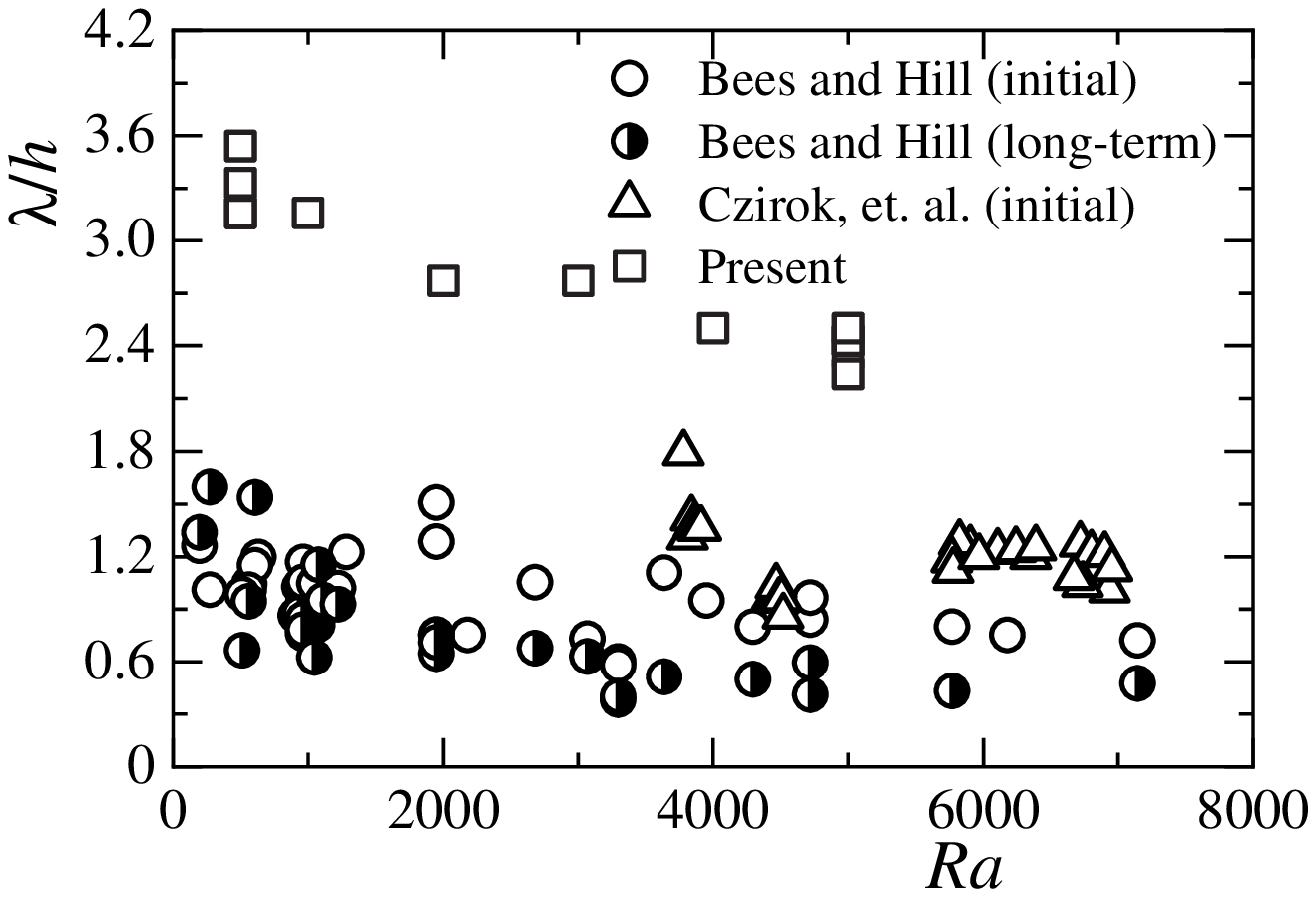} \\
(a) comparison with experiment
\end{center}
\end{minipage}
\hspace{0.01\linewidth}
\begin{minipage}{0.485\linewidth}
\begin{center}
\includegraphics[trim=2mm 0mm 0mm 2mm, clip, width=65mm]{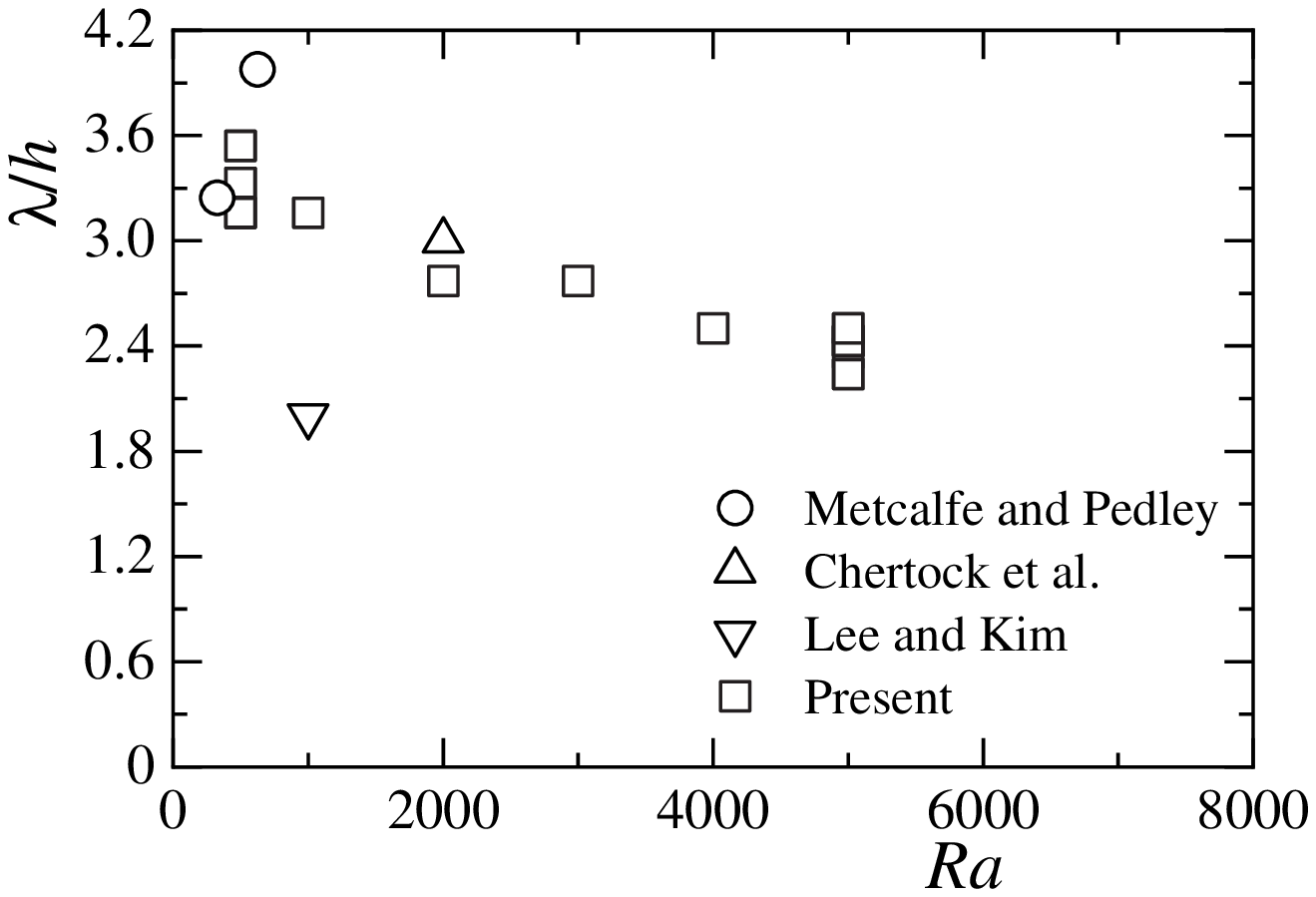} \\
(b) comparison with calculation
\end{center}
\end{minipage}
\caption{Comparison of pattern wavelength with previous results 
\citep{Bees&Hill_1997, Czirok_et_al_2000, Metcalfe&Pedley_1998, Chertock_et_al_2012, Lee&Kim_2015}.}
\label{comparison_experiment}
\end{figure}

Figure \ref{comparison_experiment} shows a comparison 
between our numerical values and experimental values 
\citep{Bees&Hill_1997, Czirok_et_al_2000} for the bioconvection pattern. 
The trend of the wavelength of the pattern to decrease 
with increasing Rayleigh number is qualitatively consistent with 
the experimental results. 
However, there is a quantitative difference between the wavelengths 
in this calculation and the experiments. 

\citet{Metcalfe&Pedley_1998} used weakly nonlinear analysis 
to determine the critical wavenumber and critical Rayleigh number 
at which a suspension becomes unstable. 
We obtained the wavelength at the critical Rayleigh number for two conditions 
($\gamma \beta=1$, $\delta=1$, $Sc=7700$) 
and ($\gamma \beta=50$, $\delta=1$, $Sc=7700$) in the existing research. 
In addition, using the previous calculations 
\citep{Chertock_et_al_2012, Lee&Kim_2015}, 
we calculate the wavelength of the bioconvection pattern generated 
when an initial disturbance is given by random numbers. 
The parameters in the calculation of 
\citet{Chertock_et_al_2012} are $\beta=4$, $\gamma=10$, $\delta=5$, and $Sc=500$, 
and the parameters in the calculation of 
\citet{Lee&Kim_2015} are $\beta=2$, $\gamma=10$, $\delta=5$, and $Sc=500$. 
We calculated the wavelength from the number of plumes 
and the length of the calculation area. 
Figure \ref{comparison_experiment} (b) compares the wavelength 
derived from the previous results with the present results. 
In all analyses using chemotactic bacteria, 
the parameters used are different, 
but the previous results are in the range of $\lambda/h=2.0-4.0$. 
These values are overestimated as compared to the experimental values 
for all Rayleigh numbers. 
Thus, further investigations into this difference are required.

Earlier studies \citep{Bees&Hill_1997, Ghorai&Hill_2002, Williams&Bees_2011} 
reported similarities between bioconvection and thermal convection. 
Therefore, the patterns observed in previous experiments for thermal convection 
\citep{Koschmieder&Switzer_1992} 
and in numerical analysis \citep{Tomita&Abe_2000} 
and the wavelengths of the patterns are compared with the present values. 
\citet{Koschmieder&Switzer_1992} showed that the number of convection cells increased 
as the temperature difference applied to the fluid increased. 
In that paper, the Rayleigh number is defined, and an increase 
in temperature difference corresponds to an increase in the Rayleigh number. 
An increase in the number of convection cells means a decrease 
in the spacing between convection cells, that is, the wavelength of the pattern. 
\citet{Tomita&Abe_2000} investigated B\'{e}nard--Marangoni convection 
and found that the wavelength of the pattern decreases 
as the parameter that combines the Rayleigh and Marangoni numbers increases. 
Increasing this parameter means increasing the Rayleigh number. 
Therefore, it can be said that the trend of the wavelength 
of the bioconvection pattern to decrease with increasing Rayleigh number 
is qualitatively consistent with the result of thermal convection. 
Furthermore, when comparing the patterns, the hexagonal convection pattern 
observed in the 
previous studies \citep{Koschmieder&Switzer_1992,Tomita&Abe_2000} is similar to 
the hexagonal-shaped plume observed in the present study. 
The wavelength of the hexagonal convection pattern \citep{Tomita&Abe_2000} 
is $2.71 \le \lambda/h \le 2.94$, 
which is similar to the wavelength of the bioconvection pattern 
observed in this study.

\subsection{Interference between plumes}

\begin{figure}
\begin{minipage}{0.485\linewidth}
\begin{center}
\includegraphics[trim=2mm 0mm 0mm 2mm, clip, width=65mm]{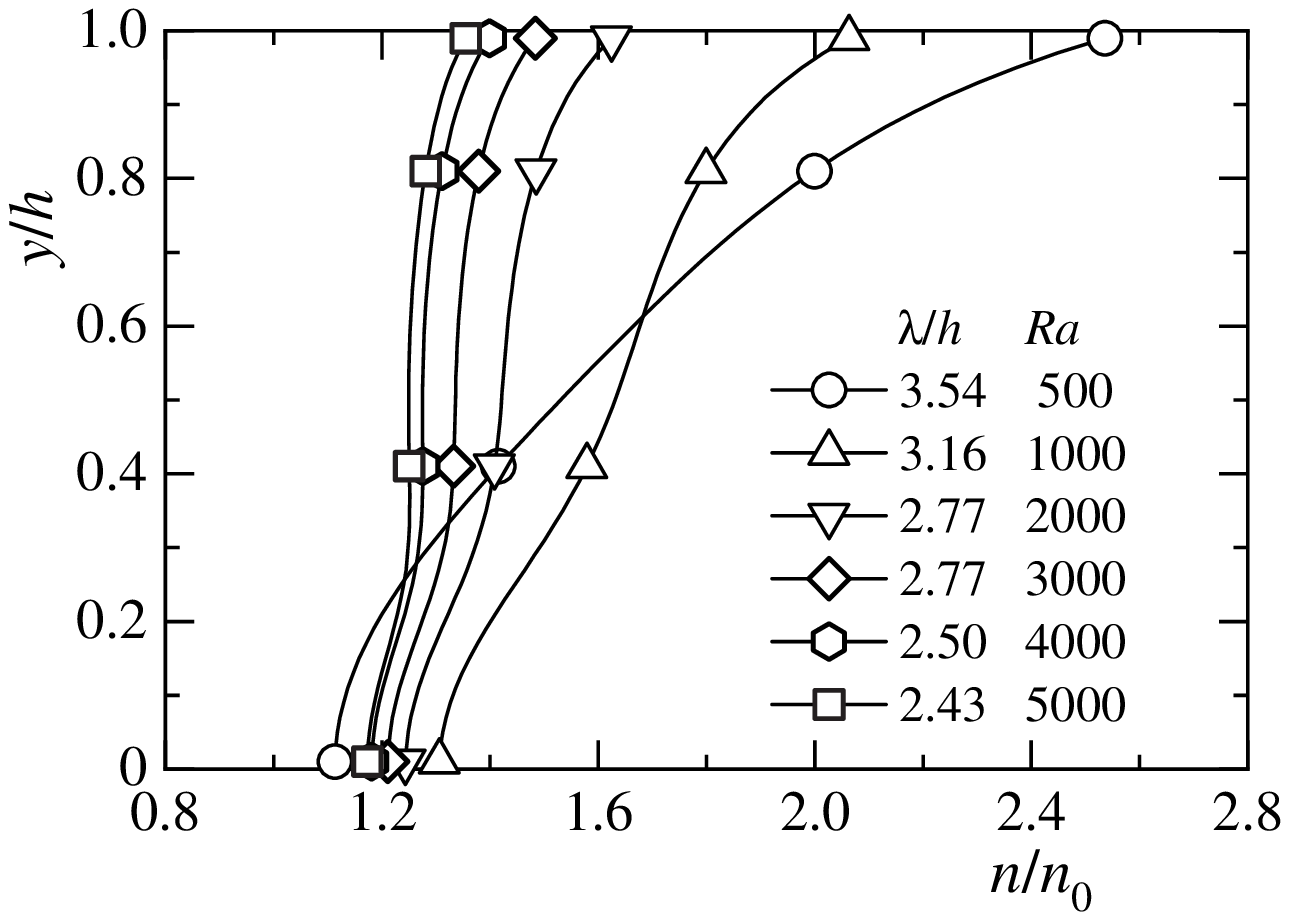} \\
(a) centre of plume \\
\end{center}
\end{minipage}
\hspace{0.01\linewidth}
\begin{minipage}{0.485\linewidth}
\begin{center}
\includegraphics[trim=2mm 0mm 0mm 2mm, clip, width=65mm]{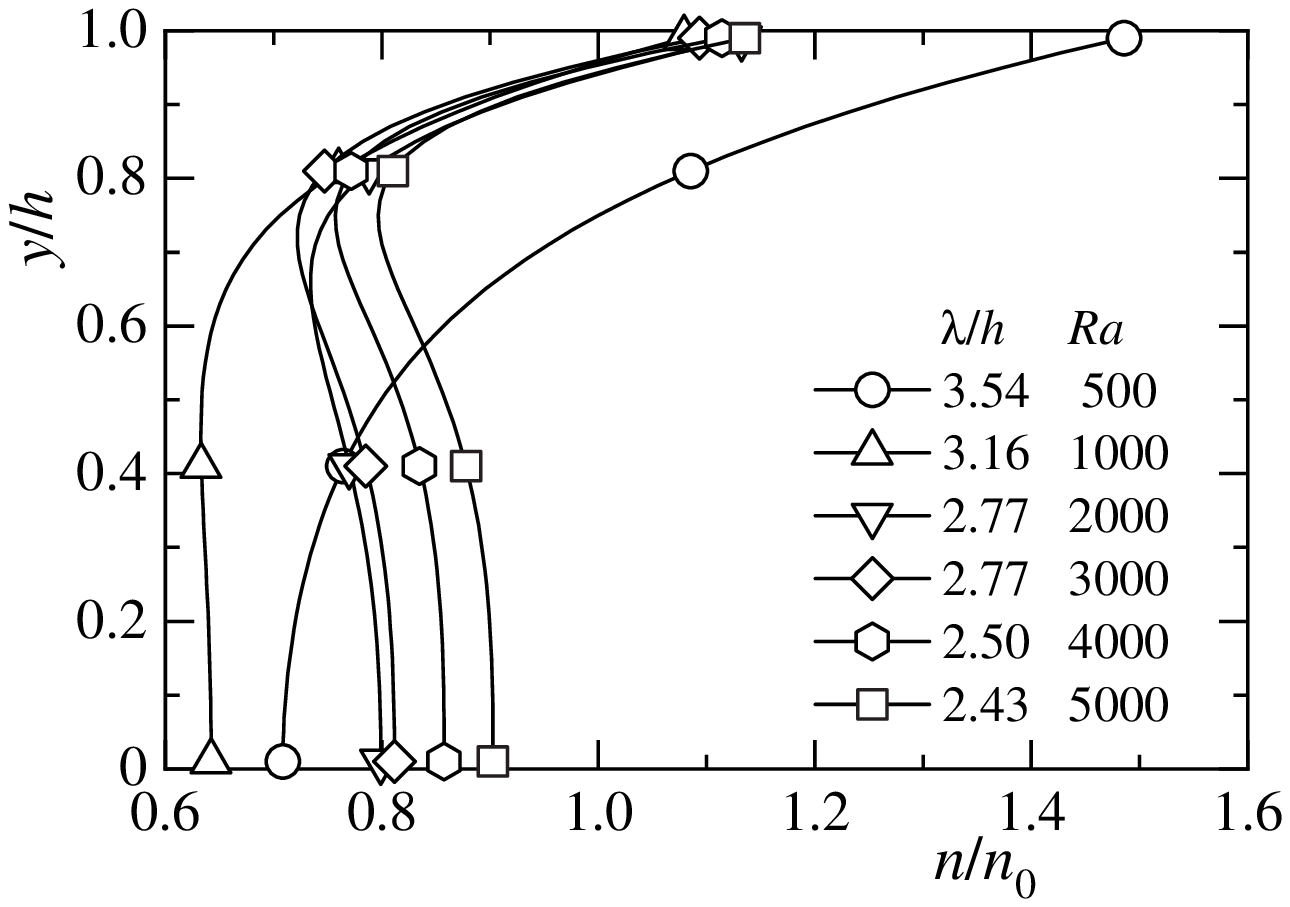} \\
(b) between plumes
\end{center}
\end{minipage}
\caption{Bacterial concentration distributions in $y$-direction 
at plume centre and between plumes.}
\label{dist_bacteria}
\end{figure}

\begin{figure}
\begin{minipage}{0.485\linewidth}
\begin{center}
\includegraphics[trim=2mm 0mm 0mm 2mm, clip, width=65mm]{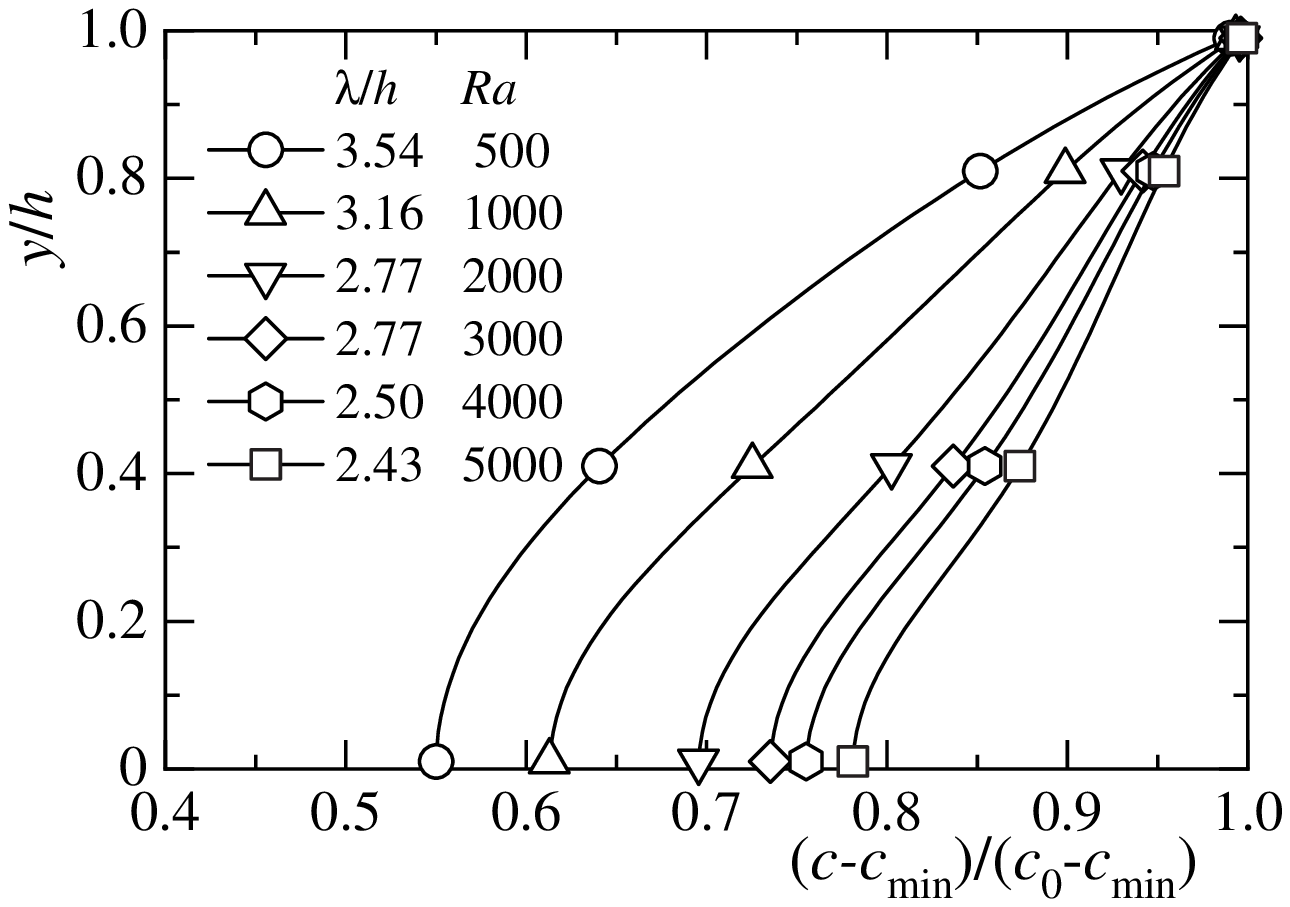} \\
(a) centre of plume \\
\end{center}
\end{minipage}
\hspace{0.01\linewidth}
\begin{minipage}{0.485\linewidth}
\begin{center}
\includegraphics[trim=2mm 0mm 0mm 2mm, clip, width=65mm]{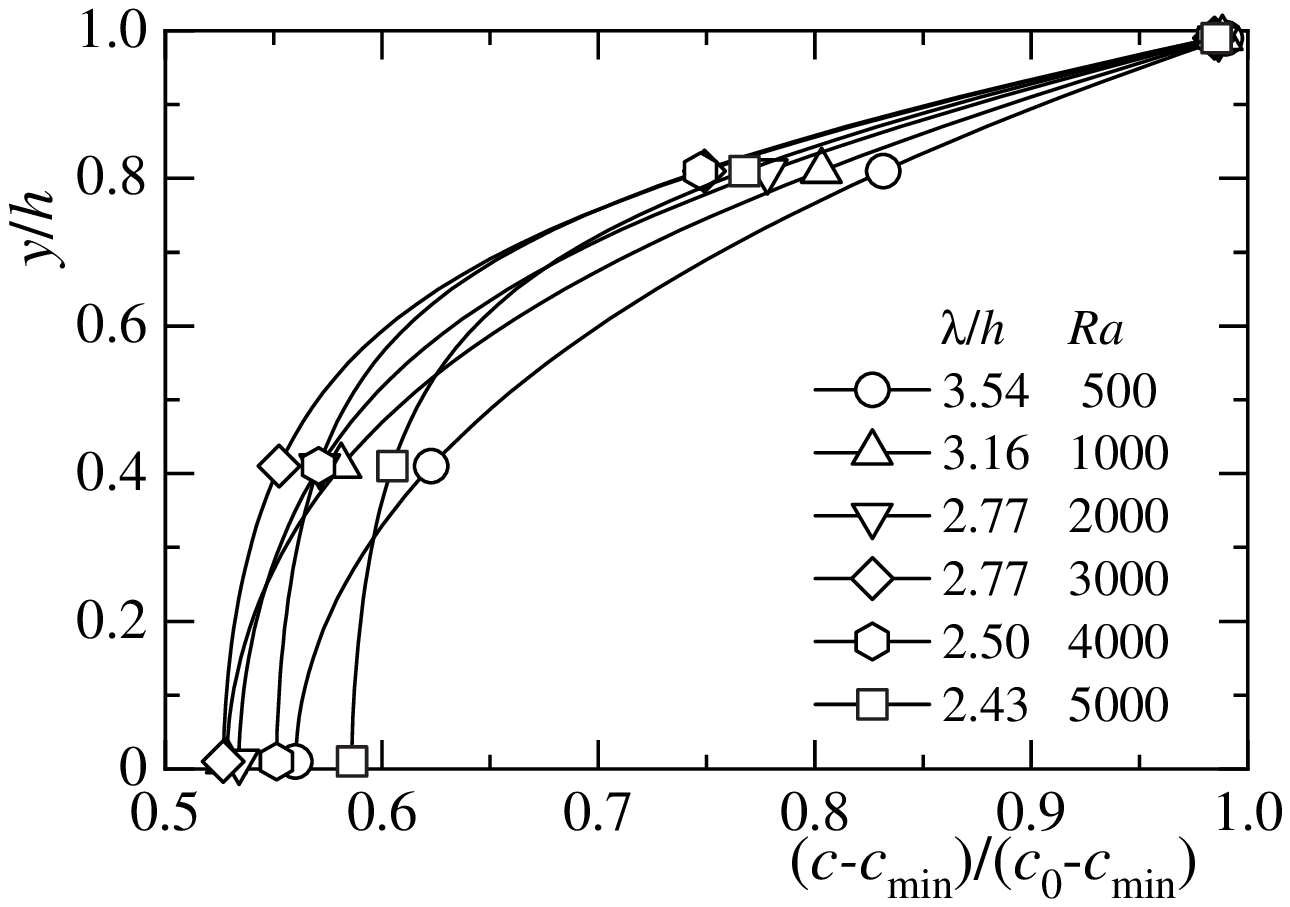} \\
(b) between plumes
\end{center}
\end{minipage}
\caption{Oxygen concentration distributions in $y$-direction 
at plume centre and between plumes.}
\label{dist_oxygen}
\end{figure}

We discuss the influence of the wavelength variation 
of the bioconvection pattern on the concentration field. 
First, figures \ref{dist_bacteria} and \ref{dist_oxygen} show 
the distribution of cell and oxygen concentrations in the $y$-direction 
at the plume centre and between plumes. 
Here, No.2 ($Ra=500$), No.11 ($Ra=1000$), No.12 ($Ra=2000$), 
No.13 ($Ra=3000$), No.14 ($Ra=4000$), 
and No.16 ($Ra=5000$) in table \ref{pattern&wavelength} are used for the results. 
All these results used the same initial disturbance. 
We consider the positions at the plume centre and between the plumes 
as being the points where the cell concentrations on the free surface 
become maximum and minimum, respectively. 
The positions of the plume centres at $Ra=500$, 1000, 2000, 3000, 4000, 5000 
are ($x/h$, $z/h$) = (4.7, 0.4), (5.8, 3.2), (8.0, 0.5), (1.9, 3.5), (1.5, 8.8), 
(2.3, 0.8), 
and the positions between the plumes are ($x/h$, $z/h$) = (3.4, 1.6), 
(5.8, 8.2), (3.0, 6.2), (6.9, 8.5), (7.7, 2.6), (6.1, 2.6). 

In the centre of the plume, as the wavelength of the pattern decreases, 
the amount of the cell transported by the downward flow increases, 
so the cell concentration near the water surface decreases. 
As can be seen from figures \ref{Ra500} and \ref{Ra5000}, 
bioconvection strengthens as the wavelength of the pattern decreases, 
and the plume extending to the lower wall diffuses to the surroundings. 
This indicates that the cell amount transported from the plume centre 
to between the plumes increases. 
Therefore, as the wavelength decreases, 
the cell concentration near the lower wall decreases 
at the plume centre, converges to a constant value, 
and increases between the plumes. 
The cell concentration near the water surface between the plumes 
does not change depending on the wavelength, except at $Ra=500$, 
and is lower than the concentration at the plume centre. 
At $Ra=500$, the bioconvection is weak, 
so the trend is different from the results at other Rayleigh numbers.

Next, we consider the oxygen concentration. 
At the plume centre, the oxygen concentration near the lower wall increases 
because the oxygen transported by the downward flow increases 
as the wavelength of the pattern decreases. 
Between the plumes, the oxygen concentration decreases at $\lambda/h>2.77$ ($Ra<3000$) 
as the wavelength decreases. 
This is because as the wavelength decreases, 
the amount of cells transported between plumes increases 
and the amount of oxygen consumed increases. 
On the other hand, the tendency of $\lambda/h \leq 2.77$ ($Ra \geq 3000$) 
is different from the result of $\lambda/h>2.77$ ($Ra<3000$), 
and the oxygen concentration increases as the wavelength decreases. 
This is because, as the wavelength decreases, 
bioconvection strengthens 
and the amount of oxygen transported between plumes exceeds 
the amount of oxygen consumed by bacteria.

\begin{figure}
\begin{minipage}{0.485\linewidth}
\begin{center}
\includegraphics[trim=2mm 0mm 0mm 2mm, clip, width=65mm]{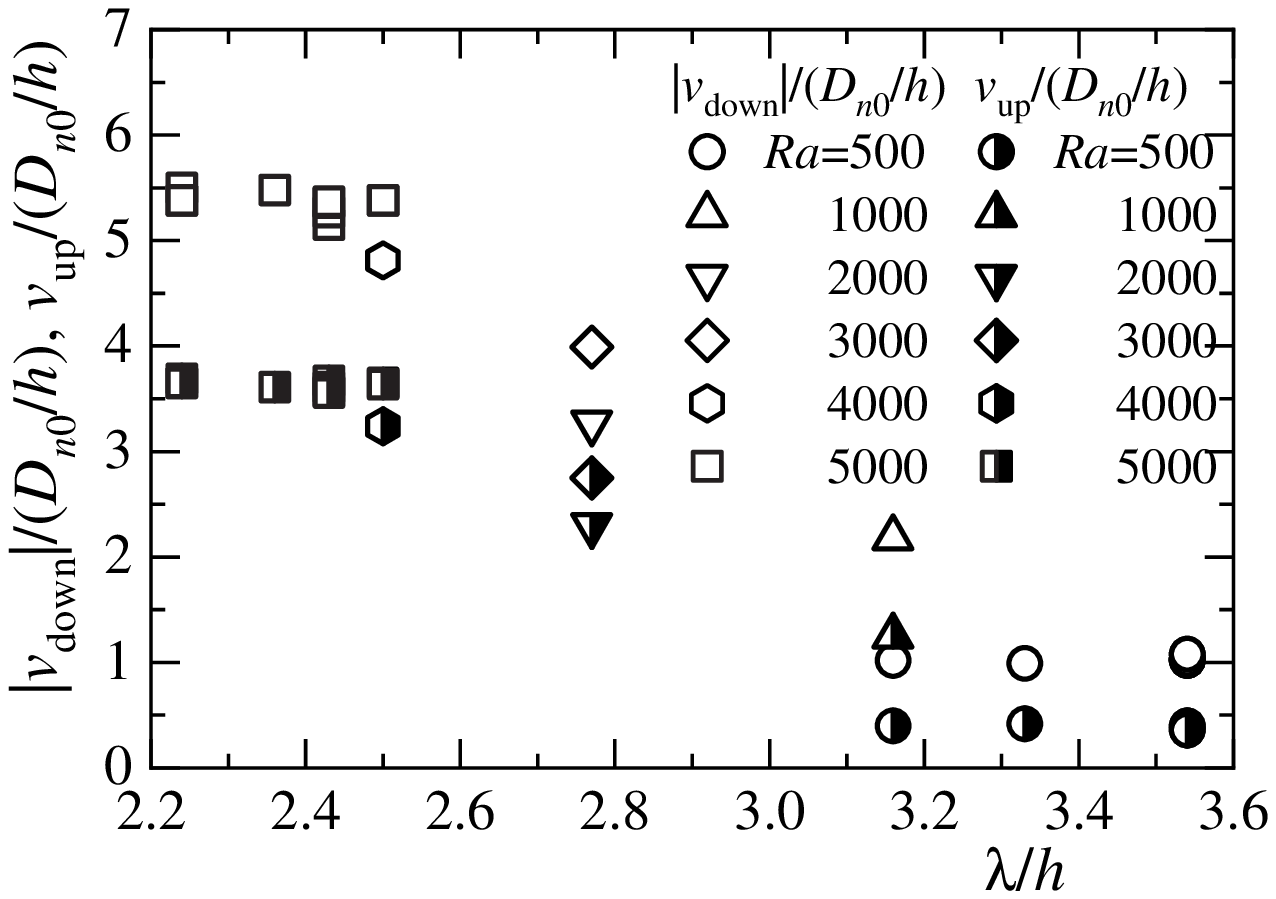} \\
(a) mean velocity \\
\end{center}
\end{minipage}
\hspace{0.01\linewidth}
\begin{minipage}{0.485\linewidth}
\begin{center}
\includegraphics[trim=2mm 0mm 0mm 2mm, clip, width=65mm]{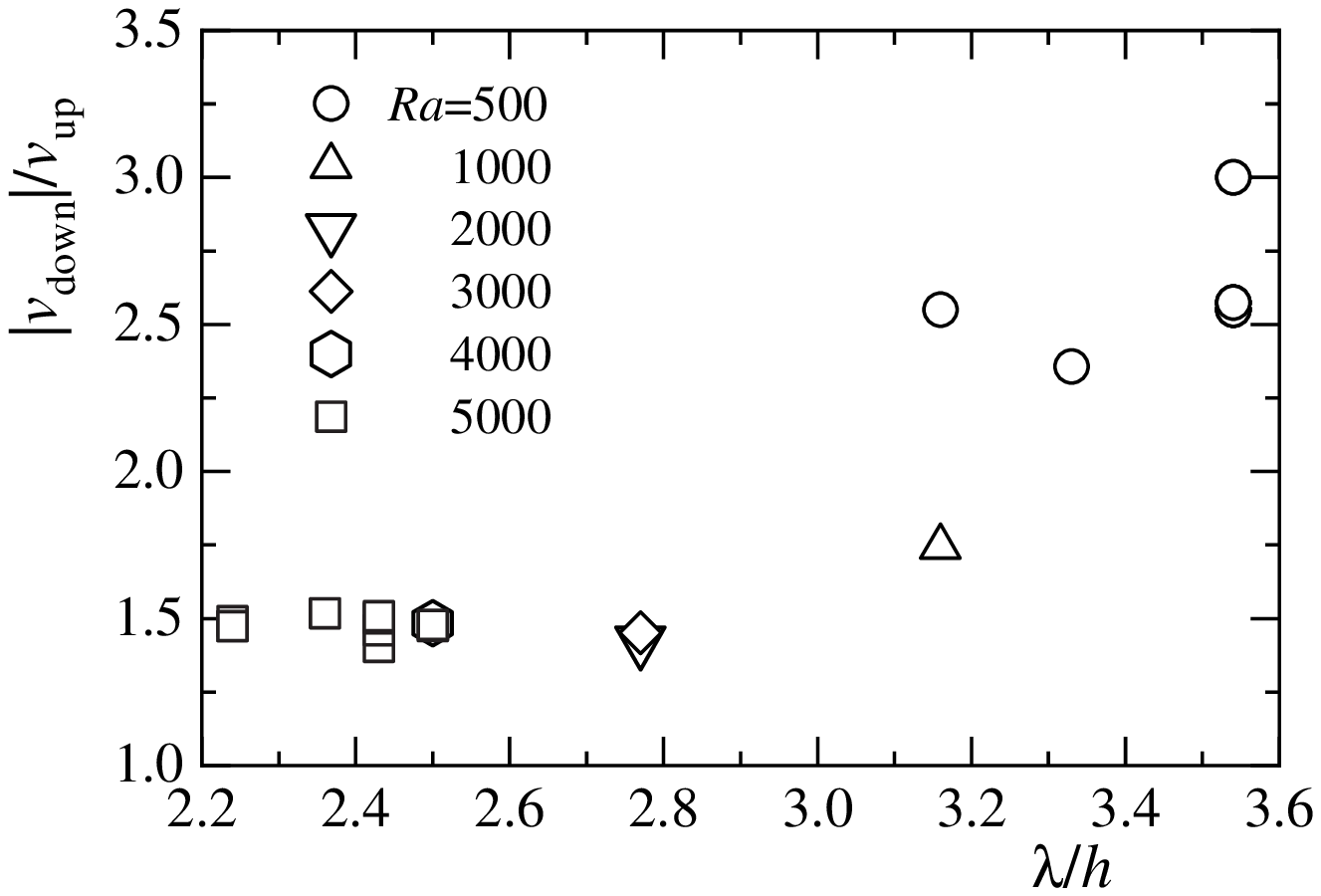} \\
(b) velocity ratio
\end{center}
\end{minipage}
\caption{Mean velocities of downward and upward flows.}
\label{wavelength_v}
\end{figure}

This study calculated the mean velocity, $v_\mathrm{down}$, 
of the downward flow at the centre of the plume, 
and the mean velocity, $v_\mathrm{up}$, of the upward flow between plumes. 
$v_\mathrm{down}$ and $v_\mathrm{up}$ are defined 
by averaging the vertical velocities such that $v < 0$ and $v > 0$, respectively. 
Figure \ref{wavelength_v} (a) shows the relationship 
between the wavelength of the pattern and mean velocities, $v_\mathrm{down}$ and $v_\mathrm{up}$. 
In the figure, the wavelength of the pattern decreases 
as the Rayleigh number increases, 
and then $|v_\mathrm{down}|$ increases at the centre of the plume. 
Also, we can observe the same tendency in $v_\mathrm{up}$ 
between the plumes shown in the figure.
This result occurs because the interference between plumes has increased. 
In other words, as the wavelength of the pattern becomes shorter, 
the plumes approach each other, and the velocities of the downward flow 
and the upward flow increase. 
In addition, we have observed in figure \ref{Ra5000} 
that the velocity of the downward flow is faster than that of the upward flow. 
The wavelengths of the patterns for $Ra=2000$ and $Ra=3000$ 
have the same values. 
However, as can be seen from $|v_\mathrm{down}|$ and $v_\mathrm{up}$, 
the flow velocities of $Ra=3000$ are faster than those of $Ra=2000$. 
This result is because the number of cells transported downward increases 
as the Rayleigh number increases, and then the bioconvection is strengthened. 

Figure \ref{wavelength_v} (b) shows the ratio of the downward velocity 
to the upward velocity. 
This ratio is also related to the volume occupied by the plumes 
and the volume occupied between the plumes. 
As can be seen in figure \ref{Ra5000}, 
the plumes with high cell concentrations are narrower 
than the areas between the plumes. 
Figure \ref{wavelength_v} (b) shows 
that if the net upward and downward transports are balanced on average, 
then the velocity of the descending flow in the plume will be faster than 
that of the upward flow, and the area occupied by the plume will become small. 
When the wavelength is large, the velocity ratio reaches three, 
and the area of influence of the downward flow in the plume on the cell transport 
becomes narrow. 
As the wavelength decreases, the velocity ratio decreases. 
This suggests that the ascending flow velocity increases 
and the area of influence of the ascending flow on the cell transport becomes narrow.

To compare the strength of convection in each pattern, 
the average kinetic energy within the calculation domain was calculated for $Ra=5000$. 
If the kinetic energy is high, it can be said that the convection is strong on average. 
The mean value obtained from the average kinetic energy of all patterns 
was $K_{av}=47.2$, and the standard deviation was 0.341. 
We find that the strength of bioconvection does not depend on the pattern.

\begin{figure}
\begin{minipage}{.328\linewidth}
\begin{center}
\includegraphics[trim=8mm 0mm 6mm 0mm, clip, width=46mm]{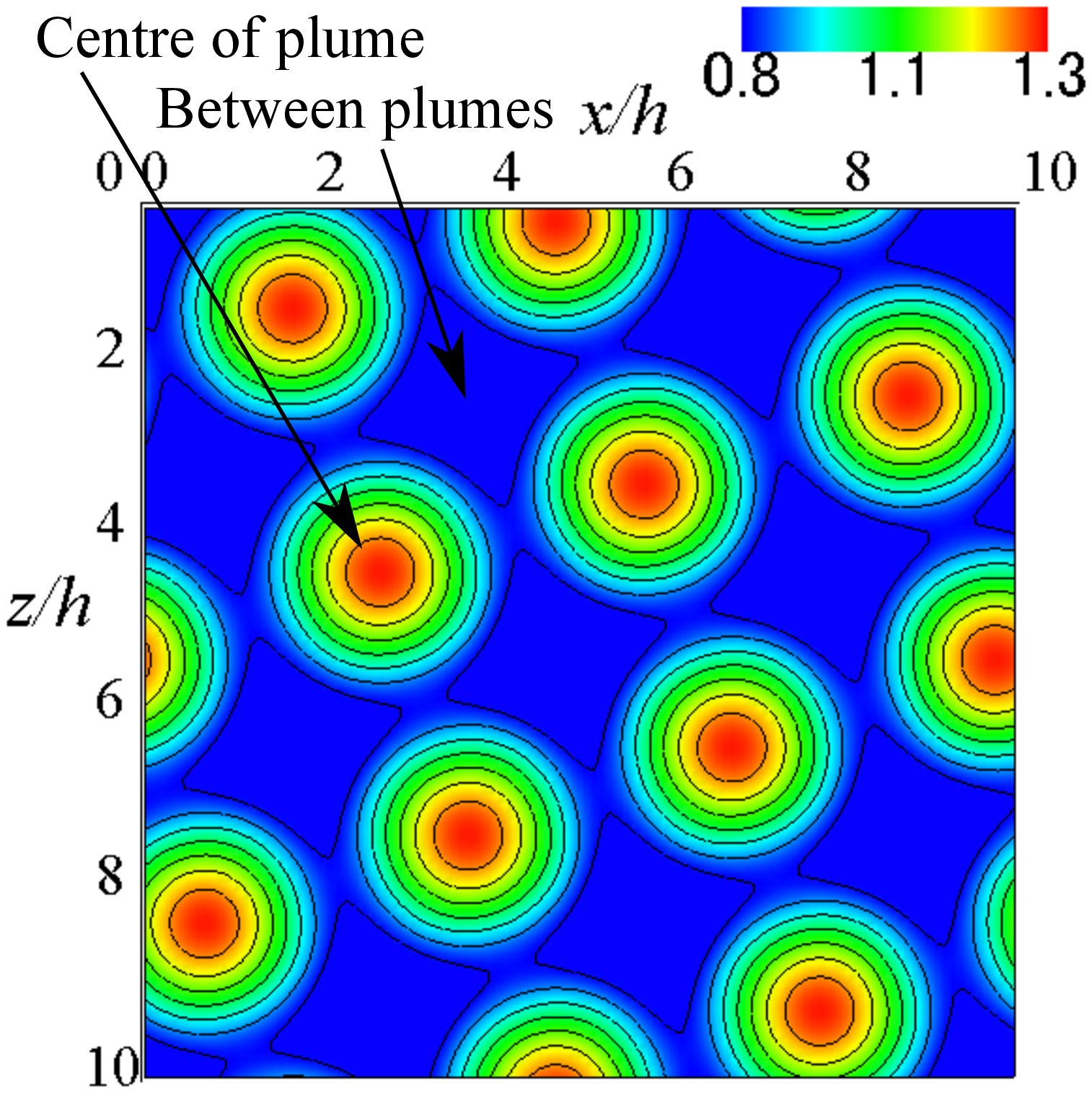} \\
(a) $\lambda/h=3.16$ ($Ra=1000$)
\end{center}
\end{minipage}
\hspace{0.002\linewidth}
\begin{minipage}{.328\linewidth}
\begin{center}
\includegraphics[trim=8mm 0mm 6mm 0mm, clip, width=46mm]{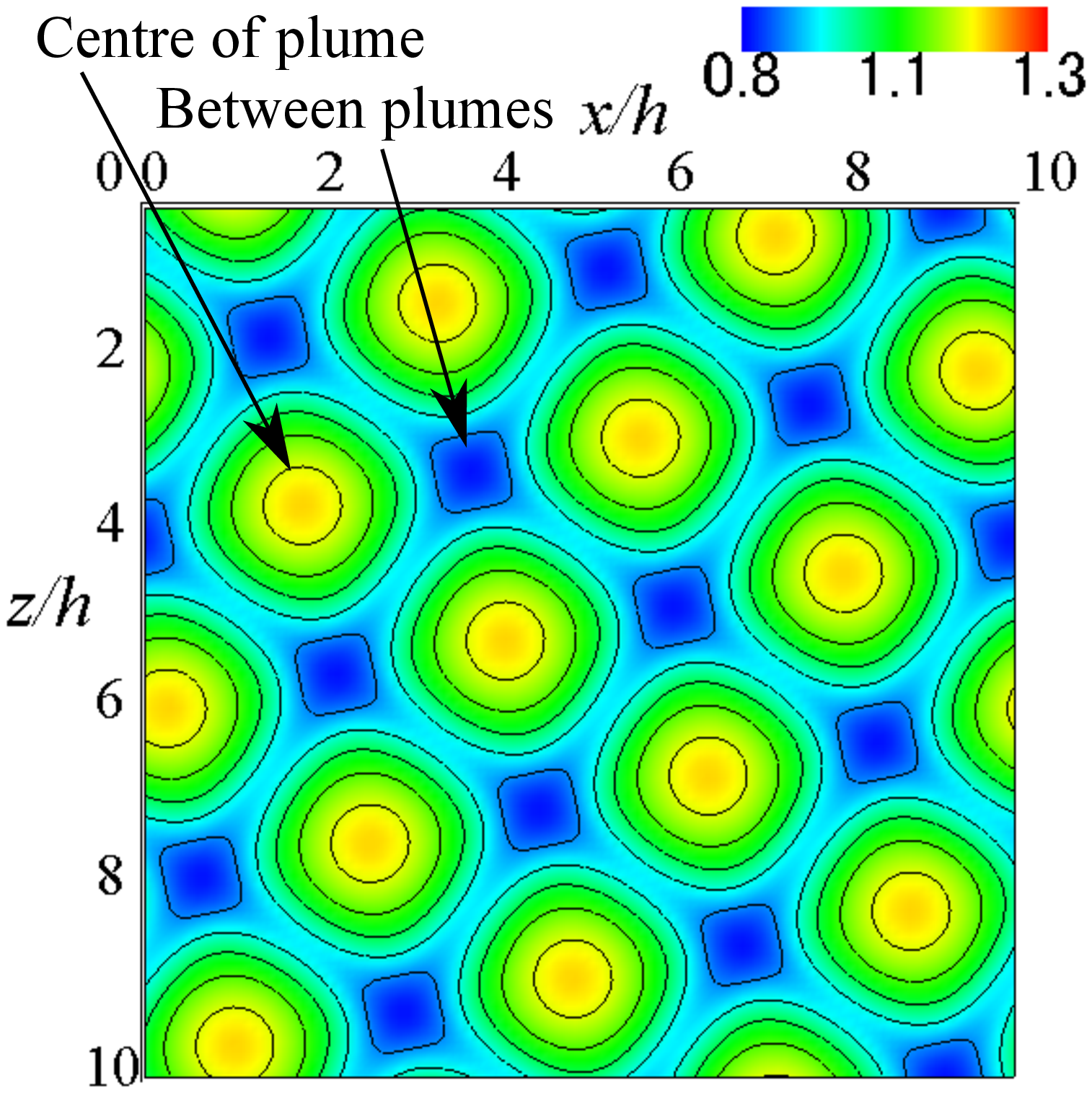} \\
(b) $\lambda/h=2.77$ ($Ra=3000$)
\end{center}
\end{minipage}
\hspace{0.002\linewidth}
\begin{minipage}{.328\linewidth}
\begin{center}
\includegraphics[trim=8mm 0mm 6mm 0mm, clip, width=46mm]{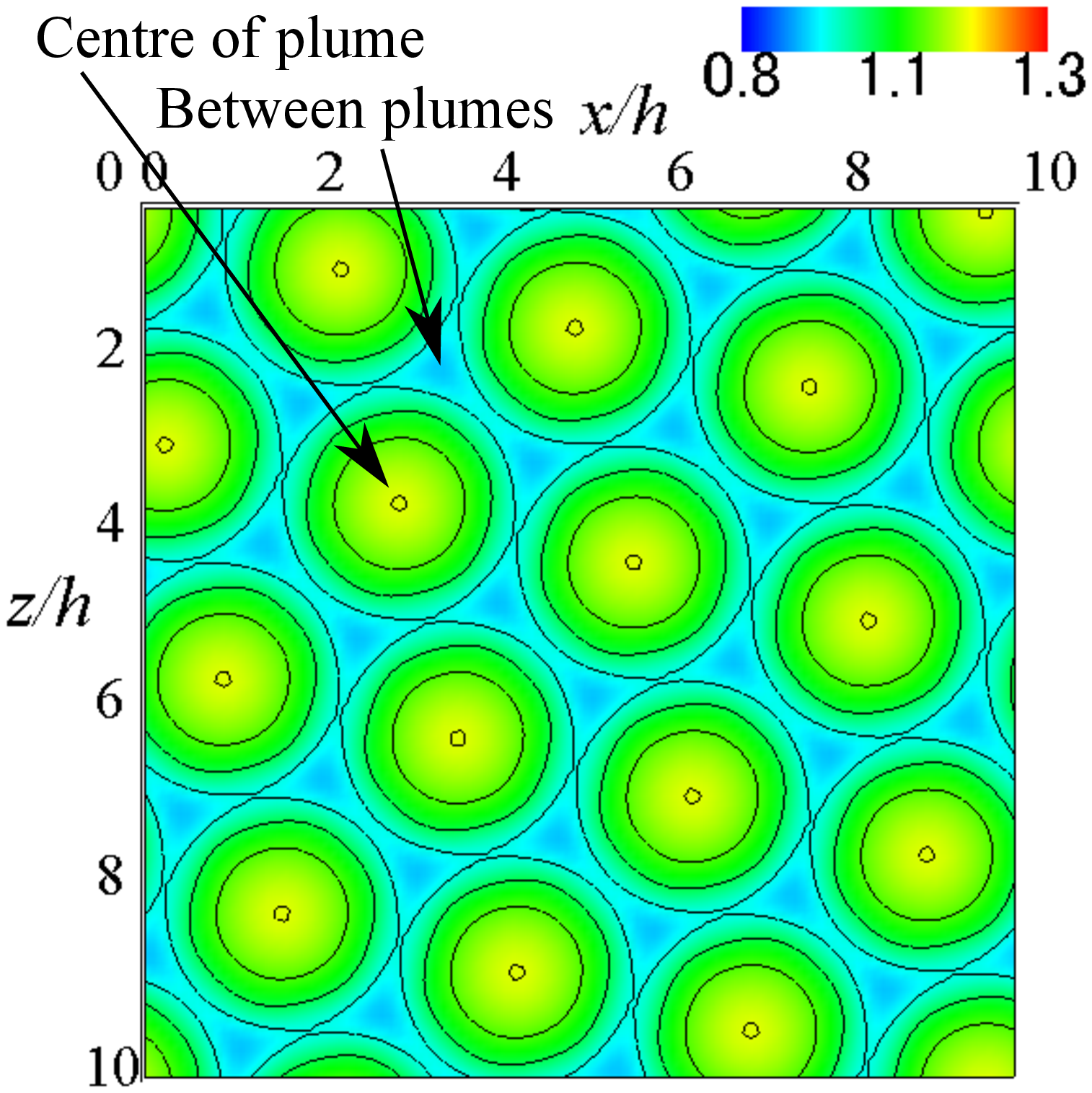} \\
(c) $\lambda/h=2.43$ ($Ra=5000$)
\end{center}
\end{minipage}
\caption{Bacterial concentration contours in $x$--$z$ plane 
at $y/h=0.01$ for $Ra=1000$, 3000, 5000.}
\label{bottom_bacteria}
\end{figure}

\begin{figure}
\begin{minipage}{.328\linewidth}
\begin{center}
\includegraphics[trim=12mm 0mm 4mm 0mm, clip, width=46mm]{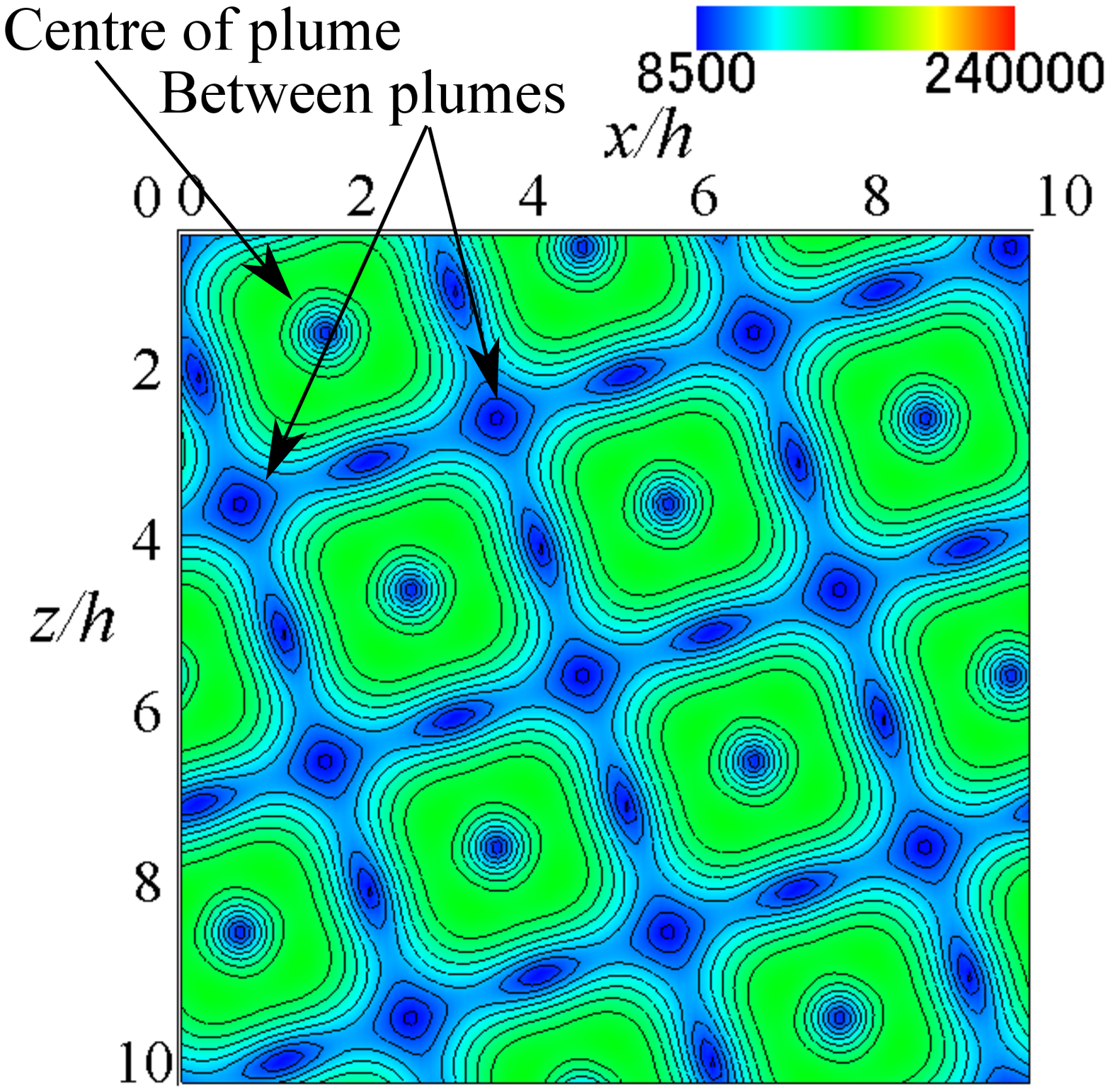} \\
(a) $\lambda/h=3.16$ ($Ra=1000$)
\end{center}
\end{minipage}
\hspace{0.002\linewidth}
\begin{minipage}{.328\linewidth}
\begin{center}
\includegraphics[trim=12mm 0mm 4mm 0mm, clip, width=46mm]{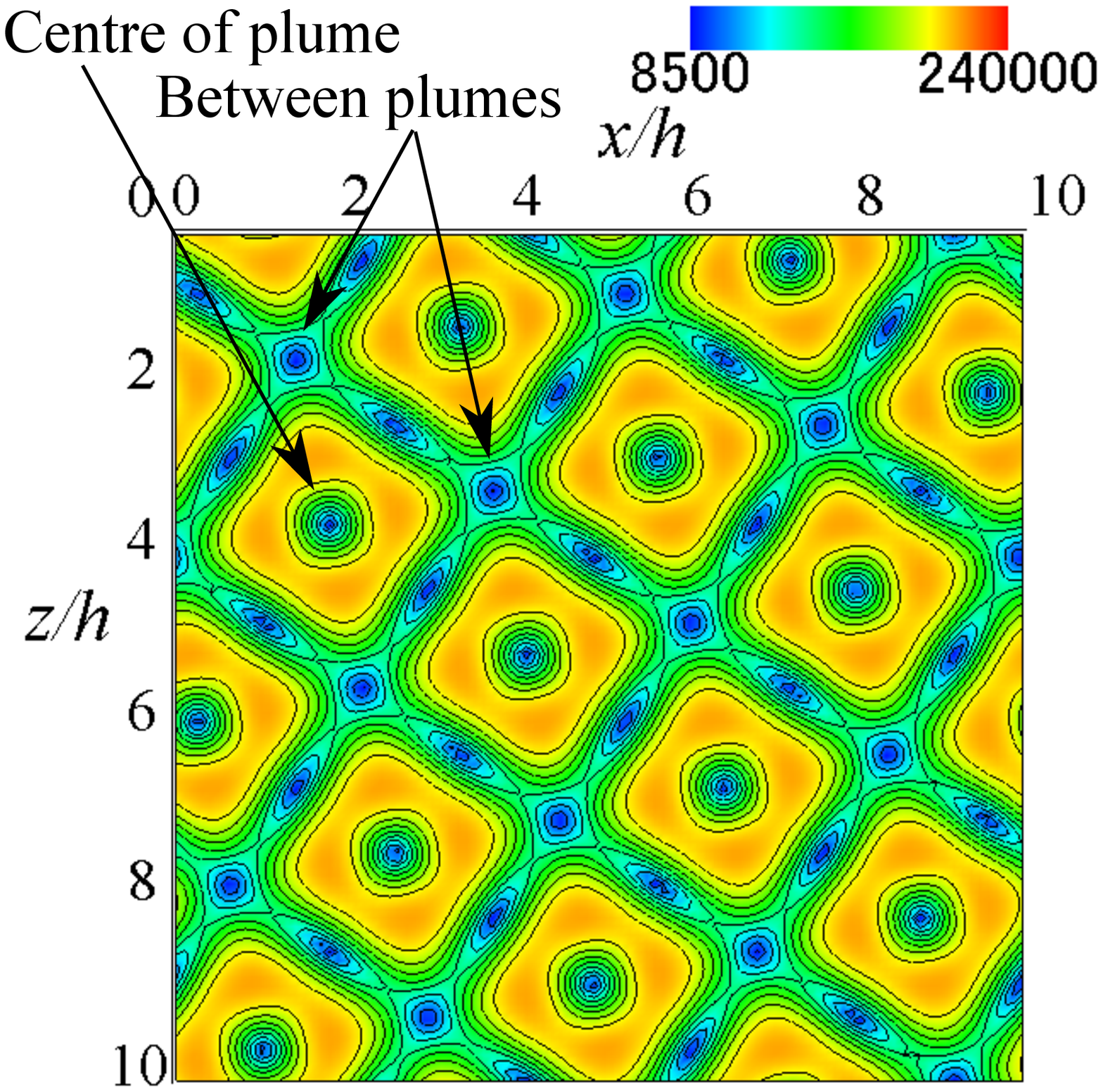} \\
(b) $\lambda/h=2.77$ ($Ra=3000$)
\end{center}
\end{minipage}
\hspace{0.002\linewidth}
\begin{minipage}{.328\linewidth}
\begin{center}
\includegraphics[trim=12mm 0mm 4mm 0mm, clip, width=46mm]{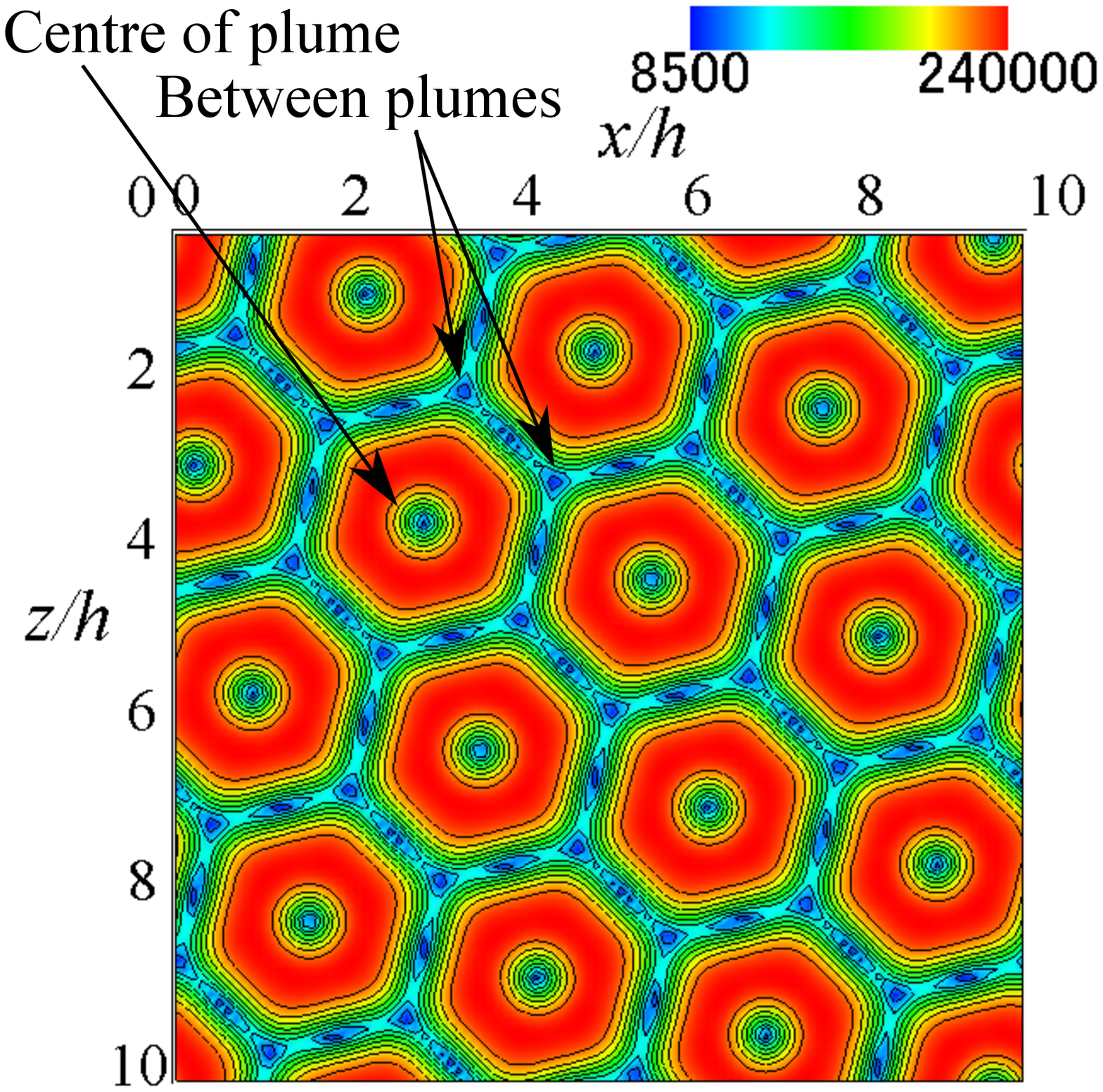} \\
(c) $\lambda/h=2.43$ ($Ra=5000$) \\
\end{center}
\end{minipage}
\caption{Surface friction coefficient contours in $x$--$z$ plane 
at the lower wall for $Ra=1000$, 3000, 5000.}
\label{cf_contour}
\end{figure}

We investigate the influence of the shear flow near the wall surface on bacteria. 
The surface friction coefficient $C_f$ on the bottom wall is defined as 
$C_f=\tau_w/[\rho (D_{n0}/h)^2/2]$, 
where $\tau_w=\mu (\partial V/\partial y)$, 
and $V$ is a composite component of the velocities in the $x$- and $z$-directions. 
Figures \ref{bottom_bacteria} and \ref{cf_contour} show the contours 
of the cell concentration at $y/h=0.01$ and the surface friction coefficient using the data of No.11 ($Ra=1000$), No.13 ($Ra=3000$), 
and No.16 ($Ra=5000$) in table \ref{pattern&wavelength}. 
The same initial disturbance is applied to all these results. 
In the region where the plumes exist, the cell concentration becomes high, 
and it can be seen that a large amount of bacteria exists under the plumes. 
Comparing the results for each Rayleigh number, 
we observe that the cell concentration decreases in the centre of a plume 
and increases between the plumes as the wavelength of the bioconvection pattern decreases. 
It is found that the surface friction coefficient is large in the region 
of a plume and that a strong shear flow occurs in this region. 
Also, the wall friction coefficient 
increases as the wavelength of the pattern decreases. 
Therefore, the cells under the plumes are affected by strong shear stress. 
\citet{Pol&Tramper_1998} investigated the influence of shear stress 
on animal cells and reported that shear stress causes damage to animal cells and cell death. 
In future research, it will be interesting to investigate the influence of shear stress on bacteria
due to bioconvection.

\subsection{Transport characteristics}

To clarify the transport characteristics of cells and oxygen, 
we investigated the variations of each flux distribution. 
As in previous studies \citep{Yanaoka_et_al_2007, Yanaoka_et_al_2008}, 
the flux of cells and oxygen is defined as
\begin{equation}
   J_{n}^{\mathrm{total}} = J_{n}^{\mathrm{conv}} + J_{n}^{\mathrm{diff}} + J_{n}^{\mathrm{swim}}
\end{equation}
\begin{equation}
   J_{c}^{\mathrm{total}} = J_{c}^{\mathrm{conv}} + J_{c}^{\mathrm{diff}}
\end{equation}
where $J_n^\mathrm{total}$ is the total flux of cells, 
$J_c^\mathrm{total}$ is the total flux of oxygen, 
$J_n^\mathrm{conv}$ is the convective flux of cells, 
$J_c^\mathrm{conv}$ is the convective flux of oxygen, 
$J_n^\mathrm{diff}$ is the diffusive flux of cells, 
$J_c^\mathrm{diff}$ is the diffusive flux of oxygen, and 
$J_n^\mathrm{swim}$ is the swimming flux of cells. 
These fluxes are given as
\begin{equation}
   J_{n}^{\mathrm{conv}} = {\bf u} n, \quad 
   J_{n}^{\mathrm{diff}} = - H(c) \nabla n, \quad 
   J_{n}^{\mathrm{swim}} = H(c) \gamma n \nabla c
   \label{flux_bac}
\end{equation}
\begin{equation}
   J_{c}^{\mathrm{conv}} = {\bf u} c, \quad 
   J_{c}^{\mathrm{diff}} = - \delta \nabla c
   \label{flux_oxy}
\end{equation}

\begin{figure}
\begin{minipage}{0.485\linewidth}
\begin{center}
\includegraphics[trim=2mm 0mm 0mm 2mm, clip, width=66mm]{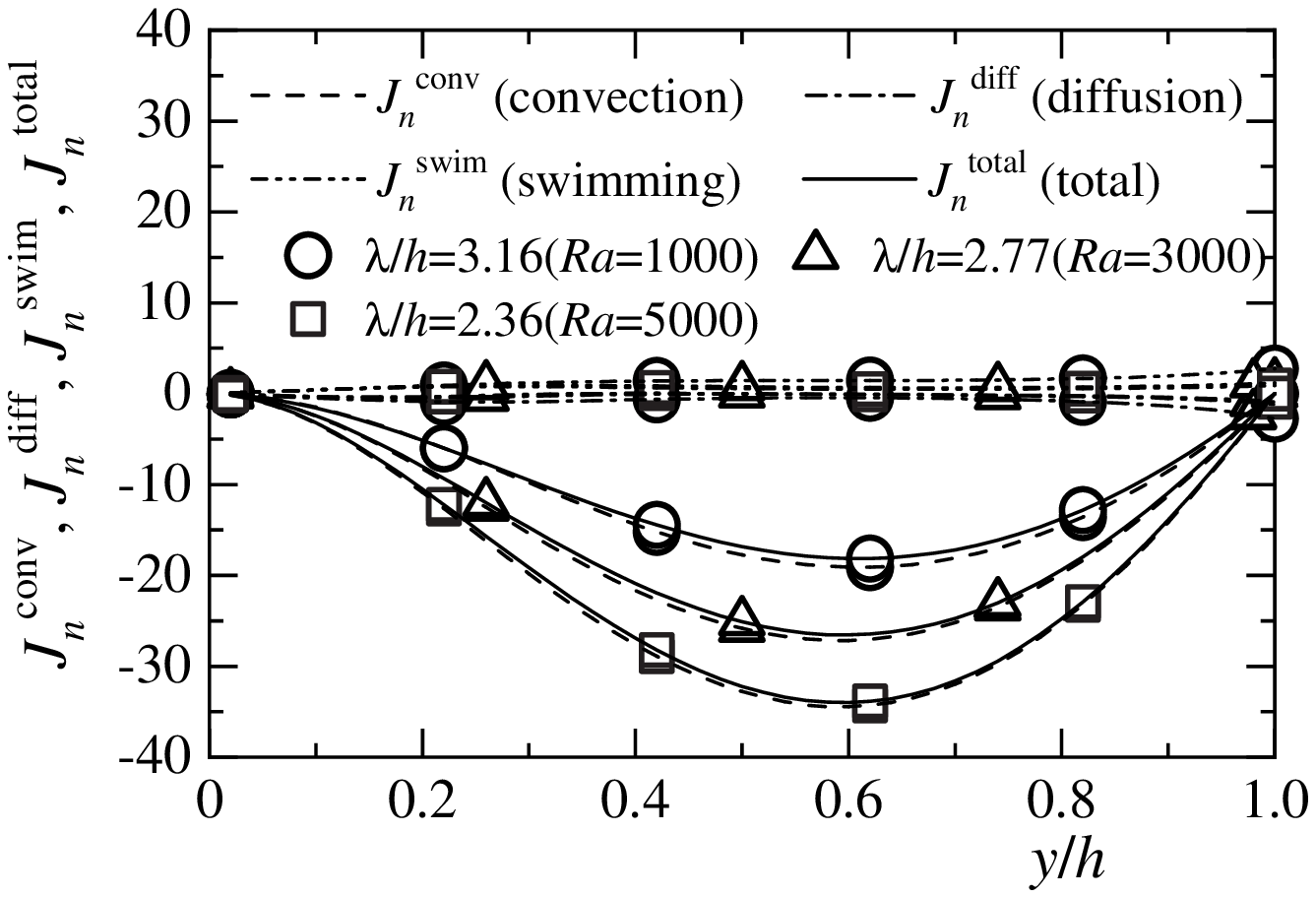} \\
(a) bacteria
\end{center}
\end{minipage}
\hspace{0.01\linewidth}
\begin{minipage}{0.485\linewidth}
\begin{center}
\includegraphics[trim=2mm 0mm 0mm 2mm, clip, width=66mm]{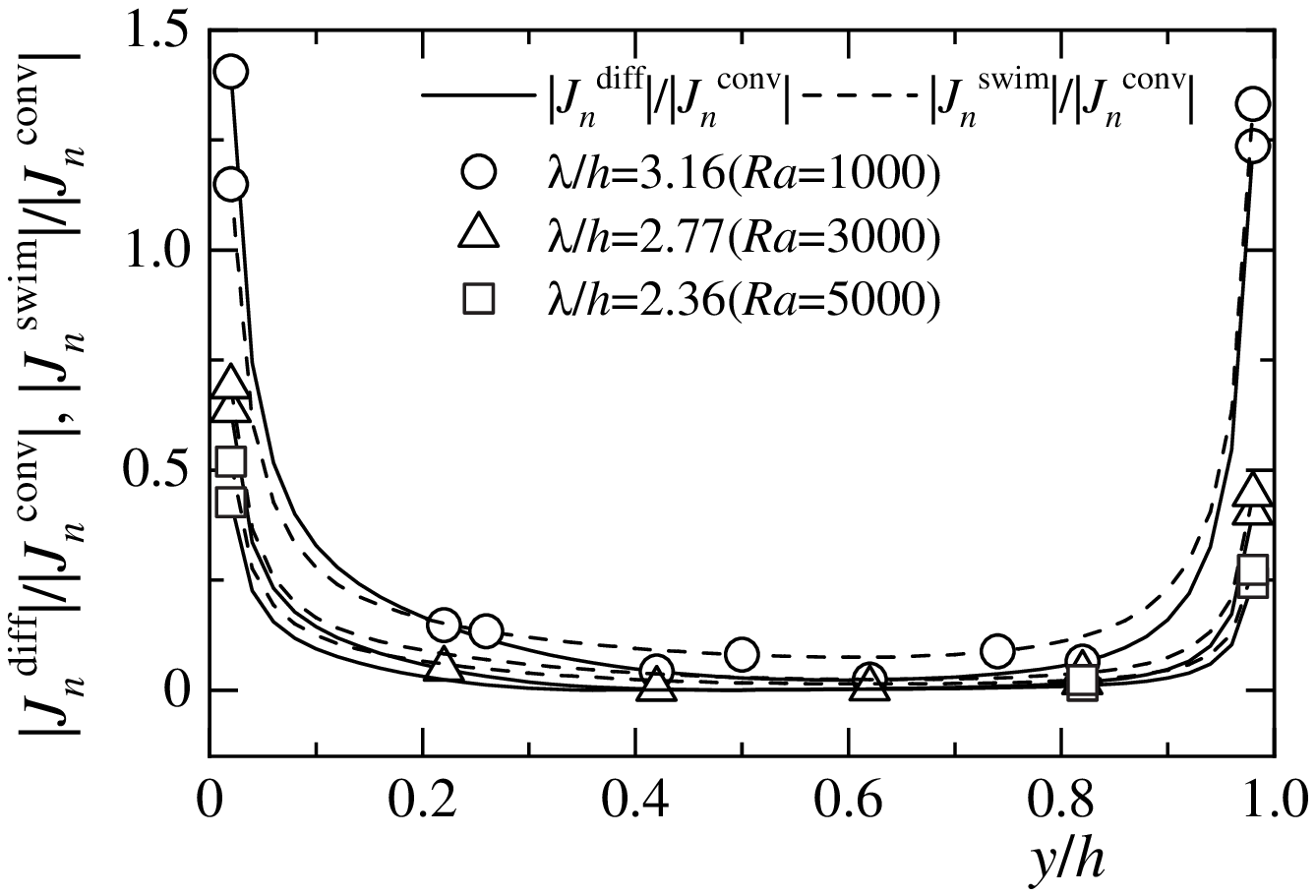} \\
(b) bacteria (flux ratio)
\end{center}
\end{minipage}

\begin{minipage}{0.485\linewidth}
\begin{center}
\includegraphics[trim=2mm 0mm 0mm 2mm, clip, width=66mm]{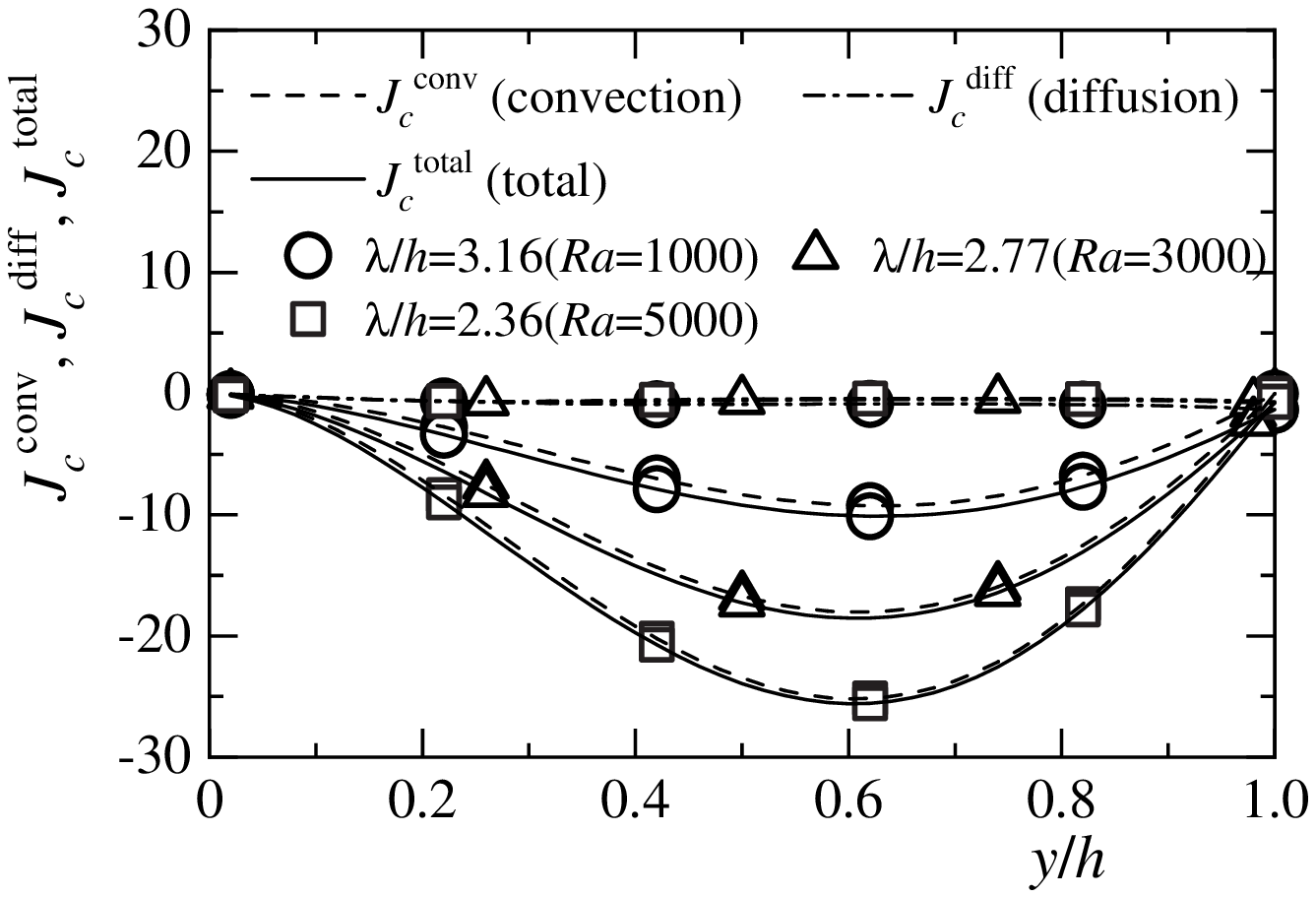} \\
(c) oxygen
\end{center}
\end{minipage}
\hspace{0.01\linewidth}
\begin{minipage}{0.485\linewidth}
\begin{center}
\includegraphics[trim=2mm 0mm 0mm 2mm, clip, width=66mm]{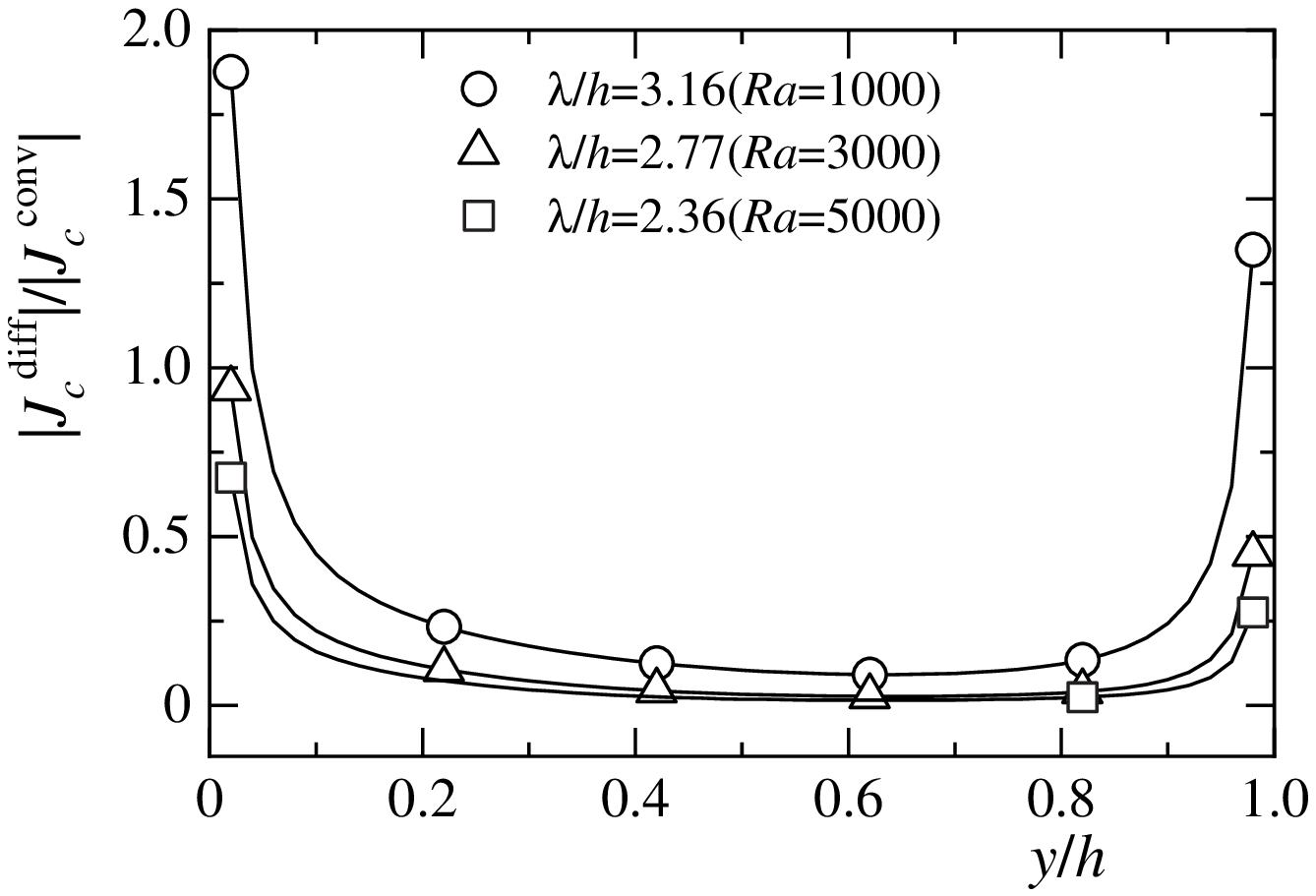} \\
(d) oxygen (flux ratio)
\end{center}
\end{minipage}
\caption{Flux distributions of bacteria and oxygen in $y$-direction 
at the centre of a plume.}
\label{center_plume_flux}
\end{figure}

\begin{figure}
\begin{minipage}{0.485\linewidth}
\begin{center}
\includegraphics[trim=2mm 0mm 0mm 2mm, clip, width=66mm]{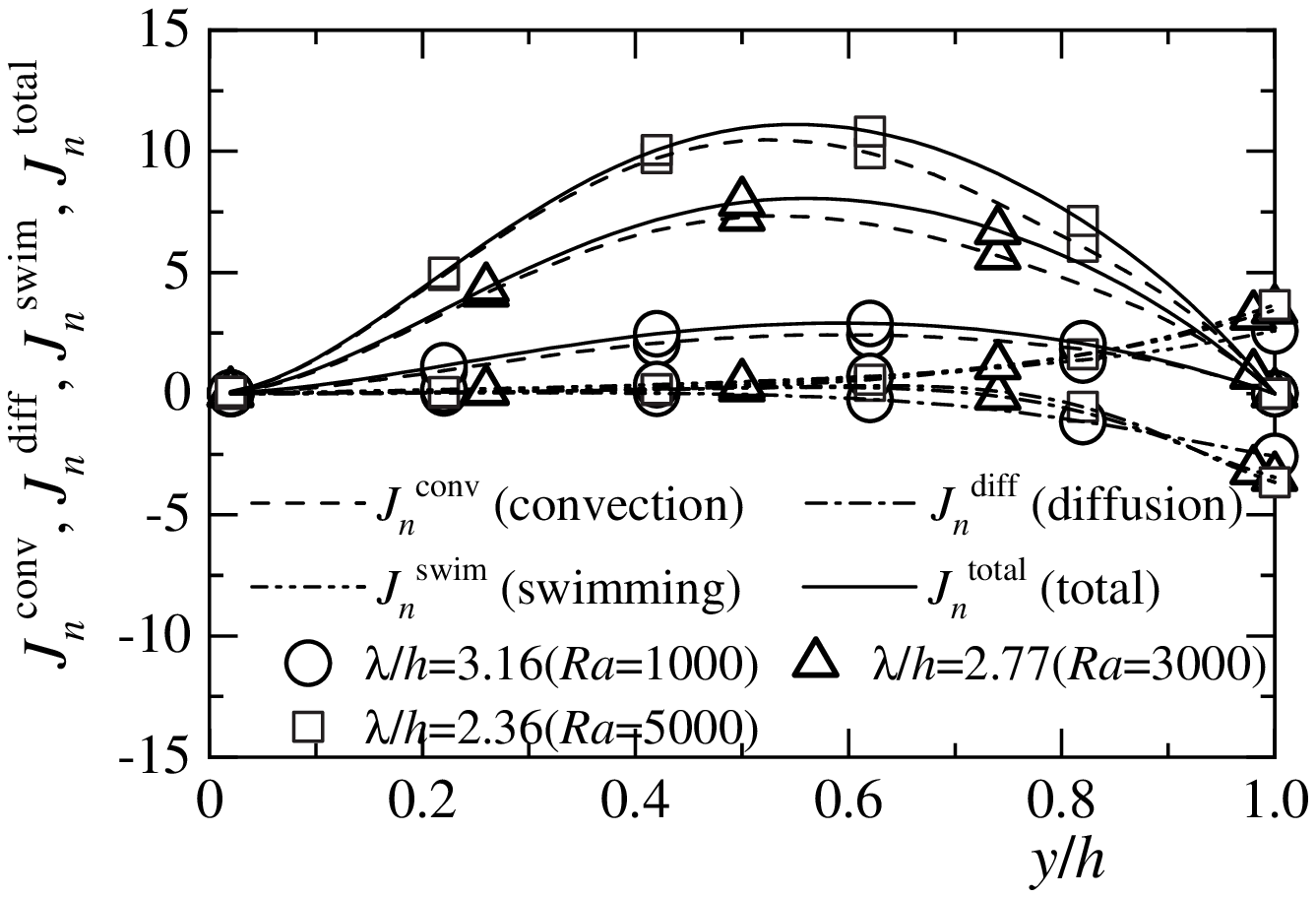} \\
(a) bacteria
\end{center}
\end{minipage}
\hspace{0.01\linewidth}
\begin{minipage}{0.485\linewidth}
\begin{center}
\includegraphics[trim=2mm 0mm 0mm 2mm, clip, width=66mm]{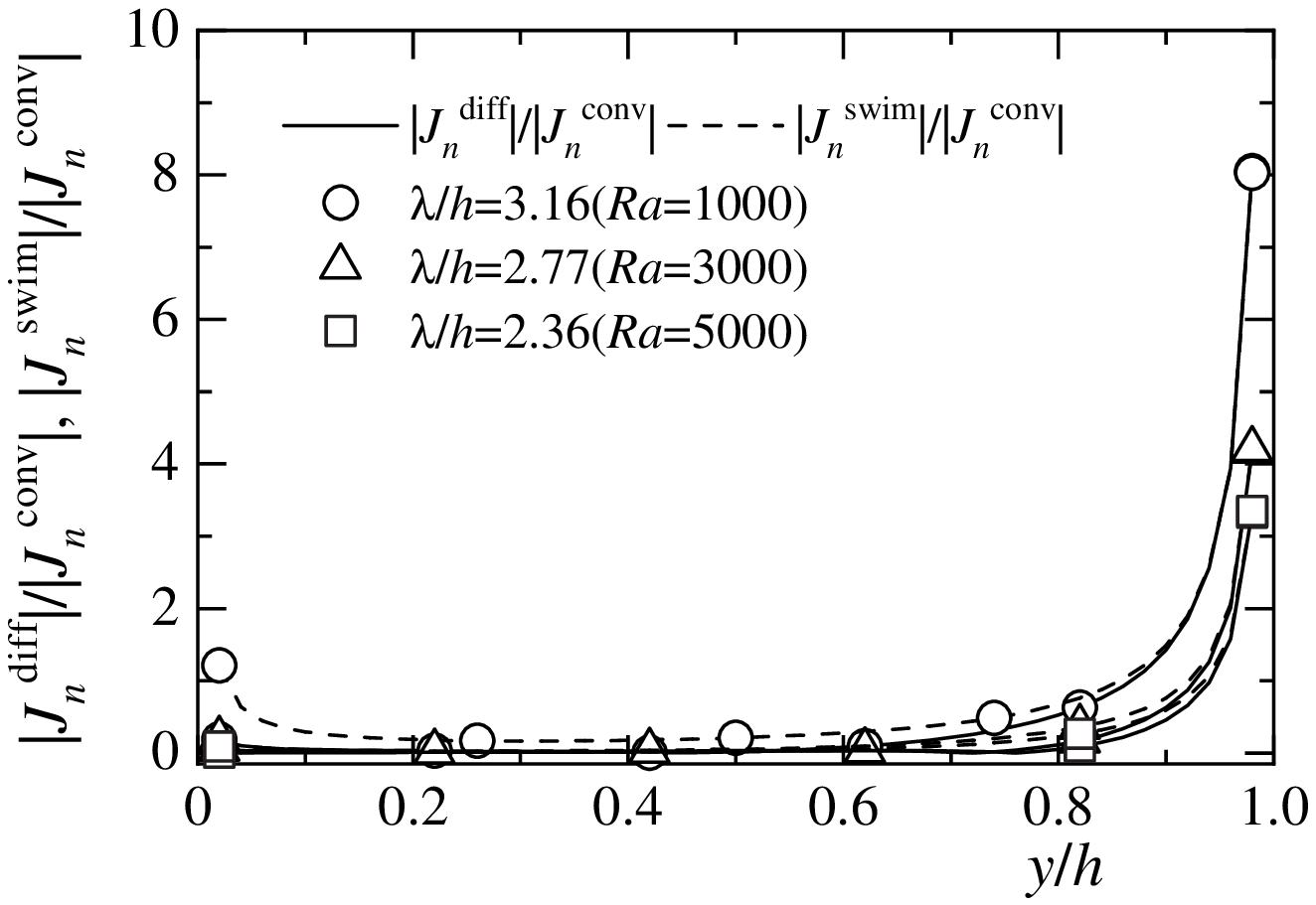} \\
(b) bacteria (flux ratio)
\end{center}
\end{minipage}

\begin{minipage}{0.485\linewidth}
\begin{center}
\includegraphics[trim=2mm 0mm 0mm 2mm, clip, width=66mm]{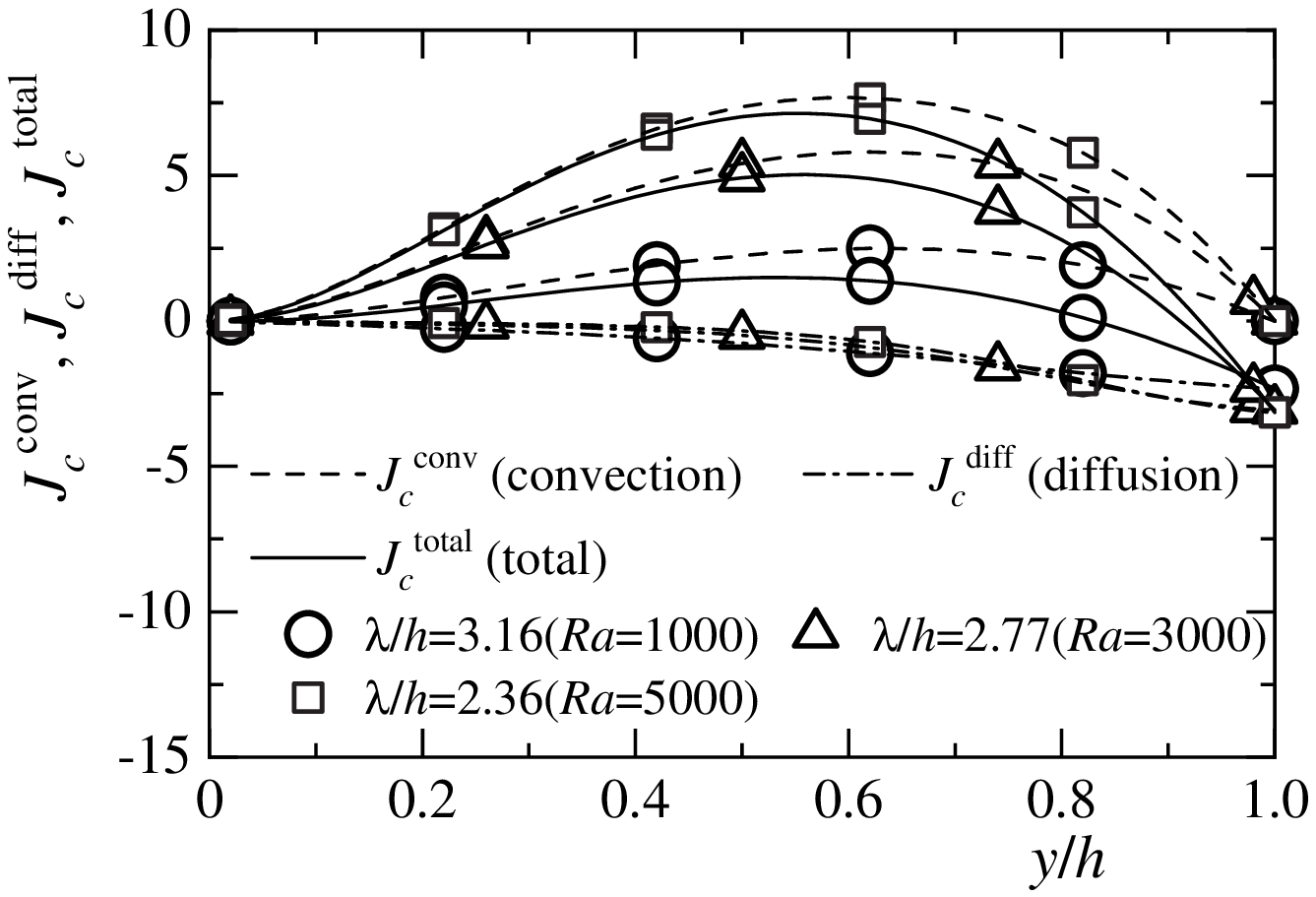} \\
(c) oxygen
\end{center}
\end{minipage}
\hspace{0.01\linewidth}
\begin{minipage}{0.485\linewidth}
\begin{center}
\includegraphics[trim=2mm 0mm 0mm 2mm, clip, width=66mm]{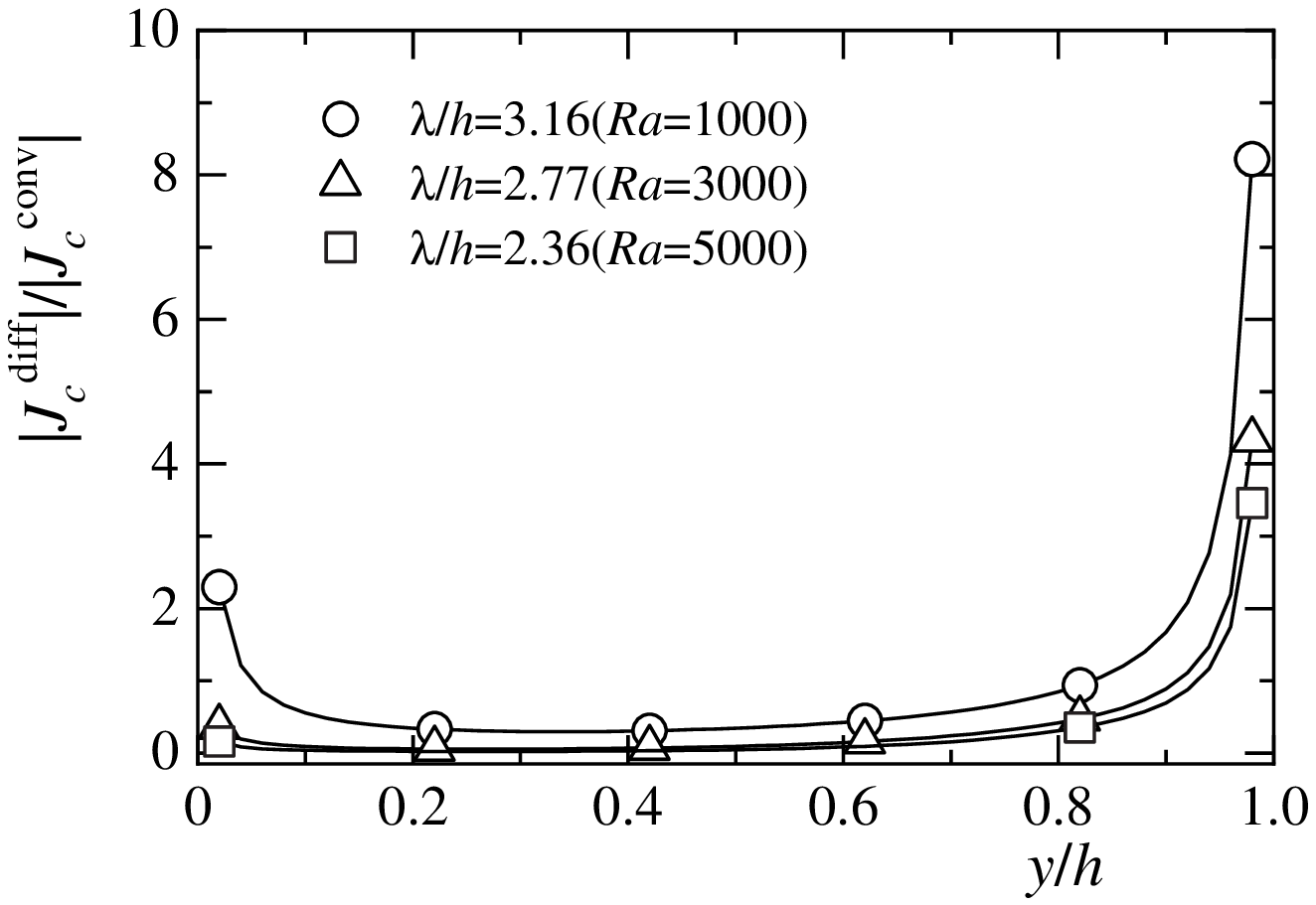} \\
(d) oxygen (flux ratio)
\end{center}
\end{minipage}
\caption{Flux distributions of bacteria and oxygen in $y$-direction between plumes.}
\label{between_plume_flux}
\end{figure}

For $Ra=1000$, 3000, and 5000, 
figures \ref{center_plume_flux} and \ref{between_plume_flux} show 
the cell and oxygen flux distributions in the $y$-direction 
at the centre of a plume and between the plumes, respectively. 
The results shown are for a lattice arrangement 
(figures \ref{pattern}(b), (c), and (e)) in which plumes are regularly arrayed. 
For $Ra=1000$, 3000, and 5000, the positions of the plume centres are 
($x/h$, $z/h$)=(5.8, 3.2), (1.9, 3.5), and (6.2, 6.0), 
and the positions between the plumes are ($x/h$, $z/h$)=(5.8, 8.2), (6.9, 8.5), 
and (1.2, 6.0). 
At the centre of a plume, the cell and oxygen transport 
due to the convection are dominant 
throughout the region. 
In the velocity vectors shown in figure \ref{Ra5000}, 
the convective velocity increased near $y/h=0.6$, 
where the flows around the plume joined together. 
Thus, the transport by the convection becomes active there. 
These trends are similar between the plumes, 
but because the convective velocity between the plumes is slower, 
the convective transport between the plumes becomes smaller than 
that at the centre of a plume. 
The flux ratios are also shown in figures \ref{center_plume_flux} and \ref{between_plume_flux}. 
The previous study \citep{Chertock_et_al_2012} also discussed 
transport characteristics using the flux ratio. 
Near the lower wall and the water surface in the centre of the plume, 
the cell transport by random swimming (diffusion) and directional swimming 
and the oxygen transport by diffusion increase compared to the transport by convection. 
At small wavelengths, the transport by directional swimming is slightly larger 
than the transport by random swimming. 
In contrast, the cell transport by diffusion and swimming is large 
near the free surface between the plumes, and the cells are actively moving. 
Here, the cell transport due to the directional and random swimming is almost in balance. 
Since the oxygen is supplied from the free surface of the suspension by diffusion, 
the oxygen transport increases toward the free surface. 
The flux ratios indicate that the transport by swimming and diffusion 
between the plumes is relatively larger than the transport by convection, 
as compared to the centre of the plume, 
because the convective transport by the upward flow is smaller than 
that by the downward flow.

Next, we compare the transport characteristics of cells and oxygen 
for each wavelength of the bioconvection pattern. 
The total flux of cells and oxygen increase with decreasing wavelength 
at plume centres and between the plumes. 
The reason is as follows. 
As the wavelength decrease enhances interference between plumes, 
increasing the downward and upward flow velocities at the plume centre 
and between the plumes. 
Therefore, the transport of cells and oxygen by convection increases. 
As can be seen from the flux ratios, 
the cell transport due to diffusion and swimming is low regardless of wavelength, 
except near the bottom wall and the water surface. 
Near the water surface between plumes, 
the flux ratios vary significantly with the wavelength. 
As the wavelength decreases, 
the cell transport by diffusion and swimming and the oxygen transport by diffusion 
decrease compared to the convective transport.

\begin{figure}
\begin{minipage}{0.485\linewidth}
\begin{center}
\includegraphics[trim=2mm 0mm 0mm 1mm, clip, width=66mm]{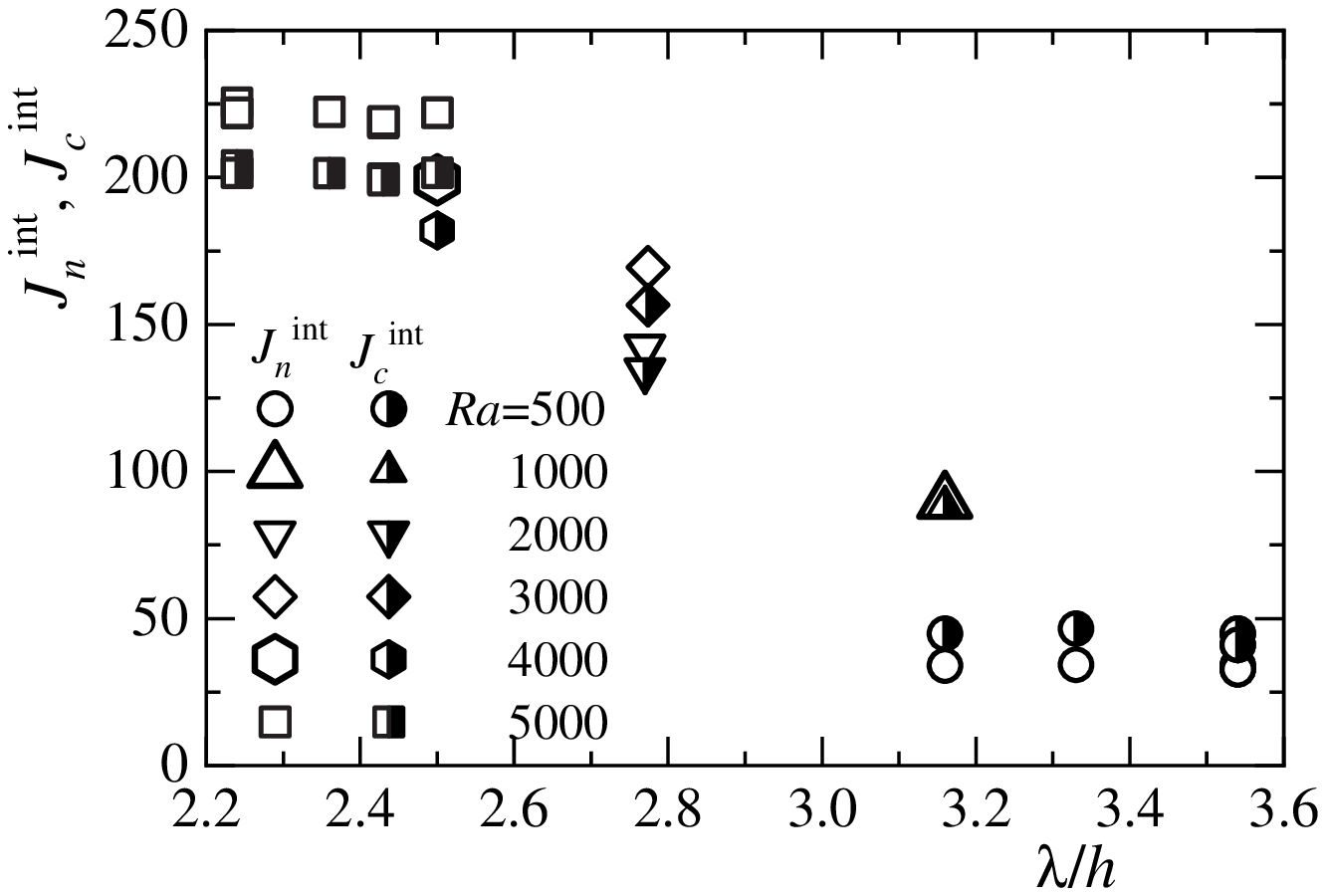} \\
(a) downward flow
\end{center}
\end{minipage}
\hspace{0.01\linewidth}
\begin{minipage}{0.485\linewidth}
\begin{center}
\includegraphics[trim=2mm 0mm 0mm 1mm, clip, width=66mm]{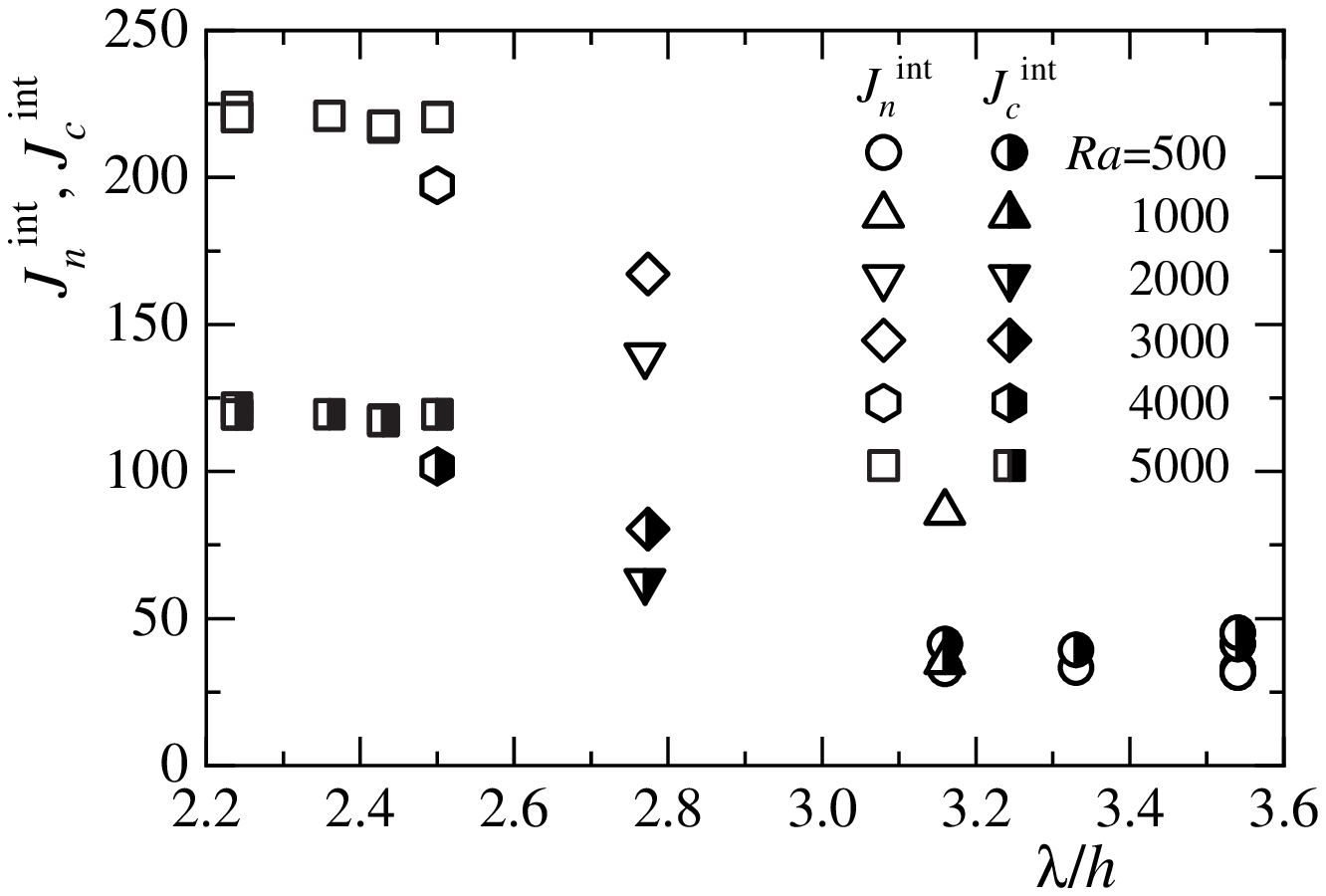} \\
(b) upward flow
\end{center}
\end{minipage}
\caption{Integral values of total flux of bacteria and oxygen in $y$-direction 
for downward flow and upward flow.}
\label{wavelength_flux}
\end{figure}

To clarify the influence of the wavelength of the pattern 
on the transport characteristics over the entire suspension, 
the integral values of the total flux of cells and oxygen, 
$J_n^\mathrm{int}$ and $J_c^\mathrm{int}$, were evaluated. 
These values are obtained by integrating $|J_n^\mathrm{total}|$ 
and $|J_c^\mathrm{total}|$ over the entire suspension. 
We consider the transport characteristics of cells and oxygen 
for the downward and upward flows. 
Figure \ref{wavelength_flux} shows the relationship between 
$J_n^\mathrm{int}$ and $J_c^\mathrm{int}$ of the total flux of cells and oxygen 
and the wavelength $\lambda/h$ of the pattern. 
As the Rayleigh number increases, the wavelength decreases 
and interference between the plumes increases, 
which improves the transport characteristics of the cells and oxygen 
due to the downward flow. 
These trends are similar in the upward flow region. 
From the above, it can be said that because the transport characteristics 
due to the downward and upward flows improve with decreasing wavelength, 
the transport characteristics over the entire suspension also improve.

\begin{figure}
\begin{minipage}{0.485\linewidth}
\begin{center}
\includegraphics[trim=2mm 0mm 0mm 1mm, clip, width=65mm]{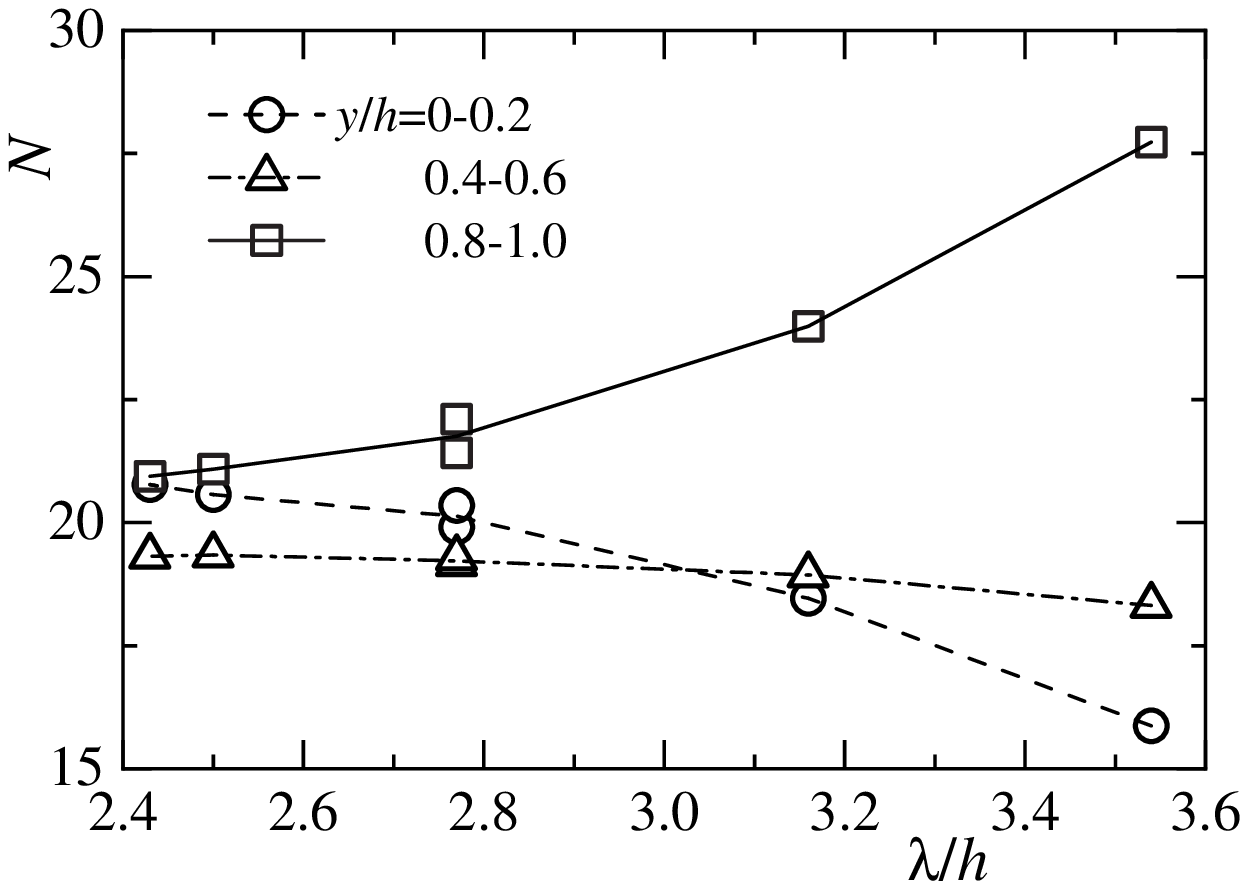} \\
(a) bacteria
\end{center}
\end{minipage}
\hspace{0.01\linewidth}
\begin{minipage}{0.485\linewidth}
\begin{center}
\includegraphics[trim=2mm 0mm 0mm 1mm, clip, width=65mm]{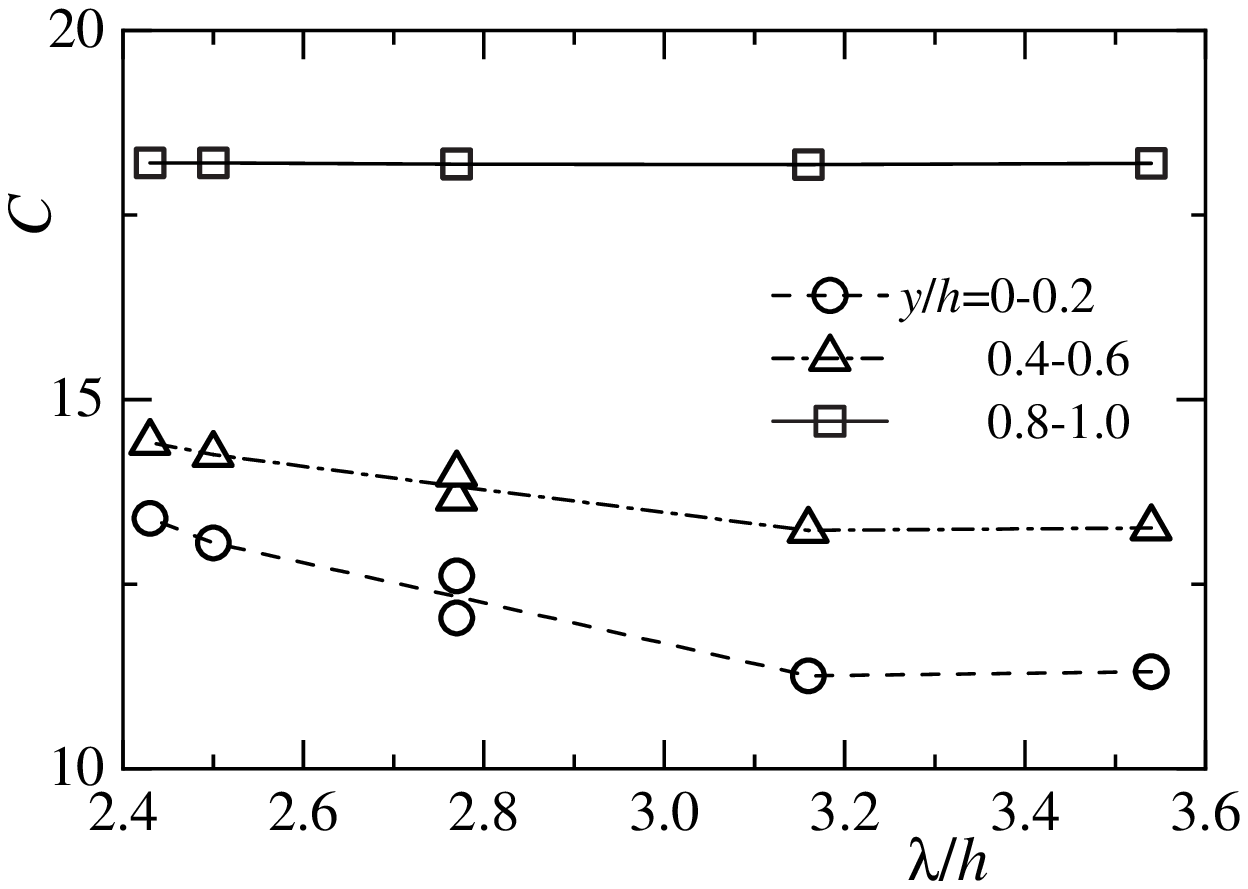} \\
(b) oxygen
\end{center}
\end{minipage}
\caption{Integral values of bacterial and oxygen concentrations.}
\label{integral_conc}
\end{figure}

To investigate the effects of pattern wavelength changes on the transport of bacteria and oxygen, 
we calculate the integral values, $N$ and $C$, of the amounts of cells and oxygen, respectively, 
near the bottom wall ($0 \leq y/h \leq 0.2$), 
in the middle ($0.4 \leq y/h \leq 0.6$), and near the water surface ($0.8 \leq y/h \leq 1.0$). 
Here, the results of No. 2 ($Ra=500$), No. 11 ($Ra=1000$), No. 12 ($Ra=2000$), 
No. 13 ($Ra=3000$), No. 14 ($Ra=4000$), and No. 16 ($Ra=5000$) in table \ref{pattern&wavelength} 
are shown in figure \ref{integral_conc}. 
These are the results when the same initial disturbance is given. 
As the wavelength decreases, 
the cell amount decreases near the water surface 
and increases near the bottom wall. 
This is because the transport characteristics improve 
as the wavelength decreases, 
and the cell amount that settles downward increases. 
At $\lambda/h=3.54$, the difference between the cell amounts 
near the bottom wall and near the water surface is large, 
but the difference decreases as the wavelength decreases. 
The cell amount near the middle increases slightly 
as the wavelength decreases, 
but at $\lambda/h<3.16$, the cell amount becomes approximately constant. 
The oxygen amount increases near the bottom wall and in the middle area 
because the oxygen transport characteristics improve as the wavelength decreases. 
The oxygen amount near the water surface is approximately constant. 
Since the bacteria that gather downward consume oxygen, 
the oxygen amount decreases toward the depth at each wavelength. 
It was found from the above results that the amounts of cells and oxygen 
near the bottom wall increase 
because the transport characteristics improve as the pattern wavelength decreases.

\subsection{Influences of computational region and boundary condition}
\label{comp_region}

\begin{figure}
\begin{center}
\includegraphics[trim=0mm 20mm 0mm 10mm, clip, width=120mm]{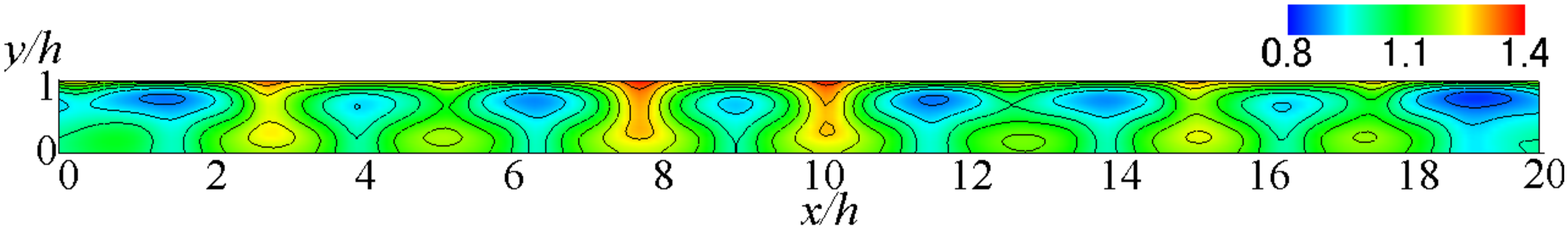} \\
(a) bacteria \\
\includegraphics[trim=0mm 20mm 0mm 10mm, clip, width=120mm]{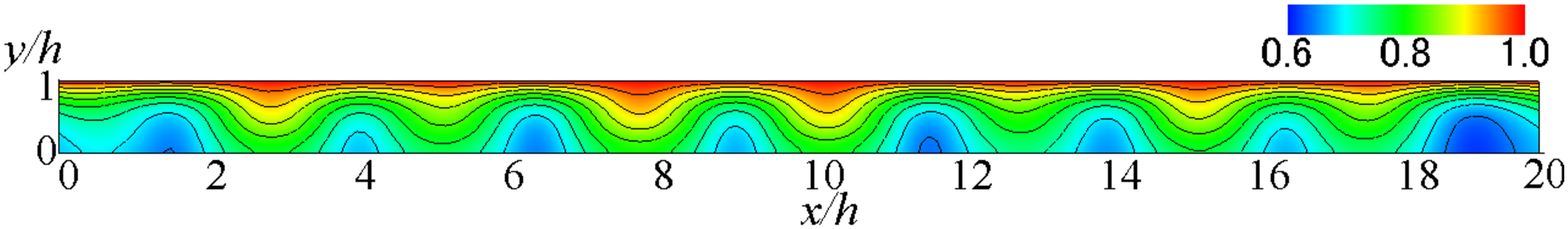} \\
(b) oxygen \\
\includegraphics[trim=0mm 20mm 0mm 15mm, clip, width=120mm]{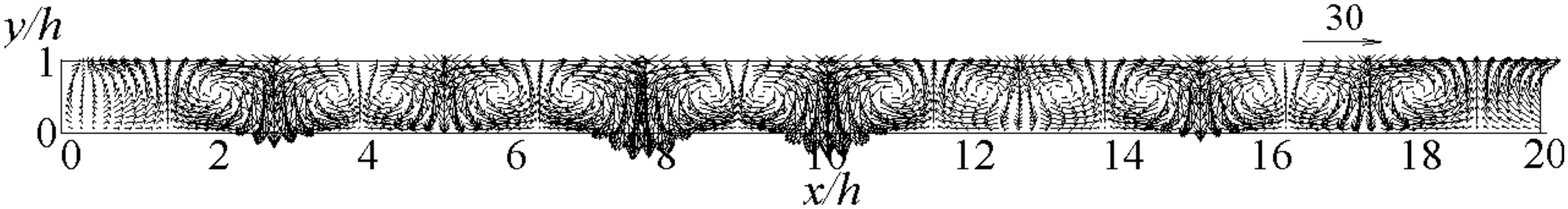} \\
(c) velocity vectors \\
\includegraphics[trim=0mm 0mm 10mm 5mm, clip, width=100mm]{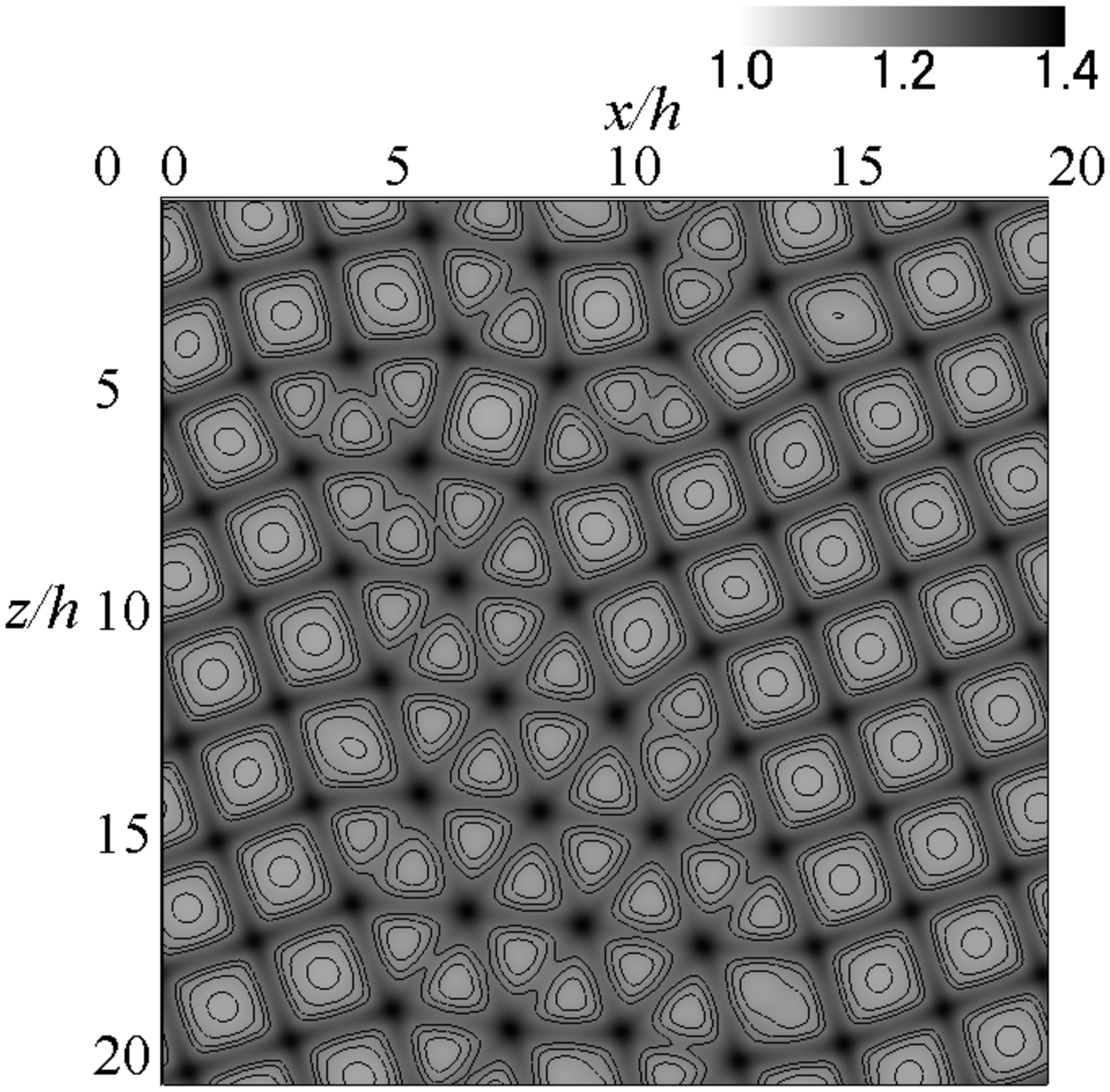} \\
(d) bacteria \\
\end{center}
\caption{Bacterial and oxygen concentration contours, 
velocity vectors at $z/h=18.8$, and
bacterial concentration contours at $y/h=1.0$ for $Ra=5000$.}
\label{L20Ra5000}
\end{figure}

To confirm that the above results do not depend on the calculation region, 
we compare the results for the computational regions of $L=10h$ and $20h$ 
for $Ra=5000$. 
The same initial disturbance for $L=10h$ 
in figure \ref{Ra5000} was set for $L=20h$. 
Figure \ref{L20Ra5000} shows the flow field and concentration field at $L=20h$. 
As in figure \ref{Ra5000}, it can be seen that bioconvection with multiple plumes occurs. 
In the bioconvection pattern in figure \ref{L20Ra5000}(d), 
a lattice arrangement in the region of $x/h \ge 10$, 
a staggered arrangement in $5 \le x/h \le 10$, $z/h \ge 10$, 
and random arrangements in $x/h \le 10$, $0 \le z/h \le 10$ appear. 
It is found from this result that the bioconvection patterns observed 
at $L=10h$ appear simultaneously. 
The wavelength of the bioconvection pattern for $L=20h$ is $\lambda/h=2.34$. 
We calculated the averaged wavelength for $L=10h$ 
to compare the results of $L=10h$ and $20h$. 
The wavelength was $\lambda/h=2.38$, 
which was obtained by averaging the wavelengths of all patterns for $L=10h$. 
The difference between the wavelength for $L=10h$ and $L=20h$ 
is approximately 1.71\%, 
and it can be said that the wavelengths of the bioconvection patterns 
observed at $L=10h$ and $20h$ are almost the same. 
As with $L=10h$, it is possible that bioconvection patterns 
with different wavelengths will occur for different initial disturbances.

Next, we investigate the transport characteristics of the cells and oxygen 
over the entire suspension for $L=20h$. 
The integral values of the total flux are 
$J_n^\mathrm{int}=1767.2$ and $J_c^\mathrm{int}=1297.9$. 
To compare the results of $L=10h$ and $20h$, 
we averaged the integrated value of the total flux of each pattern for $L=10h$, 
and the values were $J_n^\mathrm{int}=440.8$ and $J_c^\mathrm{int}=323.8$. 
To compare these values with the result for $L=20h$, which is 4 times 
the computational region of $L=10h$, 
$J_n^\mathrm{int}$ and $J_c^\mathrm{int}$ for $L=10h$ are multiplied by 4, 
and the difference between the results for $L=10h$ and $L=20h$ 
is found to be approximately 0.23\% and 0.21\%, respectively. 
The results for $L=10h$ and $20h$ are almost the same. 
Comparing the calculation results for $L=10h$ and $L=20h$, 
the same results were obtained for the wavelength and transport characteristics 
of the bioconvection pattern, even for different computational regions. 
Therefore, it can be seen that the results for $L=10h$ do not depend on 
the computational region.

\begin{figure}
\begin{center}
\includegraphics[trim=0mm 20mm 0mm 10mm, clip, width=120mm]{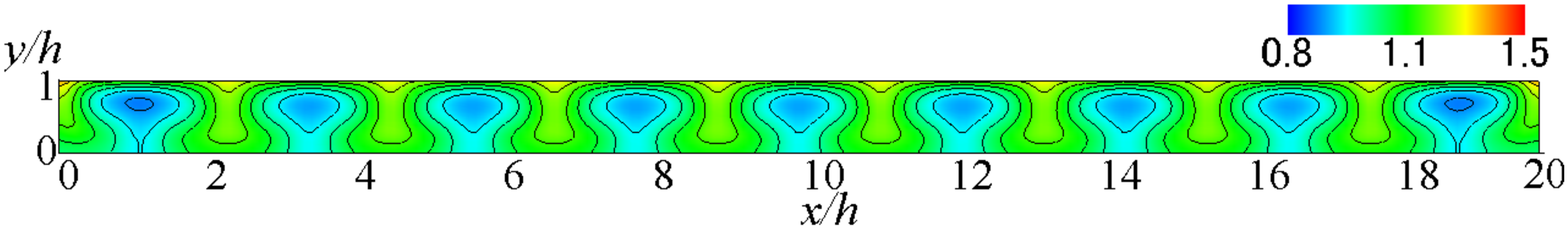} \\
(a) bacteria \\
\includegraphics[trim=0mm 20mm 0mm 10mm, clip, width=120mm]{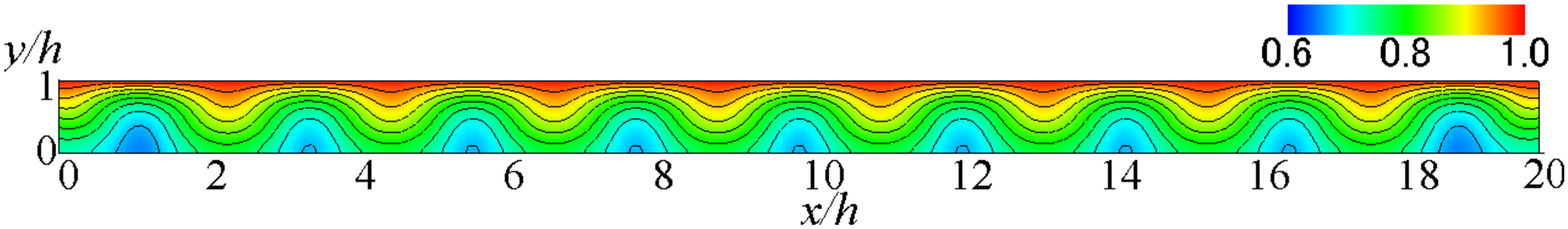} \\
(b) oxygen \\
\includegraphics[trim=0mm 20mm 0mm 15mm, clip, width=120mm]{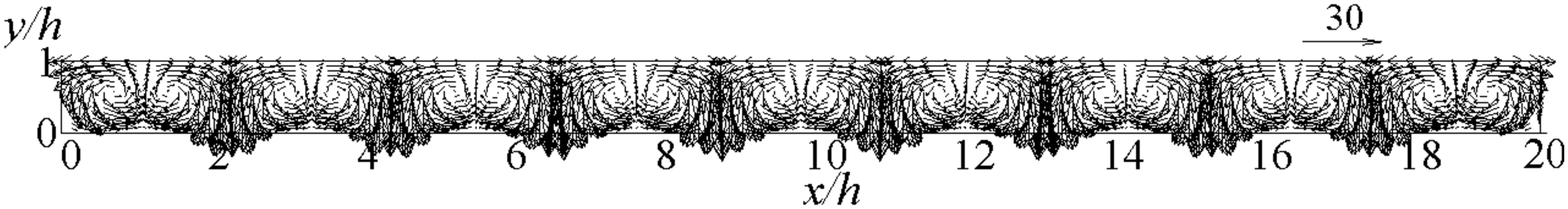} \\
(c) velocity vectors \\
\includegraphics[trim=0mm 0mm 10mm 5mm, clip, width=100mm]{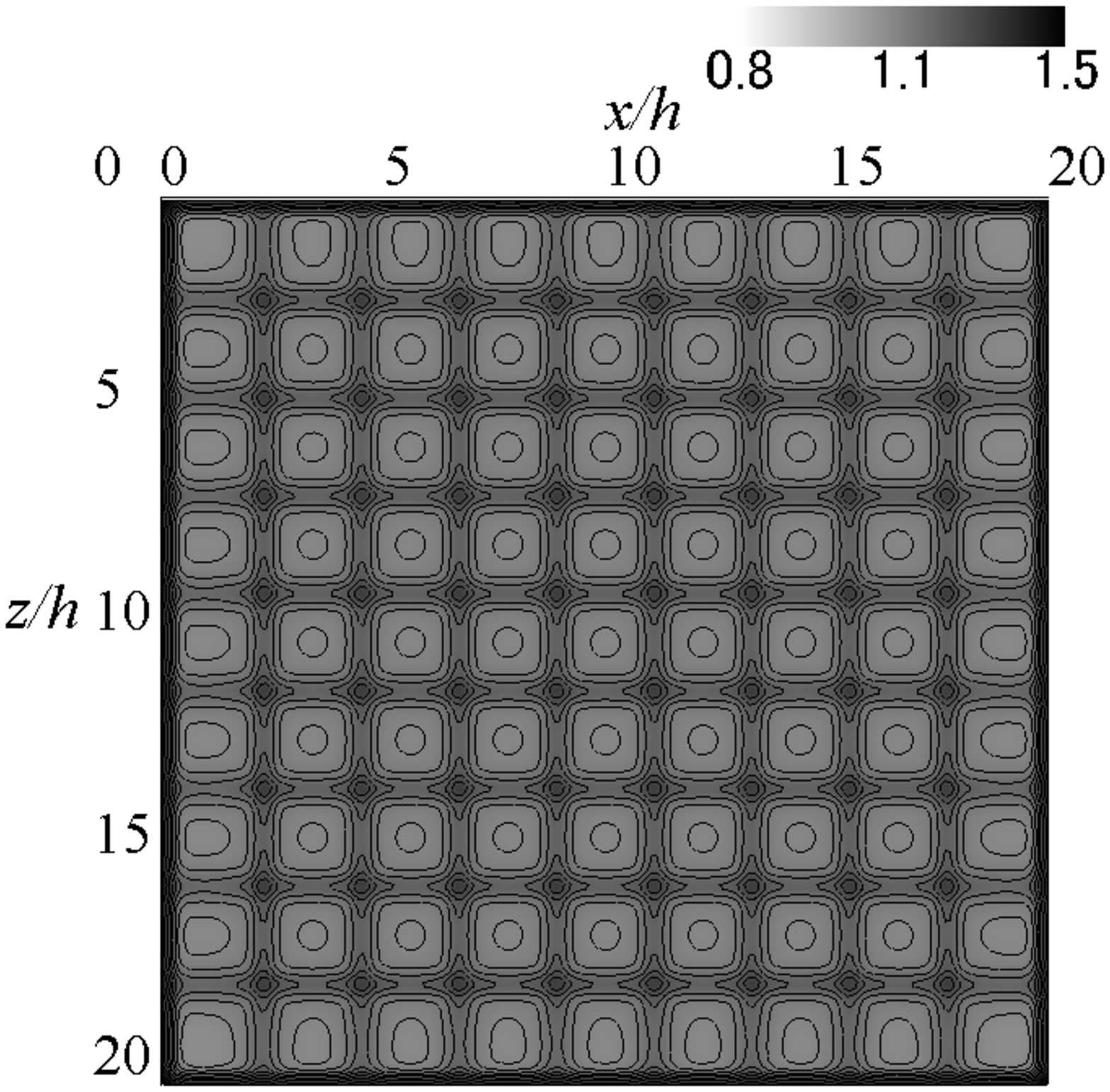} \\
(d) bacteria \\
\end{center}
\caption{Bacterial and oxygen concentration contours, 
velocity vectors at $z/h=8.9$, and 
bacterial concentration contours at $y/h=1.0$ for $Ra=5000$ in chamber with side walls.}
\label{L20Ra5000bc}
\end{figure}

The pattern of Rayleigh-B\'{e}nard convection changes 
depending on the system geometry and boundary conditions 
\citep{Cross&Hohenberg_1993, Bodenschatz_et_al_2000}. 
Similarly, it is considered that the pattern of bioconvection changes. 
\citet{Metcalfe&Pedley_1998} 
mentioned that the side wall of the Petri dish affects the pattern. 
\citet{Yamamoto_et_al_1992} 
also clarified that the vessel's wall determines the pattern generation. 
Figure \ref{L20Ra5000bc} shows the calculation result for changing the chamber boundary 
to side walls. 
In this calculation, the calculation conditions are the same as 
those using periodic boundary conditions, 
and the given initial disturbance is also the same. 
Plumes line up along the boundaries. 
A pattern with more regularity than the result in figure \ref{L20Ra5000} occurs. 
Bacteria adhere to all side walls, and plumes form near the walls. 
Therefore, downward flow occurs near all side walls. 
The same regular pattern as in figure \ref{L20Ra5000bc} was obtained 
even with different initial disturbances. 
Since bacteria gather near the water surface before bioconvection occurs, 
bacteria adhere to the side wall. 
Subsequently, the suspension becomes unstable, and plumes form. 
Since the velocity is slow near the wall due to the non-slip condition, 
the bacteria adhered to the side wall are less likely to come off, 
and even if bioconvection develops, 
the bacteria may remain attached to the side wall. 
The integrated values of the total flux are $J_n^\mathrm{int}=1789.5$ and 
$J_c^\mathrm{int}=1326.0$, and the differences from the results 
when using the periodic boundary condition are $1.3-2.2$\%. 
Therefore, there is no difference in transport characteristics due to the boundary conditions. 
\citet{Karimi&Paul_2013} 
performed a numerical analysis on gyrotactic bioconvection 
and reported that the pattern at the final stage does not depend on boundary conditions. 
This trend differs from our calculation results. 
Since the present result is a converged solution after sufficient time has passed, 
it is considered that this regular pattern does not collapse. 
\citet{Karimi&Paul_2013} reported that the final pattern did not reach a steady state, 
so this discrepancy with the existing result may be due to 
the difference between steady state and transient state but requires further investigation.

When using the periodic boundary condition, 
changing the initial disturbance can change not only the cell concentration 
on the boundary, but also the velocity and oxygen, 
so that the bioconvection pattern can also vary. 
If the side wall is set, bacteria will adhere to the wall surface. 
Therefore, it is considered that the pattern is difficult to change freely. 
Since we impose a condition of zero cell flux at the side wall, 
bacterial adhesion is a natural consequence. 
However, it is still necessary to investigate 
the cause of bacteria remaining attached to the side wall. 
In bioconvection, in which plumes interact with each other, 
the adhesion phenomenon of bacteria to walls has not been investigated in detail. 
Therefore, such investigations will be the subject of future research.

\section{Conclusions}

We conducted a three-dimensional numerical simulation on bioconvection 
generated by oxygen-reactive chemotactic bacteria. 
The bioconvection patterns, interference between plumes, 
wavelength of the bioconvection pattern, 
and transport characteristics of cells and oxygen were investigated 
when multiple plumes occurred in the suspension. 
The findings obtained are summarized as follows.
\begin{enumerate}[(1)]
\item When the Rayleigh number reaches the critical value, 
multiple three-dimensional plumes arise in the suspension. 
Three-dimensional bioconvection occurs around the plumes 
and vortex rings form around the plumes. 
Even if bioconvection at a high Rayleigh number is disturbed, 
the bioconvection is strongly stable with respect to disturbances, 
and the pattern does not change due to disturbances.
\item The ratio of the diffusion coefficient of oxygen to bacteria 
greatly affects the strength of bioconvection, 
and when the diffusion coefficient of oxygen is relatively large, 
bioconvection becomes weak. 
The rate of oxygen consumption by bacteria significantly affects 
the strength of bioconvection, 
and the stronger the oxygen consumption by bacteria, 
the stronger the bioconvection. 
This trend to strengthen bioconvection is also the same 
when the directional swimming velocity of bacteria increases.
\item Bioconvection patterns with different plume arrangements and shapes are formed 
for different Rayleigh numbers or initial disturbances of the cell concentration. 
Thus, the wavelengths of the patterns also vary. 
Comparing the arrangement and shape of the plumes, 
a staggered arrangement is observed consisting of circular 
or hexagonal plumes, a lattice arrangement constituted by tetragonal plumes, 
and a random arrangement consisting of plumes with various shapes.
\item As the Rayleigh number increases, interference between the plumes is 
strengthened by the decrease in wavelength of the pattern. 
Thus, the velocities of both the upward and downward flows increase. 
The velocity of the downward flow is faster than that of the upward flow.
\item Many cells are located under plumes, and a strong shear flow occurs 
in these regions. 
As the wavelength of the pattern decreases, the shear stress increases. 
Therefore, many cells are affected by a strong shear stress.
\item The convective transport of the entire suspension strengthens 
as the wavelength of the pattern decreases. 
Therefore, the transport characteristics of cells and oxygen improve, 
and then the amounts of cells and oxygen transported near the bottom wall 
in the suspension increases.
\item When the chamber boundary is changed to side walls, 
bacteria adhere to the wall surface, 
and plumes are regularly arranged along the side walls. 
Periodic boundary conditions can cause various patterns of bioconvection, 
while the side wall may suppress free formations of patterns.
\end{enumerate}

In the present study, we investigated the effect of the chamber side wall 
on the pattern of bioconvection and found that the side wall changed the pattern. 
This time, we used a rectangular container with side walls. 
The chamber geometry may change the pattern of bioconvection 
and accompanying transport characteristics, 
so that further investigation of bioconvection under different boundary conditions 
is necessary. 
In addition, since characteristics such as the oxygen consumption rate of bacteria 
significantly affect bioconvection, 
advanced mathematical models are required. 
More detailed experimental information will be useful 
for developing mathematical models, 
and it is thought that the investigation of transport characteristics 
by numerical analysis will develop. 
Furthermore, it is necessary to investigate the control of bioconvection 
for the engineering application of bioconvection. 
We have investigated the interference between bioconvection 
and thermal convection by heating suspension 
and the improvement of transport characteristics, 
and we plan to report the results in the future.


\vspace*{1.0\baselineskip}
\noindent
{\bf Acknowledgments.}
The numerical results in this research were obtained 
using the supercomputing resources at Cyberscience Center, Tohoku University. 
We would like to express our gratitude to Associate Professor Yosuke Suenaga 
of Iwate University for his support of my laboratory. 
The authors wish to acknowledge the time and effort of everyone involved in this study.

\vspace*{1.0\baselineskip}
\noindent
{\bf Funding.}
This work was supported by Grants-in-Aid for Scientific Research C 
from Japan Society for the Promotion of Science (grant number 24560181). 

\noindent
{\bf Declaration of Interests.}
The authors report no conflict of interest.

\vspace*{1.0\baselineskip}
\noindent
{\bf Author ORCID.} \\
Yanaoka H. https://orcid.org/0000-0002-4875-8174.

\vspace*{1.0\baselineskip}
\noindent
{\bf Author contributions.}
H. Yanaoka conceived and planned the research 
and developed the calculation method and numerical codes. 
T. Nishimura performed the simulations. 
All authors contributed equally to analysing data and reaching conclusions, 
and in writing the paper.


\bibliographystyle{arXiv_elsarticle-harv}
\bibliography{reference_bioconvection_bibfile}

\begin{thebibliography}{48}
\expandafter\ifx\csname natexlab\endcsname\relax\def\natexlab#1{#1}\fi
\providecommand{\url}[1]{\texttt{#1}}
\providecommand{\href}[2]{#2}
\providecommand{\path}[1]{#1}
\providecommand{\DOIprefix}{doi:}
\providecommand{\ArXivprefix}{arXiv:}
\providecommand{\URLprefix}{URL: }
\providecommand{\Pubmedprefix}{pmid:}
\providecommand{\doi}[1]{\href{http://dx.doi.org/#1}{\path{#1}}}
\providecommand{\Pubmed}[1]{\href{pmid:#1}{\path{#1}}}
\providecommand{\bibinfo}[2]{#2}
\ifx\xfnm\relax \def\xfnm[#1]{\unskip,\space#1}\fi
\bibitem[{Amsden and Harlow(1970)}]{Amsden&Harlow_1970}
\bibinfo{author}{Amsden, A.A.}, \bibinfo{author}{Harlow, F.H.},
  \bibinfo{year}{1970}.
\newblock \bibinfo{title}{A simplified {MAC} technique for incompressible fluid
  flow calculations}.
\newblock \bibinfo{journal}{J. Comput. Phys.} \bibinfo{volume}{6},
  \bibinfo{pages}{322--325}.
\newblock \DOIprefix\doi{https://doi.org/10.1016/0021-9991(70)90029-X}.
\bibitem[{Bees and Croze(2014)}]{Bees&Croze_2014}
\bibinfo{author}{Bees, M.A.}, \bibinfo{author}{Croze, O.A.},
  \bibinfo{year}{2014}.
\newblock \bibinfo{title}{Mathematics for streamlined biofuel production from
  unicellular algae}.
\newblock \bibinfo{journal}{Biofuels} \bibinfo{volume}{5},
  \bibinfo{pages}{53--65}.
\newblock \DOIprefix\doi{https://doi.org/10.4155/bfs.13.66}.
\bibitem[{Bees and Hill(1997)}]{Bees&Hill_1997}
\bibinfo{author}{Bees, M.A.}, \bibinfo{author}{Hill, N.A.},
  \bibinfo{year}{1997}.
\newblock \bibinfo{title}{Wavelengths of bioconvection patterns}.
\newblock \bibinfo{journal}{J. Exp. Biol.} \bibinfo{volume}{200},
  \bibinfo{pages}{1515--1526}.
\newblock \DOIprefix\doi{https://doi.org/10.1242/jeb.200.10.1515}.
\bibitem[{Berg and Brown(1972)}]{Berg&Brown_1972}
\bibinfo{author}{Berg, H.C.}, \bibinfo{author}{Brown, D.A.},
  \bibinfo{year}{1972}.
\newblock \bibinfo{title}{Chemotaxis in {\it {e}scherichia coli} analysed by
  three--dimensional tracking}.
\newblock \bibinfo{journal}{Nature} \bibinfo{volume}{239},
  \bibinfo{pages}{500--504}.
\newblock \DOIprefix\doi{https://doi.org/10.1038/239500a0}.
\bibitem[{Bodenschatz et~al.(2000)Bodenschatz, Pesch and
  Ahlers}]{Bodenschatz_et_al_2000}
\bibinfo{author}{Bodenschatz, E.}, \bibinfo{author}{Pesch, W.},
  \bibinfo{author}{Ahlers, G.}, \bibinfo{year}{2000}.
\newblock \bibinfo{title}{Recent developments in rayleigh-b\'{e}nard
  convection}.
\newblock \bibinfo{journal}{Annu. Rev. Fluid Mech.} \bibinfo{volume}{32},
  \bibinfo{pages}{709--778}.
\newblock \DOIprefix\doi{https://doi.org/10.1146/annurev.fluid.32.1.709}.
\bibitem[{Chertock et~al.(2012)Chertock, Fellner, Kurganov, Lorz and
  Markowich}]{Chertock_et_al_2012}
\bibinfo{author}{Chertock, A.}, \bibinfo{author}{Fellner, K.},
  \bibinfo{author}{Kurganov, A.}, \bibinfo{author}{Lorz, A.},
  \bibinfo{author}{Markowich, P.A.}, \bibinfo{year}{2012}.
\newblock \bibinfo{title}{Sinking, merging and stationary plumes in a coupled
  chemotaxis--fluid model: {A} high--resolution numerical approach}.
\newblock \bibinfo{journal}{J. Fluid Mech.} \bibinfo{volume}{694},
  \bibinfo{pages}{155--190}.
\newblock \DOIprefix\doi{https://doi.org/10.1017/jfm.2011.534}.
\bibitem[{Cross and Hohenberg(1993)}]{Cross&Hohenberg_1993}
\bibinfo{author}{Cross, M.C.}, \bibinfo{author}{Hohenberg, P.C.},
  \bibinfo{year}{1993}.
\newblock \bibinfo{title}{Pattern formation outside of equilibrium}.
\newblock \bibinfo{journal}{Rev. Mod. Phys.} \bibinfo{volume}{65},
  \bibinfo{pages}{851--1112}.
\newblock \DOIprefix\doi{https://doi.org/10.1103/RevModPhys.65.851}.
\bibitem[{Czir$\acute{\mbox{o}}$k et~al.(2000)Czir$\acute{\mbox{o}}$k,
  J$\acute{\mbox{a}}$nosi and Kessler}]{Czirok_et_al_2000}
\bibinfo{author}{Czir$\acute{\mbox{o}}$k, A.},
  \bibinfo{author}{J$\acute{\mbox{a}}$nosi, I.M.}, \bibinfo{author}{Kessler,
  J.O.}, \bibinfo{year}{2000}.
\newblock \bibinfo{title}{Bioconvective dynamics : dependence on organism
  behaviour}.
\newblock \bibinfo{journal}{J. Exp. Biol.} \bibinfo{volume}{203},
  \bibinfo{pages}{3345--3354}.
\newblock \DOIprefix\doi{https://doi.org/10.1242/jeb.203.21.3345}.
\bibitem[{Geng and Kuznetsov(2005)}]{Geng&Kuznetsov_2005}
\bibinfo{author}{Geng, P.}, \bibinfo{author}{Kuznetsov, A.V.},
  \bibinfo{year}{2005}.
\newblock \bibinfo{title}{Settling of bidispersed small solid particles in a
  dilute suspension containing gyrotactic micro--organisms}.
\newblock \bibinfo{journal}{Int J Eng Sci} \bibinfo{volume}{43},
  \bibinfo{pages}{992--1010}.
\newblock \DOIprefix\doi{https://doi.org/10.1016/j.ijengsci.2005.03.002}.
\bibitem[{Ghorai and Hill(2000)}]{Ghorai&Hill_2000}
\bibinfo{author}{Ghorai, S.}, \bibinfo{author}{Hill, N.A.},
  \bibinfo{year}{2000}.
\newblock \bibinfo{title}{Wavelengths of gyrotactic plumes in bioconvection}.
\newblock \bibinfo{journal}{Bull. Math. Biol.} \bibinfo{volume}{62},
  \bibinfo{pages}{429--450}.
\newblock \DOIprefix\doi{https://doi.org/10.1006/bulm.1999.0160}.
\bibitem[{Ghorai and Hill(2002)}]{Ghorai&Hill_2002}
\bibinfo{author}{Ghorai, S.}, \bibinfo{author}{Hill, N.A.},
  \bibinfo{year}{2002}.
\newblock \bibinfo{title}{Axisymmetric bioconvection in a cylinder}.
\newblock \bibinfo{journal}{J. Theor. Biol.} \bibinfo{volume}{219},
  \bibinfo{pages}{137--152}.
\newblock \DOIprefix\doi{https://doi.org/10.1006/jtbi.2002.3077}.
\bibitem[{Hart and Edwards(1987)}]{Hart&Edwards_1987}
\bibinfo{author}{Hart, A.}, \bibinfo{author}{Edwards, C.},
  \bibinfo{year}{1987}.
\newblock \bibinfo{title}{Buoyant density fluctuations during the cell cycle of
  {\it bacillus subtilis}}.
\newblock \bibinfo{journal}{Arch. Microbiol.} \bibinfo{volume}{147},
  \bibinfo{pages}{68--72}.
\newblock \DOIprefix\doi{https://doi.org/10.1007/BF00492907}.
\bibitem[{Hillesdon and Pedley(1996)}]{Hillesdon&Pedley_1996}
\bibinfo{author}{Hillesdon, A.J.}, \bibinfo{author}{Pedley, T.J.},
  \bibinfo{year}{1996}.
\newblock \bibinfo{title}{Bioconvection in suspensions of oxytactic bacteria:
  {L}inear theory}.
\newblock \bibinfo{journal}{J. Fluid Mech.} \bibinfo{volume}{324},
  \bibinfo{pages}{223--259}.
\newblock \DOIprefix\doi{https://doi.org/10.1017/S0022112096007902}.
\bibitem[{Hillesdon et~al.(1995)Hillesdon, Pedley and
  Kessler}]{Hillesdon_et_al_1995}
\bibinfo{author}{Hillesdon, A.J.}, \bibinfo{author}{Pedley, T.J.},
  \bibinfo{author}{Kessler, J.O.}, \bibinfo{year}{1995}.
\newblock \bibinfo{title}{The development of concentration gradients in a
  suspension of chemotactic bacteria}.
\newblock \bibinfo{journal}{Bull. Math. Biol.} \bibinfo{volume}{57},
  \bibinfo{pages}{299--344}.
\newblock \DOIprefix\doi{https://doi.org/10.1007/BF02460620}.
\bibitem[{Hirooka and Nagase(2003)}]{Hirooka&Nagase_2003}
\bibinfo{author}{Hirooka, T.}, \bibinfo{author}{Nagase, H.},
  \bibinfo{year}{2003}.
\newblock \bibinfo{title}{Biodegradation of endocrine disrupting chemicals and
  its application for bioremediation --utilization of photoautotrophic
  microorganisms--}.
\newblock \bibinfo{journal}{J. Environ. Biotechnol.} \bibinfo{volume}{3},
  \bibinfo{pages}{23--32}.
\newblock \bibinfo{note}{(in Japanese)}.
\bibitem[{Itoh et~al.(2001)Itoh, Toida and Saotome}]{Itoh_et_al_2001}
\bibinfo{author}{Itoh, A.}, \bibinfo{author}{Toida, H.},
  \bibinfo{author}{Saotome, Y.}, \bibinfo{year}{2001}.
\newblock \bibinfo{title}{Control of bioconvection and it's application for
  mechanical system (1st report, basic effects of electrical field on
  bioconvection)}.
\newblock \bibinfo{journal}{JSME, Ser. B} \bibinfo{volume}{67},
  \bibinfo{pages}{2449--2454}.
\newblock \DOIprefix\doi{https://doi.org/10.1299/kikaib.67.2449}.
  \bibinfo{note}{(in Japanese)}.
\bibitem[{Itoh et~al.(2006)Itoh, Toida and Saotome}]{Itoh_et_al_2006}
\bibinfo{author}{Itoh, A.}, \bibinfo{author}{Toida, H.},
  \bibinfo{author}{Saotome, Y.}, \bibinfo{year}{2006}.
\newblock \bibinfo{title}{Control of bioconvection and it's application for
  mechanical system (2nd report, pitch optimization of electrodes array and
  driving of mechanical system by controlled bioconvection)}.
\newblock \bibinfo{journal}{JSME, Ser. B} \bibinfo{volume}{72},
  \bibinfo{pages}{972--978}.
\newblock \DOIprefix\doi{https://doi.org/10.1299/kikaib.72.972}.
  \bibinfo{note}{(in Japanese)}.
\bibitem[{J$\acute{\mbox{a}}$nosi et~al.(1998)J$\acute{\mbox{a}}$nosi, Kessler
  and Horv$\acute{\mbox{a}}$th}]{Janosi_et_al_1998}
\bibinfo{author}{J$\acute{\mbox{a}}$nosi, I.M.}, \bibinfo{author}{Kessler,
  J.O.}, \bibinfo{author}{Horv$\acute{\mbox{a}}$th, V.K.},
  \bibinfo{year}{1998}.
\newblock \bibinfo{title}{Onset of bioconvection in suspensions of {\it
  {bacillus} subtilis}}.
\newblock \bibinfo{journal}{Phys. Rev. E} \bibinfo{volume}{58},
  \bibinfo{pages}{4793--4800}.
\newblock \DOIprefix\doi{https://link.aps.org/doi/10.1103/PhysRevE.58.4793}.
\bibitem[{Kage et~al.(2013)Kage, Hosoya, Baba and Mogami}]{Kage_et_al_2013}
\bibinfo{author}{Kage, A.}, \bibinfo{author}{Hosoya, C.},
  \bibinfo{author}{Baba, S.A.}, \bibinfo{author}{Mogami, Y.},
  \bibinfo{year}{2013}.
\newblock \bibinfo{title}{Drastic reorganization of bioconvection pattern of
  {\it{c}hlamydomonas}: {Q}uantitative analysis of the pattern transition
  response}.
\newblock \bibinfo{journal}{J. Exp. Biol.} \bibinfo{volume}{216},
  \bibinfo{pages}{4557--4566}.
\newblock \DOIprefix\doi{https://doi.org/10.1242/jeb.092791}.
\bibitem[{Karimi and Paul(2013)}]{Karimi&Paul_2013}
\bibinfo{author}{Karimi, A.}, \bibinfo{author}{Paul, M.R.},
  \bibinfo{year}{2013}.
\newblock \bibinfo{title}{Bioconvection in spatially extended domains}.
\newblock \bibinfo{journal}{Phys. Rev. E} \bibinfo{volume}{87},
  \bibinfo{pages}{053016}.
\newblock \DOIprefix\doi{https://link.aps.org/doi/10.1103/PhysRevE.87.053016}.
\bibitem[{Khan et~al.(2020)Khan, Salahuddin, Malik, Alqarni and
  Alqahtani}]{Khan_et_al_2020}
\bibinfo{author}{Khan, M.}, \bibinfo{author}{Salahuddin, T.},
  \bibinfo{author}{Malik, M.Y.}, \bibinfo{author}{Alqarni, M.S.},
  \bibinfo{author}{Alqahtani, A.M.}, \bibinfo{year}{2020}.
\newblock \bibinfo{title}{Numerical modeling and analysis of bioconvection on
  {MHD} flow due to an upper paraboloid surface of revolution}.
\newblock \bibinfo{journal}{Physica A} \bibinfo{volume}{553},
  \bibinfo{pages}{124231}.
\newblock \DOIprefix\doi{https://doi.org/10.1016/j.physa.2020.124231}.
\bibitem[{Koschmieder and Switzer(1992)}]{Koschmieder&Switzer_1992}
\bibinfo{author}{Koschmieder, E.L.}, \bibinfo{author}{Switzer, D.W.},
  \bibinfo{year}{1992}.
\newblock \bibinfo{title}{The wavenumbers of supercritical
  surface--tension--driven {B}$\acute{\mbox{e}}$nard convection}.
\newblock \bibinfo{journal}{J. Fluid Mech.} \bibinfo{volume}{240},
  \bibinfo{pages}{533--548}.
\newblock \DOIprefix\doi{https://doi.org/10.1017/S0022112092000181}.
\bibitem[{Kuznetsov(2005)}]{Kuznetsov_2005a}
\bibinfo{author}{Kuznetsov, A.V.}, \bibinfo{year}{2005}.
\newblock \bibinfo{title}{The onset of bioconvection in a suspension of
  negatively geotactic microorganisms with highfrequency vertical vibration}.
\newblock \bibinfo{journal}{Int. Commun. Heat Mass Transf.}
  \bibinfo{volume}{32}, \bibinfo{pages}{1119--1127}.
\newblock
  \DOIprefix\doi{https://doi.org/10.1016/j.icheatmasstransfer.2005.05.004}.
\bibitem[{Kuznetsov(2011)}]{Kuznetsov_2011}
\bibinfo{author}{Kuznetsov, A.V.}, \bibinfo{year}{2011}.
\newblock \bibinfo{title}{Nanofluid bioconvection: interaction of
  microorganisms oxytactic upswimming, nanoparticle distribution, and
  heating/cooling from below}.
\newblock \bibinfo{journal}{Theor. Comput. Fluid Dyn.} \bibinfo{volume}{26},
  \bibinfo{pages}{291--310}.
\newblock \DOIprefix\doi{https://doi.org/10.1007/s00162-011-0230-1}.
\bibitem[{Lee and Kim(2015)}]{Lee&Kim_2015}
\bibinfo{author}{Lee, H.G.}, \bibinfo{author}{Kim, J.}, \bibinfo{year}{2015}.
\newblock \bibinfo{title}{Numerical investigation of falling bacterial plumes
  caused by bioconvection in a three-dimensional chamber}.
\newblock \bibinfo{journal}{Eur. J. Mech. B Fluids} \bibinfo{volume}{52},
  \bibinfo{pages}{120--130}.
\newblock \DOIprefix\doi{https://doi.org/10.1016/j.euromechflu.2015.03.002}.
\bibitem[{Matsumoto and Nishimura(1998)}]{Matsumoto&Nishimura_1998}
\bibinfo{author}{Matsumoto, M.}, \bibinfo{author}{Nishimura, T.},
  \bibinfo{year}{1998}.
\newblock \bibinfo{title}{Mersenne twister: {A} 623--dimensionally
  equidistributed uniform pseudo--random number generator}.
\newblock \bibinfo{journal}{ACM Trans. Model. Comput. Simul.}
  \bibinfo{volume}{8}, \bibinfo{pages}{3--30}.
\newblock \DOIprefix\doi{https://doi.org/10.1145/272991.272995}.
\bibitem[{Mazzoni et~al.(2008)Mazzoni, Giavazzi, Cerbino, Giglio and
  Vailati}]{Mazzoni_et_al_2008}
\bibinfo{author}{Mazzoni, S.}, \bibinfo{author}{Giavazzi, F.},
  \bibinfo{author}{Cerbino, R.}, \bibinfo{author}{Giglio, M.},
  \bibinfo{author}{Vailati, A.}, \bibinfo{year}{2008}.
\newblock \bibinfo{title}{Mutual voronoi tessellation in spoke pattern
  convection}.
\newblock \bibinfo{journal}{Phys. Rev. Lett.} \bibinfo{volume}{100},
  \bibinfo{pages}{188104}.
\newblock
  \DOIprefix\doi{https://link.aps.org/doi/10.1103/PhysRevLett.100.188104}.
\bibitem[{Metcalfe and Pedley(1998)}]{Metcalfe&Pedley_1998}
\bibinfo{author}{Metcalfe, A.M.}, \bibinfo{author}{Pedley, T.J.},
  \bibinfo{year}{1998}.
\newblock \bibinfo{title}{Bacterial bioconvection: {W}eakly nonlinear theory
  for pattern selection}.
\newblock \bibinfo{journal}{J. Fluid Mech.} \bibinfo{volume}{370},
  \bibinfo{pages}{249--270}.
\newblock \DOIprefix\doi{https://doi.org/10.1017/s0022112098001979}.
\bibitem[{Metcalfe and Pedley(2001)}]{Metcalfe&Pedley_2001}
\bibinfo{author}{Metcalfe, A.M.}, \bibinfo{author}{Pedley, T.J.},
  \bibinfo{year}{2001}.
\newblock \bibinfo{title}{Falling plumes in bacterial bioconvection}.
\newblock \bibinfo{journal}{J. Fluid Mech.} \bibinfo{volume}{445},
  \bibinfo{pages}{121--149}.
\newblock \DOIprefix\doi{https://doi.org/10.1017/s0022112001005547}.
\bibitem[{Naseem et~al.(2017)Naseem, Shafiq, Zhao and
  Naseem}]{Naseem_et_al_2017}
\bibinfo{author}{Naseem, F.}, \bibinfo{author}{Shafiq, A.},
  \bibinfo{author}{Zhao, L.}, \bibinfo{author}{Naseem, A.},
  \bibinfo{year}{2017}.
\newblock \bibinfo{title}{{MHD} biconvective flow of {P}owell {E}yring
  nanofluid over stretched surface}.
\newblock \bibinfo{journal}{AIP Adv.} \bibinfo{volume}{7},
  \bibinfo{pages}{065013}.
\newblock \DOIprefix\doi{https://doi.org/10.1063/1.4983014}.
\bibitem[{Noever and Matsos(1991a)}]{Noever&Matsos_1991a}
\bibinfo{author}{Noever, D.A.}, \bibinfo{author}{Matsos, H.C.},
  \bibinfo{year}{1991}a.
\newblock \bibinfo{title}{A bioassay for monitoring cadmium based on
  bioconvective patterns}.
\newblock \bibinfo{journal}{J. Environ. Sci. Health A} \bibinfo{volume}{26},
  \bibinfo{pages}{273--286}.
\newblock \DOIprefix\doi{https://doi.org/10.1080/10934529109375633}.
\bibitem[{Noever and Matsos(1991b)}]{Noever&Matsos_1991b}
\bibinfo{author}{Noever, D.A.}, \bibinfo{author}{Matsos, H.C.},
  \bibinfo{year}{1991}b.
\newblock \bibinfo{title}{Calcium protection from cadmium poisoning:
  {B}ioconvective indicators in {\it {t}etrahymena}}.
\newblock \bibinfo{journal}{J. Environ. Sci. Health A} \bibinfo{volume}{26},
  \bibinfo{pages}{1105--1113}.
\newblock \DOIprefix\doi{https://doi.org/10.1080/10934529109375689}.
\bibitem[{Noever et~al.(1992)Noever, Matsos and Looger}]{Noever_et_al_1992}
\bibinfo{author}{Noever, D.A.}, \bibinfo{author}{Matsos, H.C.},
  \bibinfo{author}{Looger, L.L.}, \bibinfo{year}{1992}.
\newblock \bibinfo{title}{Bioconvective indicators in {\it tetrahymena}:
  {N}ickel and copper protection from cadmium poisoning}.
\newblock \bibinfo{journal}{J. Environ. Sci. Health A} \bibinfo{volume}{27},
  \bibinfo{pages}{403--417}.
\newblock \DOIprefix\doi{https://doi.org/10.1080/10934529209375735}.
\bibitem[{Omori et~al.(2003)Omori, Habe, Yoshida, Horinouchi, Saiki and
  Nojiri}]{Omori_et_al_2003}
\bibinfo{author}{Omori, T.}, \bibinfo{author}{Habe, H.},
  \bibinfo{author}{Yoshida, T.}, \bibinfo{author}{Horinouchi, M.},
  \bibinfo{author}{Saiki, Y.}, \bibinfo{author}{Nojiri, H.},
  \bibinfo{year}{2003}.
\newblock \bibinfo{title}{Applied ecological microbiology}.
\newblock \bibinfo{publisher}{Shoko--do}.
\newblock \bibinfo{note}{(in Japanese)}.
\bibitem[{Omori et~al.(2002)Omori, Nojiri, Horinouchi and
  Kasuga}]{Omori_et_al_2000}
\bibinfo{author}{Omori, T.}, \bibinfo{author}{Nojiri, H.},
  \bibinfo{author}{Horinouchi, M.}, \bibinfo{author}{Kasuga, K.},
  \bibinfo{year}{2002}.
\newblock \bibinfo{title}{Environmental {B}iotechnology}.
\newblock \bibinfo{edition}{3} ed., \bibinfo{publisher}{Shoko--do}.
\newblock \bibinfo{note}{(in Japanese)}.
\bibitem[{Pedley et~al.(1988)Pedley, Hill and Kessler}]{Pedley_et_al_1988}
\bibinfo{author}{Pedley, T.J.}, \bibinfo{author}{Hill, N.A.},
  \bibinfo{author}{Kessler, J.O.}, \bibinfo{year}{1988}.
\newblock \bibinfo{title}{The growth of bioconvection patterns in a uniform
  suspension of gyrotactic micro--organisms}.
\newblock \bibinfo{journal}{J. Fluid Mech.} \bibinfo{volume}{195},
  \bibinfo{pages}{223--237}.
\newblock \DOIprefix\doi{https://doi.org/10.1017/s0022112088002393}.
\bibitem[{Pedley and Kessler(1992)}]{Pedley&Kessler_1992}
\bibinfo{author}{Pedley, T.J.}, \bibinfo{author}{Kessler, J.O.},
  \bibinfo{year}{1992}.
\newblock \bibinfo{title}{Hydrodynamic phenomena in suspensions of swimming
  microorganisms}.
\newblock \bibinfo{journal}{Annu. Rev. Fluid Mech.} \bibinfo{volume}{24},
  \bibinfo{pages}{313--358}.
\newblock
  \DOIprefix\doi{https://www.annualreviews.org/doi/abs/10.1146/annurev.fl.24.010192.001525}.
\bibitem[{Platt(1961)}]{Platt_1961}
\bibinfo{author}{Platt, J.R.}, \bibinfo{year}{1961}.
\newblock \bibinfo{title}{Bioconvection patterns in cultures of free--swimming
  organisms}.
\newblock \bibinfo{journal}{Science} \bibinfo{volume}{133},
  \bibinfo{pages}{1766--1767}.
\newblock \DOIprefix\doi{https://doi.org/10.1126/science.133.3466.1766}.
\bibitem[{Shi et~al.(2021)Shi, Hamid, Khan, Kumar, Gowda, Prasannakumara, Shah,
  Khan,  and Chung}]{Shi_et_al_2021}
\bibinfo{author}{Shi, Q.H.}, \bibinfo{author}{Hamid, A.},
  \bibinfo{author}{Khan, M.I.}, \bibinfo{author}{Kumar, R.N.},
  \bibinfo{author}{Gowda, R.J.P.}, \bibinfo{author}{Prasannakumara, B.C.},
  \bibinfo{author}{Shah, N.A.}, \bibinfo{author}{Khan, S.U.}, ,
  \bibinfo{author}{Chung, J.D.}, \bibinfo{year}{2021}.
\newblock \bibinfo{title}{Numerical study of bio-convection flow of
  magneto-cross nanofluid containing gyrotactic microorganisms with activation
  energy}.
\newblock \bibinfo{journal}{Sci. Rep.} \bibinfo{volume}{11},
  \bibinfo{pages}{16030}.
\newblock \DOIprefix\doi{https://doi.org/10.1038/s41598-021-95587-2}.
\bibitem[{Tomita and Abe(2000)}]{Tomita&Abe_2000}
\bibinfo{author}{Tomita, H.}, \bibinfo{author}{Abe, K.}, \bibinfo{year}{2000}.
\newblock \bibinfo{title}{Numerical simulation of pattern formation in the
  {B}{\'{e}}nard{\textendash}{M}arangoni convection}.
\newblock \bibinfo{journal}{Phys. Fluids} \bibinfo{volume}{12},
  \bibinfo{pages}{1389--1400}.
\newblock \DOIprefix\doi{https://doi.org/10.1063/1.870390}.
\bibitem[{Tuval et~al.(2005)Tuval, Cisneros, Dombrowski, Wolgemuth, Kessler and
  Goldstein}]{Tuval_et_al_2005}
\bibinfo{author}{Tuval, I.}, \bibinfo{author}{Cisneros, L.},
  \bibinfo{author}{Dombrowski, C.}, \bibinfo{author}{Wolgemuth, C.W.},
  \bibinfo{author}{Kessler, J.O.}, \bibinfo{author}{Goldstein, R.E.},
  \bibinfo{year}{2005}.
\newblock \bibinfo{title}{Bacterial swimming and oxygen transport near contact
  lines}.
\newblock \bibinfo{journal}{Proc. Natl. Acad. Sci. U.S.A.}
  \bibinfo{volume}{102}, \bibinfo{pages}{2277--2282}.
\newblock \DOIprefix\doi{https://doi.org/10.1073/pnas.0406724102}.
\bibitem[{Uddin et~al.(2016)Uddin, Kabir and B\'{e}g}]{Uddin_et_al_2016}
\bibinfo{author}{Uddin, M.J.}, \bibinfo{author}{Kabir, M.N.},
  \bibinfo{author}{B\'{e}g, O.A.}, \bibinfo{year}{2016}.
\newblock \bibinfo{title}{Computational investigation of {S}tefan blowing and
  multiple-slip effects on buoyancy-driven bioconvection nanofluid flow with
  microorganisms}.
\newblock \bibinfo{journal}{Int. J. Heat Mass Transf.} \bibinfo{volume}{95},
  \bibinfo{pages}{116--130}.
\newblock
  \DOIprefix\doi{http://dx.doi.org/10.1016/j.ijheatmasstransfer.2015.11.015}.
\bibitem[{{{V}an der Pol} and Tramper(1998)}]{Pol&Tramper_1998}
\bibinfo{author}{{{V}an der Pol}, L.}, \bibinfo{author}{Tramper, J.},
  \bibinfo{year}{1998}.
\newblock \bibinfo{title}{Shear sensitivity of animal cells from a
  culture--medium perspective}.
\newblock \bibinfo{journal}{Trends Biotechnol.} \bibinfo{volume}{16},
  \bibinfo{pages}{323--328}.
\newblock \DOIprefix\doi{https://doi.org/10.1016/S0167-7799(98)01209-8}.
\bibitem[{Williams and Bees(2011)}]{Williams&Bees_2011}
\bibinfo{author}{Williams, C.R.}, \bibinfo{author}{Bees, M.A.},
  \bibinfo{year}{2011}.
\newblock \bibinfo{title}{A tale of three taxes: {P}hoto-gyro-gravitactic
  bioconvection}.
\newblock \bibinfo{journal}{J. Exp. Biol.} \bibinfo{volume}{214},
  \bibinfo{pages}{2398--2408}.
\newblock \DOIprefix\doi{https://doi.org/10.1242/jeb.051094}.
\bibitem[{Yamamoto et~al.(1992)Yamamoto, Okayama, Sato and
  Takaoki}]{Yamamoto_et_al_1992}
\bibinfo{author}{Yamamoto, Y.}, \bibinfo{author}{Okayama, T.},
  \bibinfo{author}{Sato, K.}, \bibinfo{author}{Takaoki, T.},
  \bibinfo{year}{1992}.
\newblock \bibinfo{title}{Relation of pattern formation to external conditions
  in the flagellate, {\it chlamydomonas reinhardtii}}.
\newblock \bibinfo{journal}{Eur. J. Protistol.} \bibinfo{volume}{28},
  \bibinfo{pages}{415--420}.
\newblock \DOIprefix\doi{https://doi.org/10.1016/S0932-4739(11)80005-2}.
\bibitem[{Yanaoka et~al.(2007)Yanaoka, Inamura and Suzuki}]{Yanaoka_et_al_2007}
\bibinfo{author}{Yanaoka, H.}, \bibinfo{author}{Inamura, T.},
  \bibinfo{author}{Suzuki, K.}, \bibinfo{year}{2007}.
\newblock \bibinfo{title}{Numerical analysis of bioconvection generated by
  chemotactic bacteria}.
\newblock \bibinfo{journal}{JSME, Ser. B} \bibinfo{volume}{73},
  \bibinfo{pages}{575--580}.
\newblock \DOIprefix\doi{https://doi.org/10.1299/kikaib.73.575}.
  \bibinfo{note}{(in Japanese)}.
\bibitem[{Yanaoka et~al.(2008)Yanaoka, Inamura and Suzuki}]{Yanaoka_et_al_2008}
\bibinfo{author}{Yanaoka, H.}, \bibinfo{author}{Inamura, T.},
  \bibinfo{author}{Suzuki, K.}, \bibinfo{year}{2008}.
\newblock \bibinfo{title}{Three dimensional numerical analysis of bioconvection
  generated by chemotactic bacteria}.
\newblock \bibinfo{journal}{JSME, Ser. B} \bibinfo{volume}{74},
  \bibinfo{pages}{135--141}.
\newblock \DOIprefix\doi{https://doi.org/10.1299/kikaib.74.135}.
  \bibinfo{note}{(in Japanese)}.
\bibitem[{Zadeha et~al.(2020)Zadeha, Mehryanb, Sheremetc, Izadid and
  Ghodrate}]{Zadeha_et_al_2020}
\bibinfo{author}{Zadeha, S.M.H.}, \bibinfo{author}{Mehryanb, S.},
  \bibinfo{author}{Sheremetc, M.A.}, \bibinfo{author}{Izadid, M.},
  \bibinfo{author}{Ghodrate, M.}, \bibinfo{year}{2020}.
\newblock \bibinfo{title}{Numerical study of mixed bio-convection associated
  with a micropolar fluid}.
\newblock \bibinfo{journal}{Therm. Sci. Eng. Prog.} \bibinfo{volume}{18},
  \bibinfo{pages}{100539}.
\newblock \DOIprefix\doi{https://doi.org/10.1016/j.tsep.2020.100539}.

\end{thebibliography}

\end{document}